\documentclass{iopart}
\usepackage{graphicx}
\usepackage{harvard}
\usepackage{url}
\newcommand{\xmax}{\ensuremath{X_{\rm max}}}

\begin{document}
\topical{Cosmic Rays: The Second Knee and Beyond}
\author{Douglas R Bergman}
\address{Rutgers - The State University of New Jersey, Department of
  Physics and Astronomy, Piscataway, New Jersey, USA}
\ead{bergman@physics.rutgers.edu}

\author{John W Belz}
\address{University of Utah, Department of Physics, Salt Lake City,
  Utah, USA}
\ead{belz@cosmic.utah.edu}

\begin{abstract}
  We conduct a review of experimental results on Ultra-High Energy
  Cosmic Rays (UHECR's) including measurements of the features of the
  spectrum, the composition of the primary particle flux and the
  search for anisotropy in event arrival direction.  We find that
  while there is a general consensus on the features in the spectrum
  --- the Second Knee, the Ankle, and (to a lesser extent) the GZK
  Cutoff --- there is little consensus on the composition of the
  primaries that accompany these features.  This lack of consensus on
  the composition makes interpretation of the agreed upon features
  problematic. There is also little direct evidence about potential
  sources of UHECRs, as early reports of arrival direction
  anisotropies have not been confirmed in independent measurements.
\end{abstract}

\section{Introduction}
Ultra-High Energy Cosmic Rays (UHECR's) are the most energetic form of
radiation known to hit the earth.  At these energies, above $10^{18}$
eV, one would like to understand the workings of the astrophysical
accelerators which able to produce such high energies, energies many
orders of magnitude higher than what is available at terrestrial
accelerators.  These high energy particles also have, perhaps, much to
tell us about the regions in which they were accelerated and the vast
spaces through which they passed on their way to us.  However, before
we can understand, we must measure: what is the energy spectrum, what
kind of particles are they, where do they come from?

This will be a review of the latest experimental results from which
one may hope to understand UHECR's.  There has been significant
experimental activity since the last experimental review appeared in
this journal \cite{Yoshida-Dai-1998-JPG-24-905}, and there has been
some movement towards a consensus on the existence and energies of
various features in the UHECR spectrum.  There in not yet, however, a
consensus on the best way to interpret those features.

Cosmic rays were discovered by \citeasnoun{Hess-1912-ZP-13-1084},
using the fact that the ionization of air increases with altitude.
This cosmic radiation was later directly observed in cloud chambers.
Pierre Auger \cite{Auger-1939-RMP-11-288} observed coincident hits over
a wide area with his detectors, showing that the primary cosmic rays
induce a cascade of particles, known as an extensive air shower (EAS),
when they encounter the atmosphere.  From the numbers of particles
involved in these showers, Auger was able to estimate the energy of
some showers, and show that some must be very energetic.  The
phenomenology of EAS was worked out by Heitler
\cite{Heitler-1938-PRSL-166-529} and others.

Several features have been identified or proposed in the spectrum of
cosmic rays.  The first of these features to be identified was a
softening of the spectrum at an energy of about $3\times10^{15}$ eV,
an energy below the range we consider in paper.  Since the flux bends
\emph{down} at this point, the feature was christend the Knee.  When
later another downturn at higher energy, about $4\times10^{17}$ eV,
was observed, it was naturally called the Second Knee.  This is at the
lower end of the energy range we consider in this review, the Ultra
High Energy range.

About a order-of-magnitude higher in energy than the Second Knee, the
spectrum becomes harder again.  Continuing with the anatomic analogy,
this is called the Ankle, because the bend is in the opposite sense
than the one seen at the Knee.  Finally, at yet another
order-of-magnitude higher in energy, one expects a drastic reduction
in the flux, the GZK Cutoff
\cite{Greisen-1966-PRL-16-748,Zatsepin-1966-JETPL-4-78}, due to energy
losses of the the cosmic rays in the cosmic microwave background
radiation during their long propagation to us.

The first dedicated experiment to measure the energy and flux of
UHECRs was built by \citeasnoun{Linsley-1963-PRL-10-146} at Volcano
Ranch.  Data from Volcano Ranch and other early experiments in covered
in other reviews on this subject.  In this review, we will cover only
those experimental results which were released since the review of
\citeasnoun{Yoshida-Dai-1998-JPG-24-905}, or those which still have a
significant impact on the world data set at a given energy.  This
includes results from Haverah Park, the HiRes Prototype/MIA hybrid
experiment, the Akeno 1 km Array, Yakutsk, Fly's Eye, the Akeno Giant
Air Shower Array (AGASA), the High Resolution Fly's Eye (HiRes), and
the Pierre Auger Observatory.

\section{The Experiments}
The observable characteristics of an UHECR are its energy, the type of
particle it is and the direction from which it came.  All experiments
try to measure these characteristics, directly or indirectly, with
various degrees of precision.  Since the flux of UHECRs is so low,
direct measurement of these properties is impractical, so one must
measure the properties of the Extensive Air Shower (EAS) created by
the cosmic ray when it enters the atmosphere.  In the case of
measuring the primary energy and direction, the properties of the EAS
are a reasonable proxy for the properties of the primary cosmic ray.
All the energy of the primary goes into the shower and most of the
shower energy is deposited in the atmosphere.  Likewise, the momentum
of the primary particle is so much greater than any transverse
momentum generated in the shower that the shower points in the same
direction as the primary.  However, the primary particle type must be
inferred from the way the shower develops, which makes it hard to
determine the particle type on a shower-by-shower basis even for the
best measurements of EAS development.

There are two principal techniques used to detect and measure EAS's:
measure the density of shower particle at the ground, or measure the
amount of fluorescence light emitted by the atmosphere as the shower
passes through it. The two techniques can also be used together for
hybrid measurements.

Ground arrays sample the shower front at one level, with a sparse
array of detectors on the ground.  The detectors are typically either
slabs of scintillator or tanks of water, which have different
sensitivity to the particle components of the shower.  Scintillators
are primarily sensitive to electrons and photons, but appropriate
shielding can allow one to separate out the muon and hadron component
as well as differentiate the electrons from the photons.  Water tanks,
which detect shower particles by the Cerenkov radiation they emit
while passing through the detector, are much more sensitive to muons
than electromagnetic particles.  The shower direction is inferred from
the relative timing of the various detector elements as the shower
front sweeps across them, but the energy and composition of the shower
must be inferred indirectly from the size and shape of the shower
footprint.  The shower energy is usually determined by the shower
density at a given distance from the core (600 to 1000 m).  The energy
is roughly proportional to this density and the shower-to-shower
fluctuations are reduced at this distance.  However, one must take
into account the attenuation of the shower as showers at different
zenith angles have traversed different amounts of atmosphere.  Both
the normalization and the attenuation correction contribute to
systematic uncertainties and are dependent on shower development
modeling.  One can gain more information by having separate ground
stations which are sensitive to electrons or to muons or to photons,
but the model dependence is hard to avoid.  On the other hand, shower
arrays can run continuously and are mostly independent of the weather.

Fluorescence detectors collect the fluorescence light generated as the
shower particles excite the nitrogen in the air.  The amount of light
produced in this way is proportional to the primary energy, so this
technique provides a calorimetric measurement of the shower energy.
The direction of shower can be determined from the relative timing of
light arriving at the detector, though there are large correlations in
fitting between the distance to a shower and the angle it makes with
respect to the viewer.  A much better determination of the shower
geometry is available when one views a shower with two detectors
simultaneously which is known as stereo observation (as opposed to
monocular observation where one uses timing).  In this case, the
geometry of the shower is determined uniquely by the two
shower-detector planes.  The composition of the primary cosmic ray
determines the longitudinal development of the shower.  This
development is directly visible to the fluorescence detector and is
often abbreviated by measuring the depth in the atmosphere at which
the shower reached its maximum size, \xmax.  The draw back to using a
fluorescence detector is that it can only be operated on dark,
moonless nights.  In addition, one must control the level and
variation of aerosols in the atmosphere by choosing an appropriate
(desert) site.

\begin{figure}[h]
  \includegraphics[width=\columnwidth]{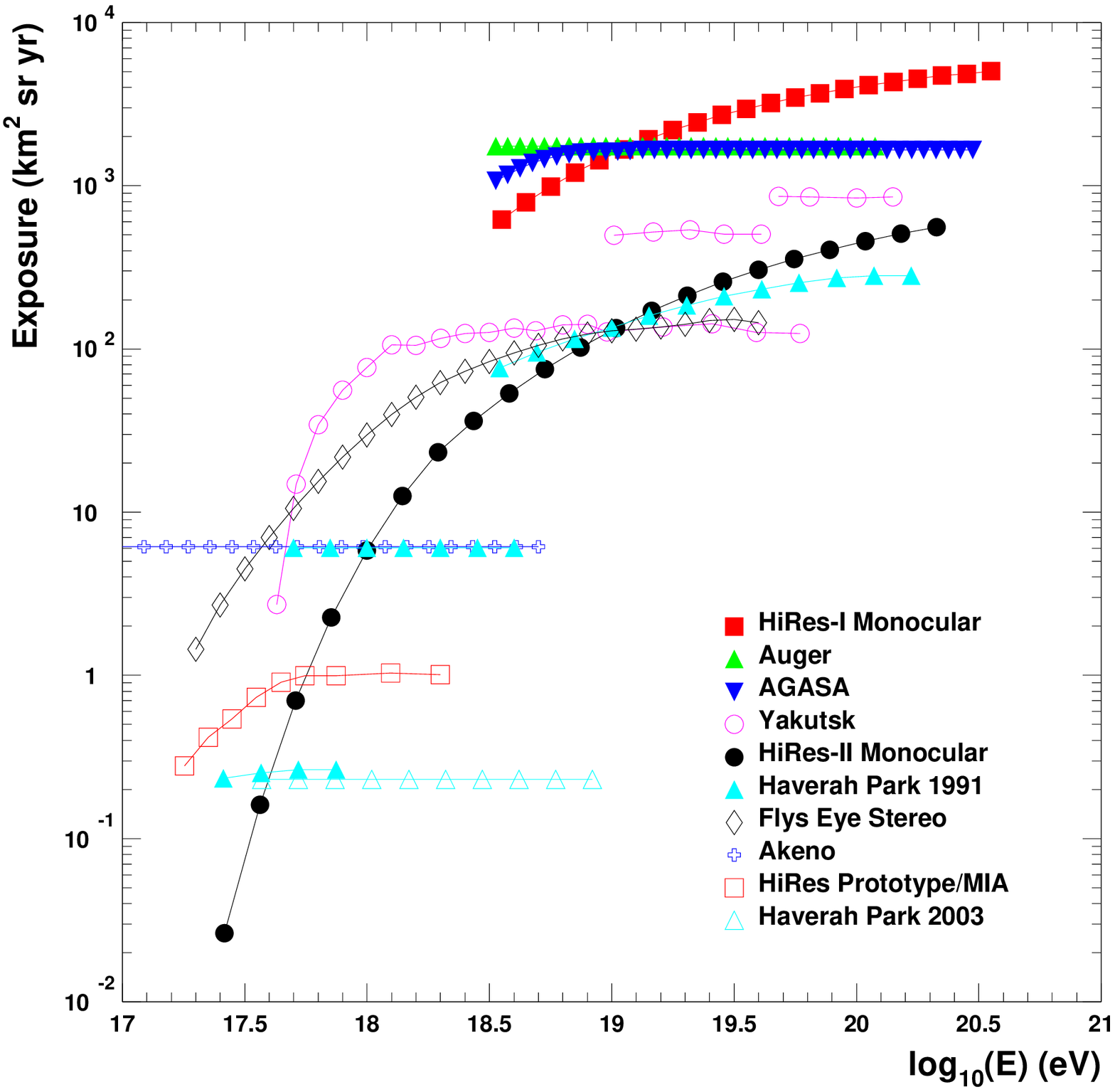}
  \caption{The exposures of all the experiments discussed in the
    text.  The data for the exposures come from the following papers:
    Haverah Park \cite{Lawrence-1991-JPG-17-733,Ave-2003-APP-19-47},
    HiRes Prototype/MIA \cite{AbuZayyad-2001-ApJ-557-686}, Akeno
    \cite{Nagano-1992-JPG-18-423}, Fly's Eye
    \cite{Bird-1994-ApJ-424-491}, Yakutsk (all arrays)
    \cite{Egorova-2004-NPBps-136-3}, AGASA
    \cite{Takeda-2003-APP-19-447}, Auger
    \cite{Sommers-2005-ICRC-abs1}, HiRes (both monocular measurements)
    \cite{Abbasi-2007-PRL}.}
  \label{all-exposures}
\end{figure}

We now discuss the various detectors roughly in order of exposure.
The exposures of all these experiments is shown as a function of
energy in Figure~\ref{all-exposures}.  In the cases where exposures
were not explicitly published by the experimental groups, they have
been inferred from the published flux values and uncertainties.

\subsection{Haverah Park}
\begin{figure}[h]
  \begin{center}
  \includegraphics[width=0.5\columnwidth]{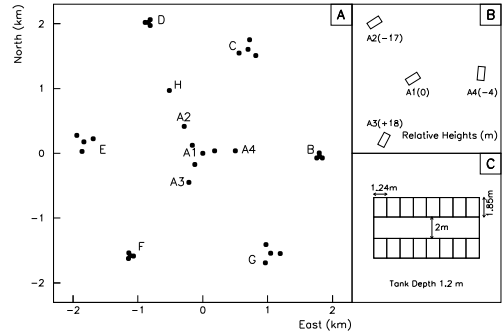}
  \end{center}
  \caption{Layout of the Haverah Park array. (A) The whole array.  (B)
    The orientation and relative heights of the detector huts A1--A4.
    (C) The arrangement of water tanks within one of the four main
    A-site huts.  Figure taken from \citeasnoun{Ave-2003-APP-19-61}.}
  \label{hp-array}
\end{figure}

Haverah Park was operated from 1962 to 1987, and was the pioneer in
using Cerenkov water tanks to sample the shower front at the ground.
The array was located at 54.0$^\circ$ N, 1.6$^\circ$ W, near Leeds
University, at an atmospheric depth of 1016 g/cm$^2$.  The tanks were
2.29 m$^2$ in area an 1.2 m deep \cite{Lawrence-1991-JPG-17-733}, with
the Cerenkov light collected by a single 5 inch diameter
photomultiplier tube suspended so that the photocathode was just in
contact with the water.  The tanks were then grouped together to make
large detection areas able to detect relatively small fluxes of
charged particles.  At the center of the array (see
Figure~\ref{hp-array}), a set of $4\times34{\rm\ m^2}$ (A1--A4)
detectors was used for triggering the rest of the detector.  Signals
above threshold (0.3 vem/m$^2$) in the central detector (A1) and at
least two of the other three A detectors (at 500 m) were required to
form a trigger.

Six groups of tanks (each $4\times13.5{\rm\ m}^2$ were placed in a
ring around the center tanks.  These tanks were used to constrain the
core position in large showers, and thus increase the aperture of the
entire detector.  An infill array of 30 1 m$^2$ tanks with a spacing of
about 150 m was also operated for several years in the area between
the central A tanks.  Data acquisition consisted of both photography
of oscilloscope traces (which allowed for pulse shape and rise time
analyses) and digital methods.

\citeasnoun{Yoshida-Dai-1998-JPG-24-905} discuss the ``final''
analysis of the Haverah Park array \cite{Lawrence-1991-JPG-17-733}.
Since that time, however, the Haverah Park data has been re-analyzed
using shower simulation code unavailable in 1991.  Results from this
re-analysis were published in \citeasnoun{Hinton-1999-ICRC-26-3-288},
\citeasnoun{Ave-2000-PRL-85-2244}, \citeasnoun{Ave-2001-ICRC-27-381},
\citeasnoun{Ave-2002-PRD-65-063007}, \citeasnoun{Ave-2003-APP-19-47},
and \citeasnoun{Ave-2003-APP-19-61}.

The new analysis uses QGSJet \cite{Kalmykov-1997-NPBps-52b-17} and
Corsika \cite{Heck-1998-FZKA-6019} to model the showers and GEANT
\cite{GEANT-1993} to model the detector response.  The re-analysis led
to a change in the relation between $\rho(600)$, the density of shower
muons at 600 m from the core, and the primary energy.  The new
relation leads to a reduction in the energy of about 30\%.  A change
was also made in the attenuation length used to convert the
$\rho(600)$ of inclined showers to that of the reference angle.  The
new attenuation length was lower because only events with
$\theta<45^\circ$ were used (as opposed to $\theta<60^\circ$ in the
original analysis).  More vertical showers have a larger electron
component, which implies a smaller attenuation length.

\subsection{The SUGAR Array}
\begin{figure}[h]
  \begin{center}
  \includegraphics[width=0.5\columnwidth]{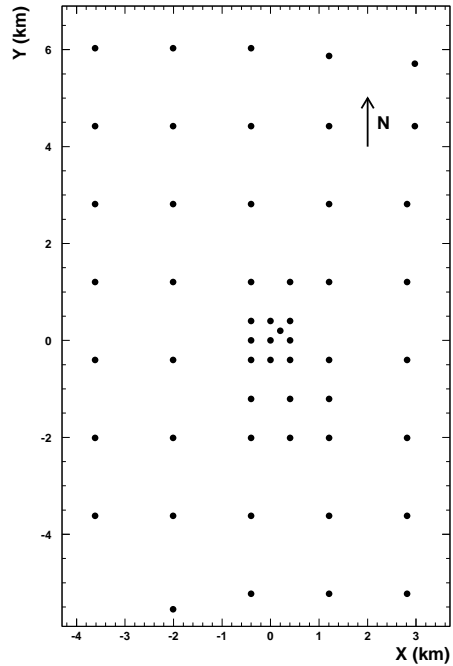}
  \end{center}
  \caption{Layout of the SUGAR Array.  Figure adapted from
    \citeasnoun{Winn-1986-JPG-12-653} The origin in the figure is
    located at (30.5$^\circ$ S, 149.6$^\circ$ E).}
  \label{sugar-layout}
\end{figure}

The Sydney University Giant Air shower Recorder (SUGAR) was opperated
in in Australia at (30.5$^\circ$ S, 149.6$^\circ$ E), from 1968 to
1979\cite{Winn-1986-JPG-12-653,Winn-1986-JPG-12-675}.  It consisted of
pairs of buried liquid scintillator tanks, with the pair separated by
50 m, on a mile (1600 m) square grid.  Each scintillator had an
effective area of 6.0 m$^2$.  We only comment on the SUGAR
measurements of anisotropy, because of problems with afterpulsing in
the photomulitplier tubes \cite{Nagano-Watson-2000-RMP-72-689}.  The
layout of the detector sites is shown in Figure~\ref{sugar-layout}.

\subsection{The HiRes Prototype/MIA Hybrid}
While the HiRes-Prototype/MIA experiment ran for only a relatively
short time, and has a relatively small exposure, it was the first
experiment to use both fluorescence and ground array measurements
simultaneously \cite{AbuZayyad-2000-NIMA-450-253}.  This hybrid
measurement pointed the way towards later experiments such as Auger.
It was formed by the fortuitous juxtiposition of the CASA-MIA array
nearing decomissioning, and the nascent High Resolution Fly's Eye
detector (see Section~\ref{review-exp-hires}).

\begin{figure}
  \begin{minipage}[t]{0.49\columnwidth}
    \includegraphics[width=\columnwidth]{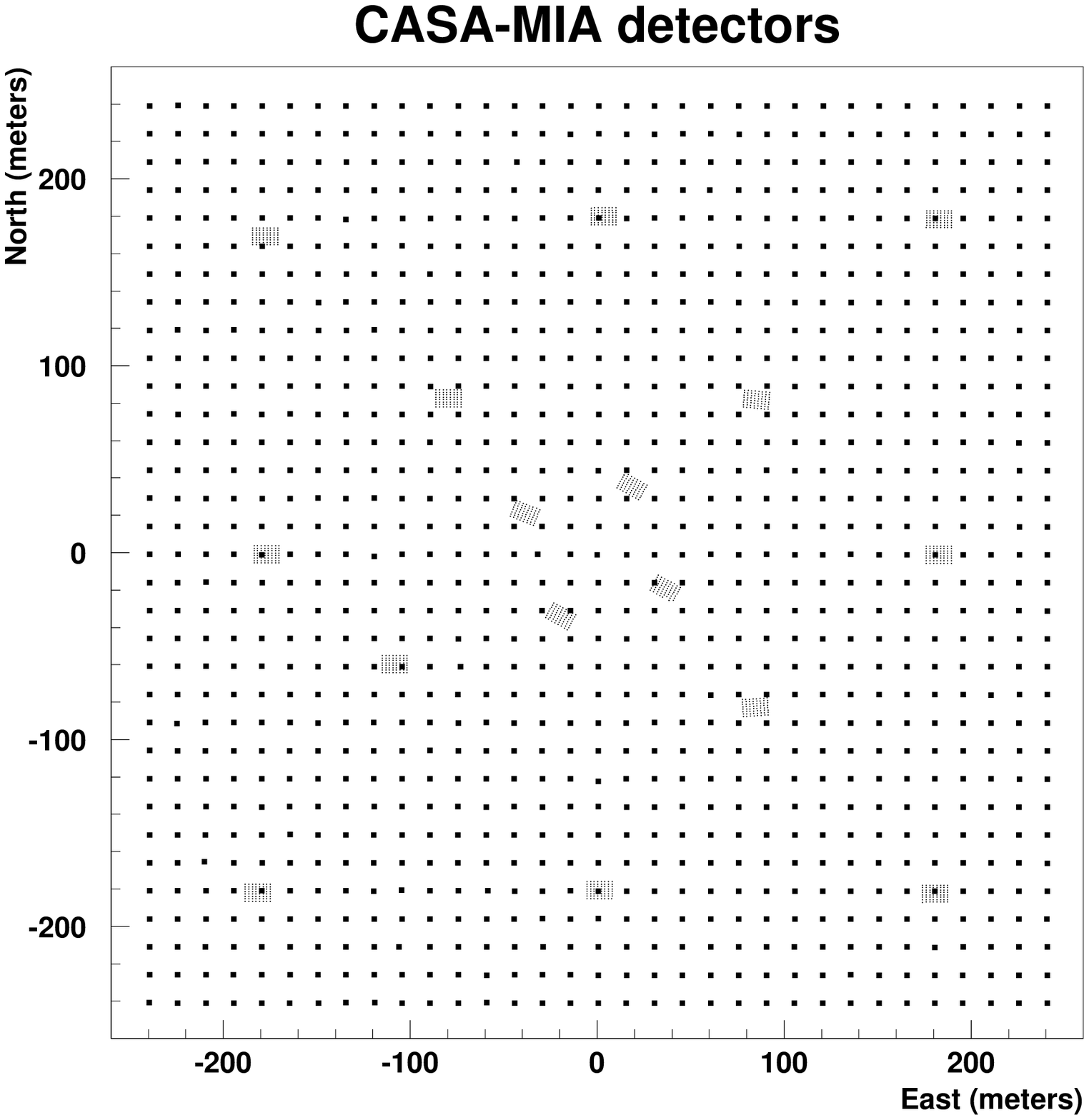}
  \end{minipage}
  \begin{minipage}[t]{0.49\columnwidth}
    \includegraphics[width=\columnwidth]{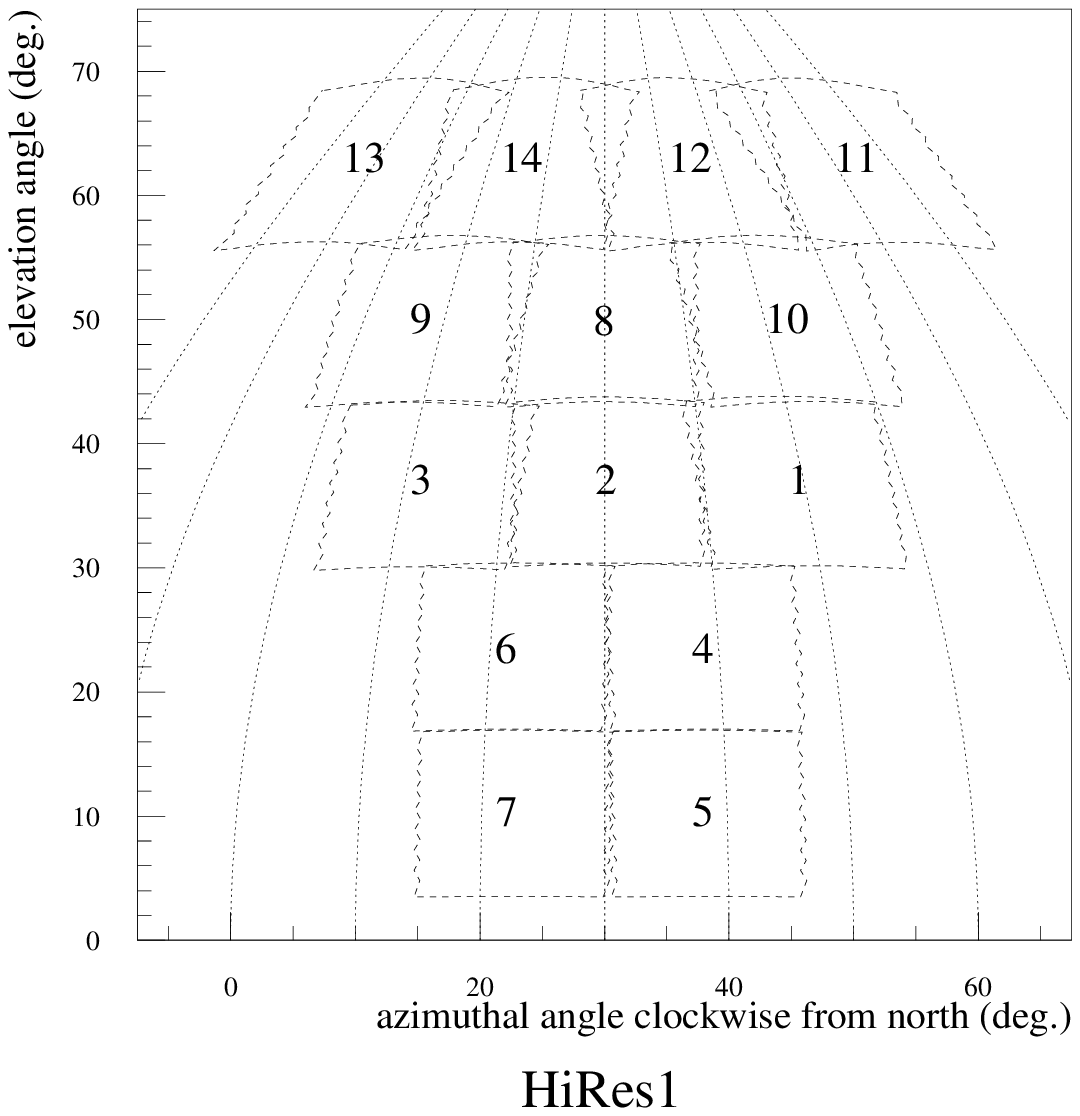}
  \end{minipage}
  \caption{Left: Layout of MIA scintillation patches (black
    rectangles) among the CASA detectors (small squares).  Right:
    Arrangement of the HiRes-Prototype mirrors as they view the sky.
    The MIA array is just below the junction between mirrors 5 and 7
    and extends for 4$^\circ$ on either side.}
  \label{hrmia-layout}
\end{figure}

The MIA array was an array of 16 patches of 64 scintillation counters
(see Figure~\ref{hrmia-layout}, left).  Each counter was $1.9{\rm\ 
  m}\times1.3{\rm\ m}$ and buried 3 m below the ground, making the
counter primarily sensitive to muons.  The total area covered by the
array was less than a quarter of a km$^2$, which limited the total
exposure available in hybrid \cite{Borione-1994-NIMA-346-329}.

The HiRes Prototype was an array of 14 fluorescence mirrors arranged
so the sky coverage formed a tower overlooking the MIA array, which
was 3.5 km to the NE (see Figure~\ref{hrmia-layout}, right).  Each
HiRes Prototype mirror covered a $16^\circ\times14^\circ$ of the sky
with 256 pixels.  Each pixel observed about one square degree on the
sky.  These mirrors were later rearranged and become the first part of
the HiRes detector (Section~\ref{review-exp-hires}).  The two
detectors were located at (40.2$^\circ$ N, 112.8$^\circ$ W), on Dugway
Proving Grounds in Utah, USA, at an atmospheric depth of 860 g/cm$^2$.
The detectors operated in hybrid mode from August 1993 until May 1996.

\subsection{Akeno}
\begin{figure}[h] 
  \begin{center}
    \includegraphics[width=\columnwidth]{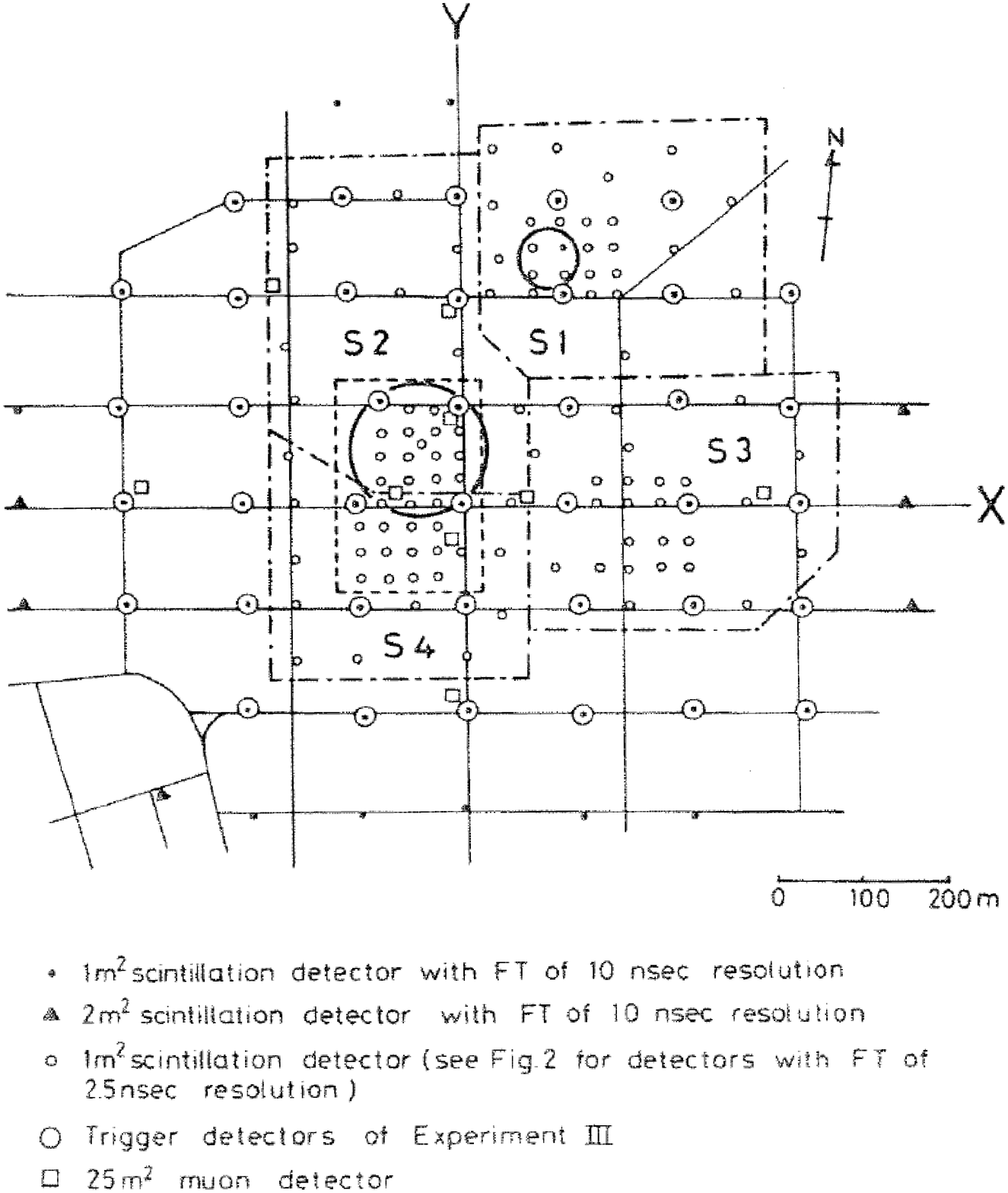}
  \end{center}
  \caption{The layout of scintillators in the Akeno 1 km array.  Figure
    taken from \citeasnoun{Nagano-1984-JPSJ-53-1667}.}
  \label{akeno-layout}
\end{figure}

The Akeno 1 km$^2$ array was an array of 156 scintillation counters,
each with an area of 1 or 2 m$^2$.  The primary spacing between the
counters was 120 m, but three regions had a smaller spacing of 30 m
(see Figure~\ref{akeno-layout}).  These closely spaced regions allowed
the measurement of showers over a wide range of energies.  The total
area of the array was about 1 km$^2$, but a fiducial area of only
$700{\rm\ m }\times 600{\rm\ m}$ was used for the spectrum
calculation.  The array was located near the Akeno Observatory
(35.8$^\circ$ N, 138.5$^\circ$ E) in Japan, at an atmospheric depth of
550 g/cm$^2$ \cite{Nagano-1992-JPG-18-423}, and was operated from the
late 1970's through 1990.

\subsection{Fly's Eye}
\begin{figure}[h] 
  \begin{center}
    \includegraphics[width=0.67\columnwidth]{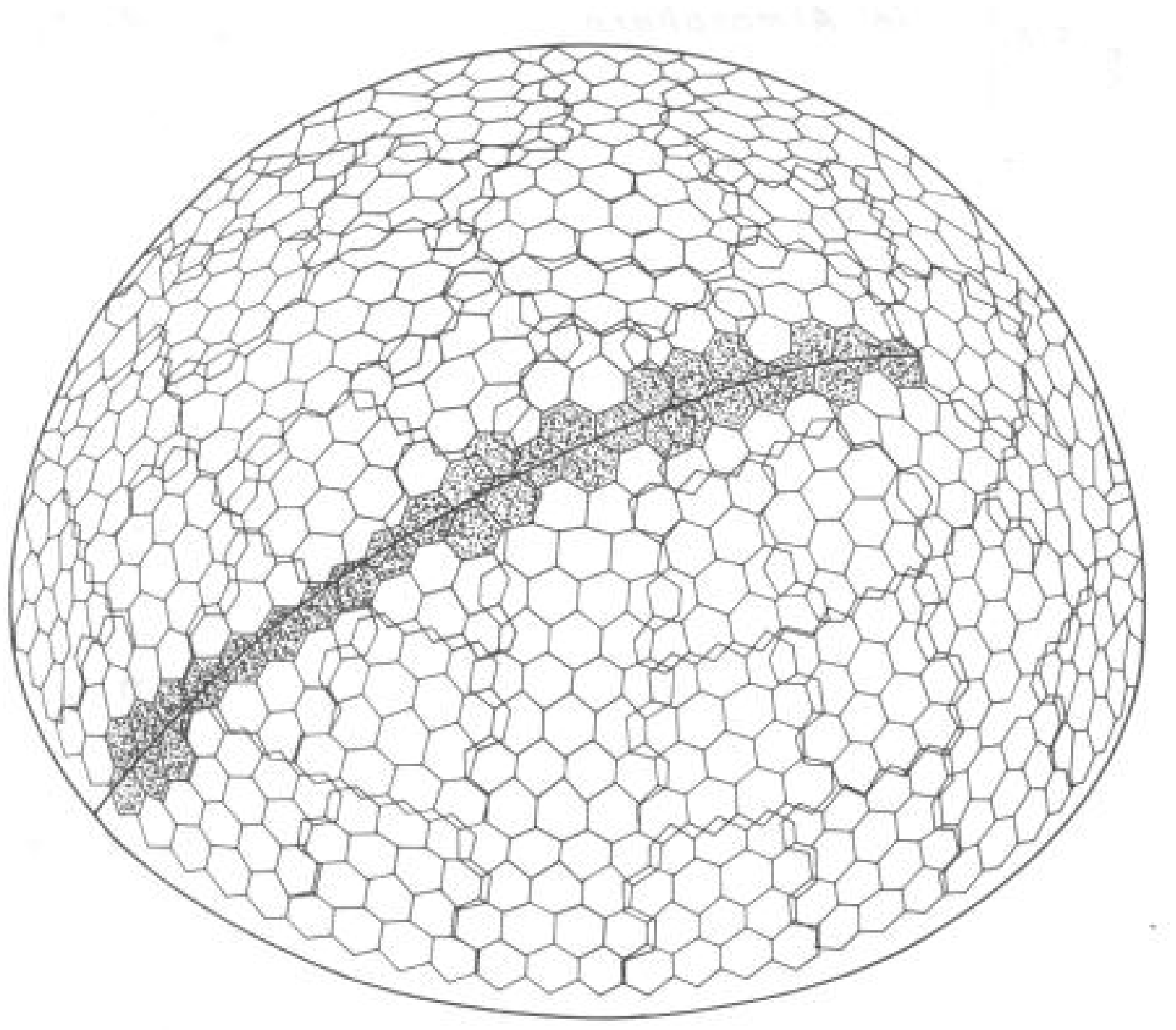}
  \end{center}
  \caption{The arrangement of Fly's Eye pixels as projected onto the
    dome of the sky with the track of an event superimposed.  Figure
    taken from \citeasnoun{Sokolsky-1989}}
  \label{flyseye-layout}
\end{figure}

While the Fly's Eye experiment was fully covered in
\citeasnoun{Yoshida-Dai-1998-JPG-24-905}, it continues to have
significant exposure in some energy ranges above the Second Knee.
Fly's Eye was the pioneering fluorescence detector, with full-sky
coverage at one site (see Figure~\ref{flyseye-layout}), and about
half-sky (with full elevation) coverage from a second site for stereo
observation \cite{Baltrusaitis-1985-NIMA-240-410}.  The fluorescence
light from the EAS was collected by 67 mirrors, each with an area of
1.95 m$^2$.  Each tube viewed a spot on the sky about $5^\circ$ in
diameter.  The two sites were separated by a distance of 3.3 km and
located at (40.2$^\circ$ N, 112.8$^\circ$ W), on Dugway Proving
Grounds in Utah, USA, at an atmospheric depth of 860 g/cm$^2$.  The
Fly's Eye Detector was operated from 1981 through 1992
\cite{Bird-1994-ApJ-424-491}.

\subsection{Yakutsk}
\begin{figure}[h]
  \begin{center}
  \includegraphics[width=\columnwidth]{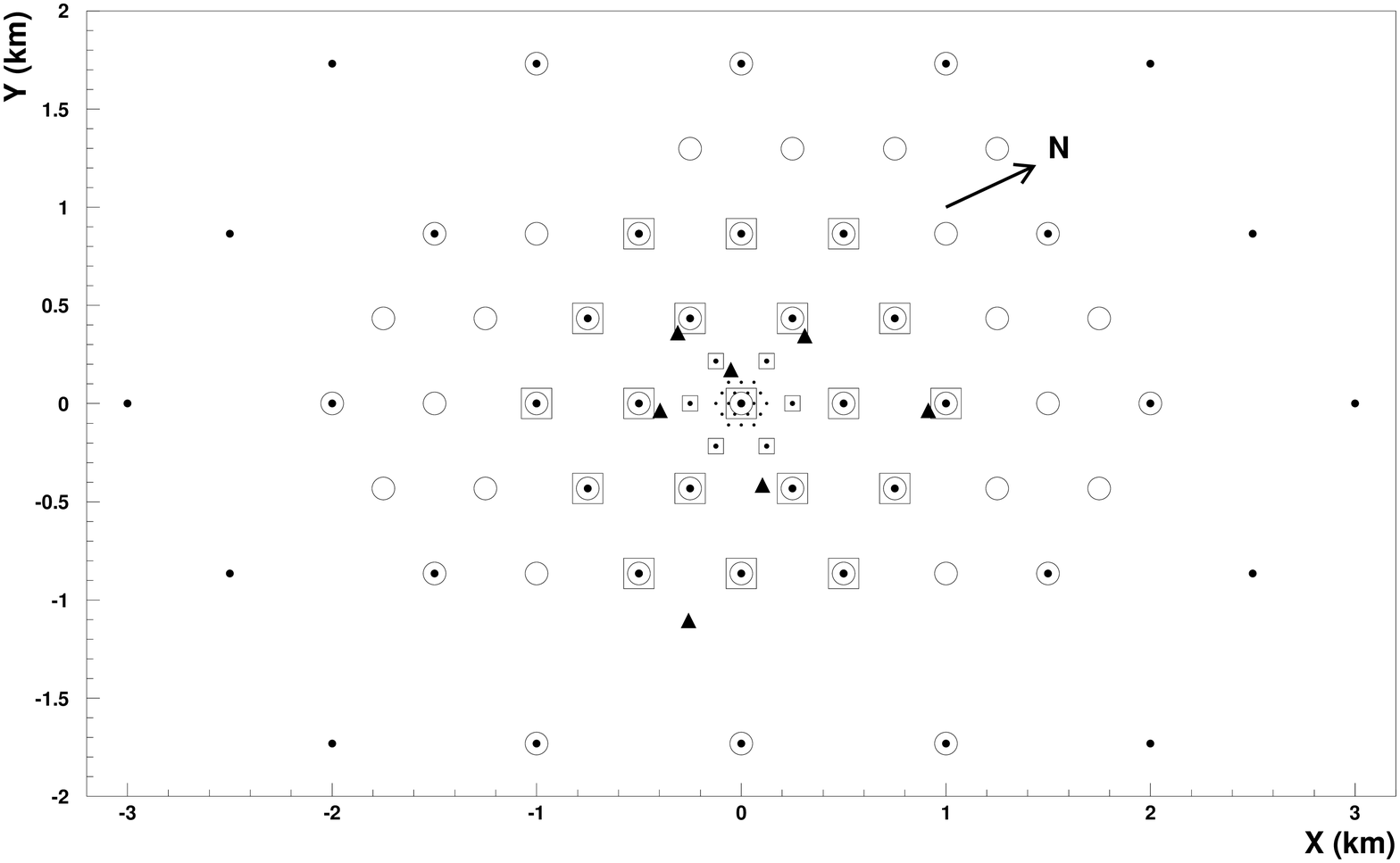}
  \end{center}
  \caption{Layout of the Yakutsk array as it existed prior to 1990.
    Figure adapted from \citeasnoun{Afanasiev-1993-Tokyo-35} and
    \citeasnoun{Ivanov-2003-NPBps-122-226}.  Filled circles represent
    the 2 $m^2$ scintillator detectors in the arrangement before 1990.
    Open squares represent these detectors after 1992.  Open squares
    represent the position of \v{C}erenkov detectors, while triangles
    the position of shielded muon detectors.}
  \label{yakutsk-layout}
\end{figure}

The Yakutsk Array is a set of three nested arrays.  At the center is a
closely spaced array of 19 scintillation counters arranged in a
hexagonal pattern with 62 m spacing covering an area of 0.026 km$^2$.
Each of these detectors has an area of 0.25 m$^2$.  This is surrounded
by more widely spaced counters, each 2 m$^2$ in area, on triangular
grids of spacing 500 m and 1 km.  Three counters on the corners of a
triangle were required to fire in coincidence to form a trigger for
the rest of the array
\cite{Afanasiev-1993-Tokyo-35,Pravdin-1999-ICRC-26-3-292}.  Prior to
1990, there were 19 counters at a spacing of 500 m, forming 24 trigger
triangles and covering an area of 2.5 km$^2$.  The surrounding 1 km
array contained 29 counters (with some overlap with the 500 m array),
40 trigger triangles and covering 16 km$^2$.  This layout is shown as
the filled circles in Figure~\ref{yakutsk-layout}.  Between 1990 and
1992, the detectors were rearranged, removing 10 counters from the 1
km array, but adding 18 counters to the 500 m array.  This increased
the size of the 500 m array substantially at the expense of the 1 km
array.  In this arrangement, there are 63 triangles in the 500 m
array, covering 7.2 km$^2$, but only 24 triangles in the 1 km array,
covering 10 km$^2$.  This layout is shown by the open circles in
Figure~\ref{yakutsk-layout} \cite{Ivanov-2003-NPBps-122-226}.

In addition to the scintillators for detecting charged particles,
there is an array of \v{C}erenkov detectors, consisting of large
photocathode photomultiplier tubes collecting EAS light directly.  The
PMTs have effecting collecting areas of either 176 or 530 cm$^2$
\cite{Ivanov-2003-NPBps-122-226}.  The positions of these detectors is
indicated by open squares in Figure~\ref{yakutsk-layout}.  Finally,
there are a number of buried muon detectors, shown as filled triangles
in Figure~\ref{yakutsk-layout}.

The Yakutsk array has been in continuous operation since 1970
\cite{Egorova-2001-JPSJ-70-supB-9}.  It is located at (61.7$^\circ$ N,
129.4$^\circ$ E) \cite{Uchihori-2000-APP-13-151}.

\subsection{The Akeno Giant Air Shower Array (AGASA)}
\begin{figure}[h] 
  \begin{center}
    \includegraphics[width=0.5\columnwidth]{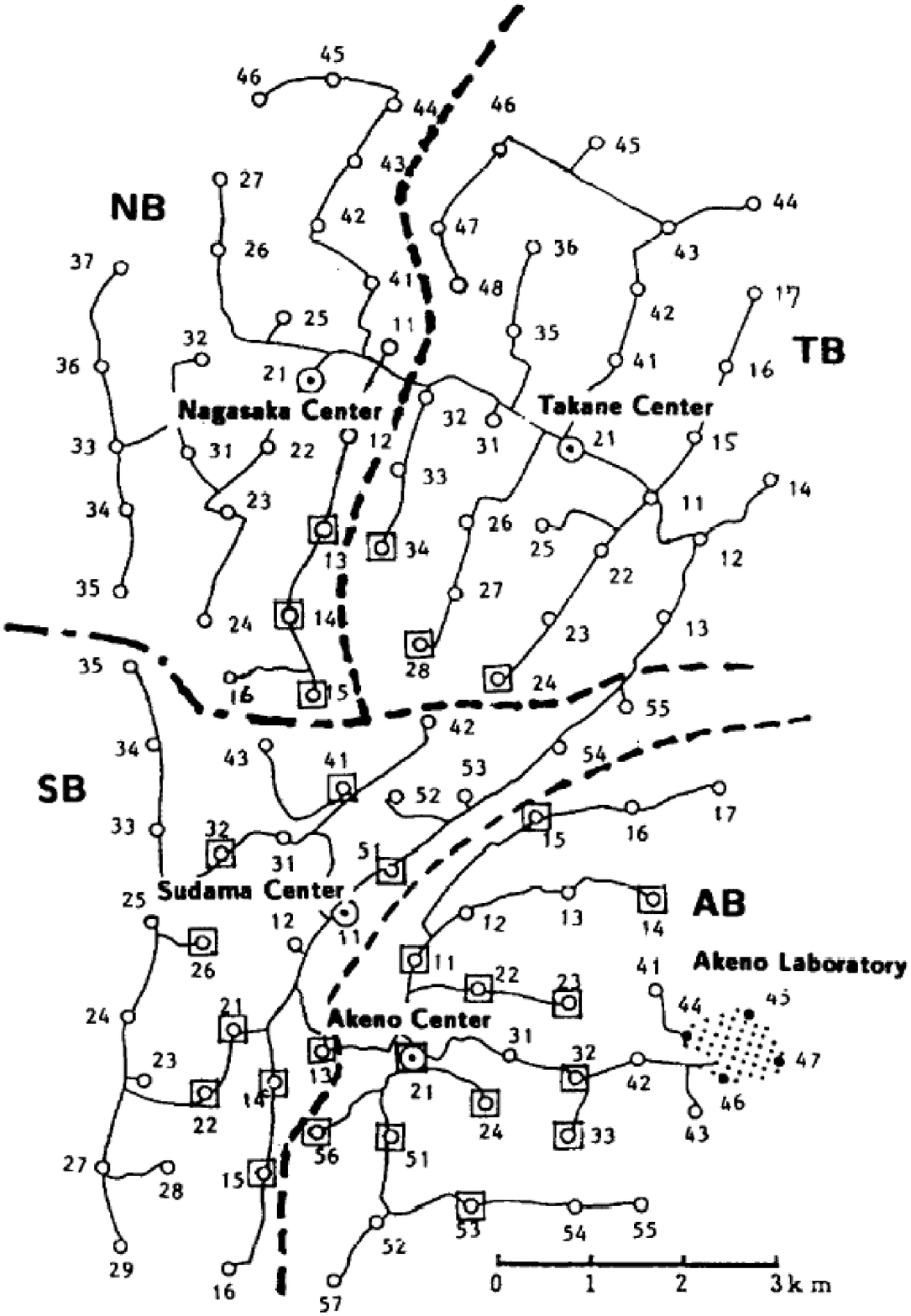}
  \end{center}
  \caption{The layout of scintillator detectors in AGASA.  Figure
    taken from \citeasnoun{Ohoka-1997-NIMA-385-268}.  Circles
    represent the positions of 2.2 m$^2$ detectors, squares the
    positions of shielded muon detectors, and lines the communication
    network.}
  \label{agasa-layout}
\end{figure}

The Akeno Giant Air Shower Array (AGASA) grew out of the Akeno 1
km$^2$ Array at the same site (35.8$^\circ$ N, 138.5$^\circ$ E).
AGASA is an array of 111 scintillation counters, each with an area of
2.2 m$^2$, on a roughly square grid with a spacing of about 1 km (see
Figure~\ref{agasa-layout}.  It covered a total area of about 100
km$^2$ \cite{Nagano-1992-JPG-18-423,Ohoka-1997-NIMA-385-268}.  It was
operated from the mid 1980's (as the Akeno 20 km Array; AB in
Figure~\ref{agasa-layout}) through 2004.  AGASA was the first detector
to have substantial exposure in the region above the expected GZK
Cutoff.

The signal in any particular counter was digitized by means of a
logarithmic amplifier.  The charge from the photomultiplier tube was
stored on a capacitor, and the capacitor was discharged with given RC
time constant.  In this way, the time-over-threshold is proportional
to the logarithm of the signal size.  This system provides a large
dynamic range but is susceptible to large errors by coincident
 signals
arriving during the discharge period.  The data from all the detectors
in a particular branch was read out when a trigger of five or more
detectors had signal within 25 $\mu$s \cite{Chiba-1992-NIMA-311-338}.
The branches were unified in 1995 \cite{Ohoka-1997-NIMA-385-268}, so
that coincidences between five detectors between two branches would
also trigger the system.

\subsection{The High Resolution Fly's Eye (HiRes) Detector}
\label{review-exp-hires}

The High Resolution Fly's Eye (HiRes) detector was designed using the
experience gained in operating the Fly's Eye detector.  Using larger
mirrors (~4 m$^2$ effective area) and smaller pixels (~1$^\circ$) it
was able to increase the usable aperture by a factor of 10 over Fly's
Eye.  It was deployed on two desert hills separated by 13 km at
(40.2$^\circ$ N, 112.8$^\circ$ W) and (40.1$^\circ$ N, 113.0$^\circ$
W) , on Dugway Proving Grounds in Utah, USA.  HiRes was operated from
May, 1997 to April, 2006.

\begin{figure}[h] 
  \begin{minipage}[t]{0.49\columnwidth}
    \includegraphics[width=\columnwidth]{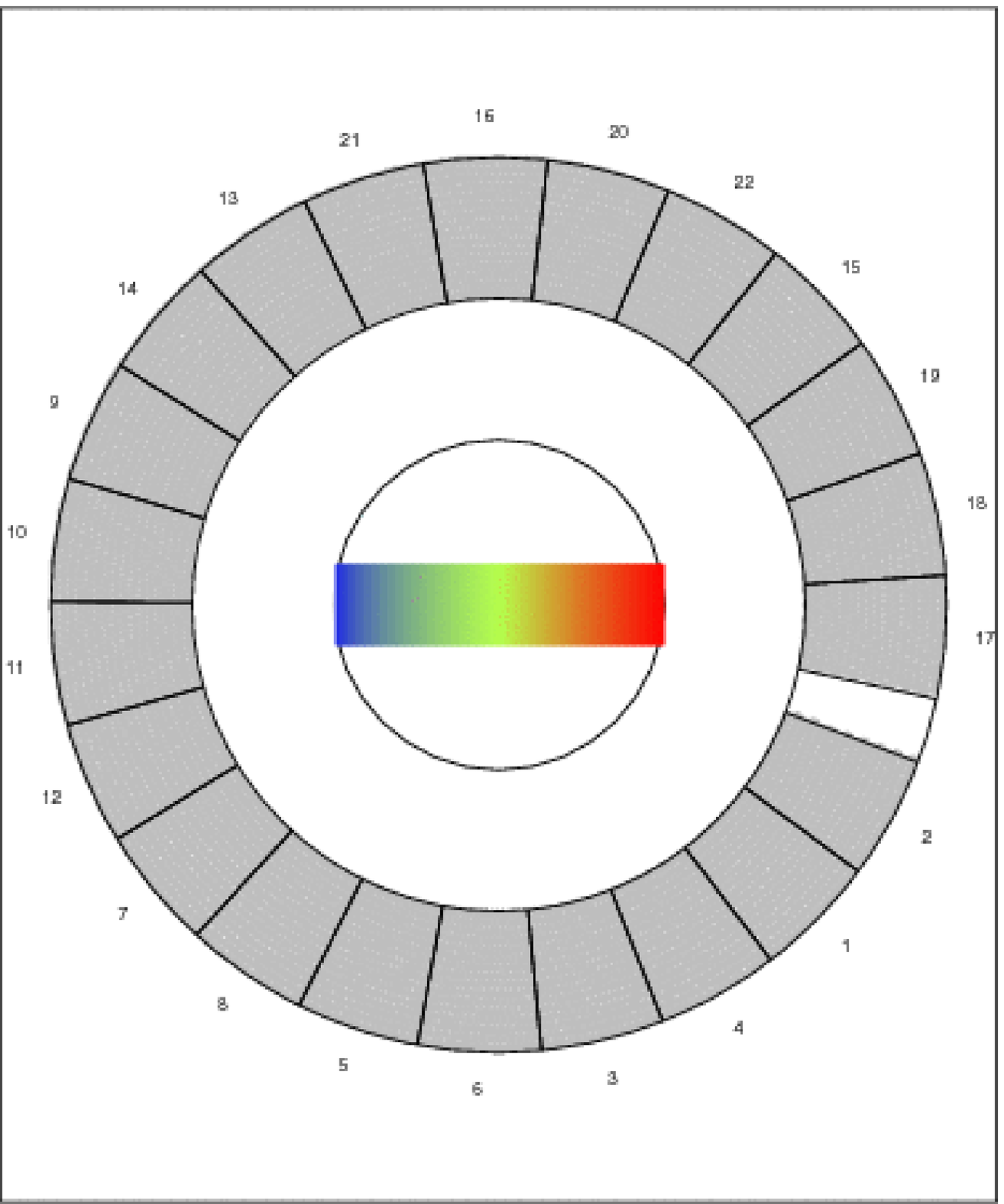}
  \end{minipage}
  \begin{minipage}[t]{0.49\columnwidth}
    \includegraphics[width=\columnwidth]{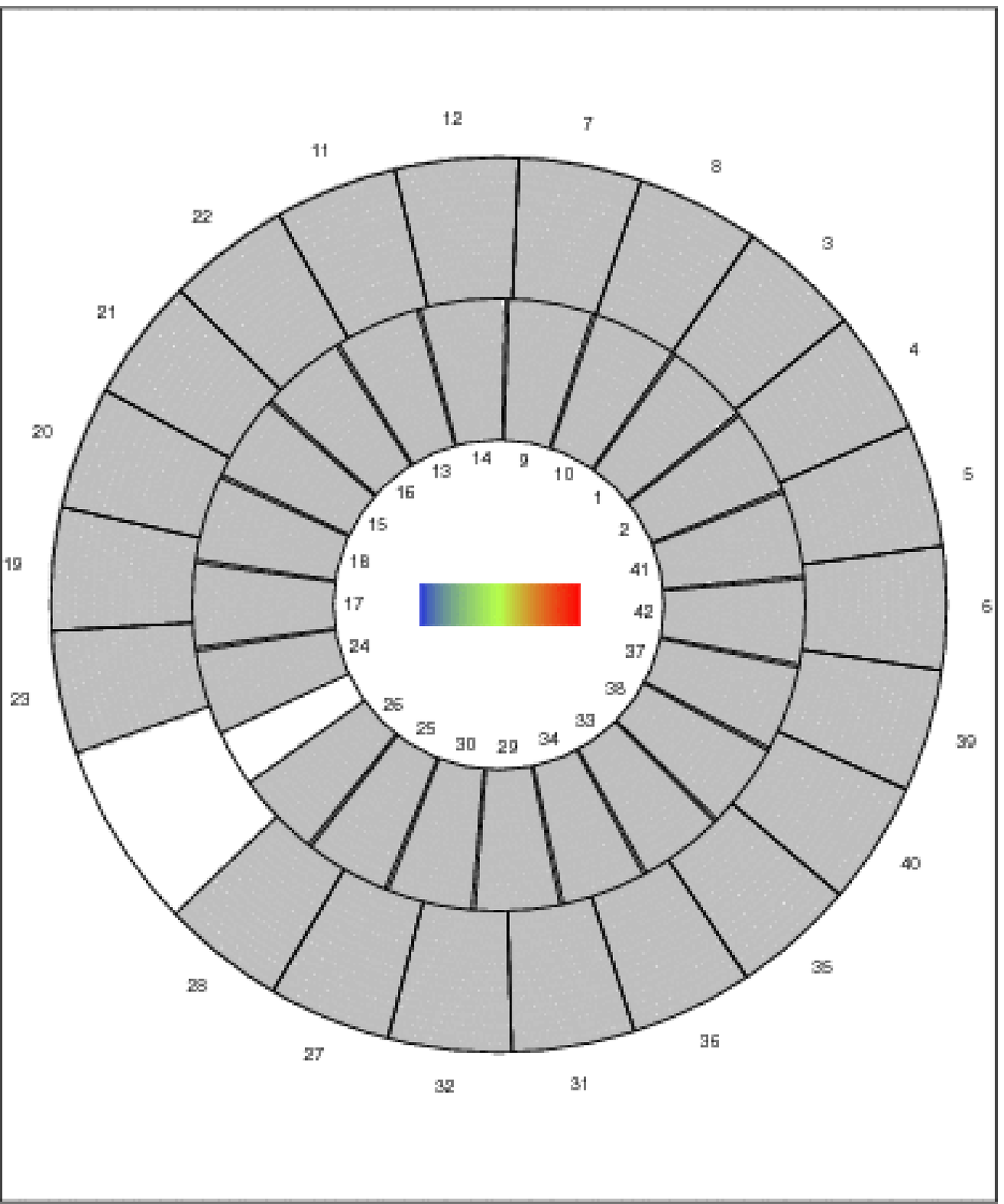}
  \end{minipage}
  \caption{The configuration of the mirrors at HiRes-I (left) and
    HiRes-II (right).  The outer ring covers 3$^\circ$ to 17$^\circ$
    in elevation, the inner ring 17$^\circ$ to 31$^\circ$.  Figure
    taken from \citeasnoun{Reil-2002-Utah-Thesis}.}
  \label{hires-layout}
\end{figure}

The two sites allow for stereo observation of events, which gives very
good geometrical reconstruction of the showers.  HiRes-I was located
on the same hill where the full sky Fly's Eye detector had been
located.  It operated as a stand-alone site from May, 1997 till the
end of 1999.  It consisted of one ring of mirrors viewing elevation
angles from 3$^\circ$ to 17$^\circ$ degrees (see
Figure~\ref{hires-layout}).  The limited elevation coverage limited
the aperture for energies below $1\times10^{18}$ eV but not at higher
energies.  Very energetic events are visible at large distances, where
the limited elevation angle covers most of the volume of atmosphere
where the shower occurs.  At lower energies, the detector cannot see
showers as far away, so most of the available atmosphere lies above
the portion observed by the detector.  This becomes significant for
HiRes-I at the energy given above.  HiRes-I recorded the time and
pulse height of each tube, and a simple coincidence trigger of three
tubes in each of two, adjacent $4\times4$ clusters of tubes within a
mirror, was required to record the event.

The second site, HiRes-II, became operational in December 1999.  It
consisted of two rings of mirrors viewing elevation angles from
3$^\circ$ to 31$^\circ$ degrees (see Figure~\ref{hires-layout}).  The
larger elevation angle coverage allows a lower threshold of about
$2\times10^{17}$ eV.  In addition, HiRes-II had a flash ADC (FADC)
data acquisition system, which recorded the voltage of each phototube
every 100 ns \cite{Boyer-2002-NIMA-482-457}.  The sum of the signals
in each row and column of tubes was also digitized with an FADC for
triggering and recording large signals at low gain.  A trigger bit was
set whenever three of five consecutive rows or columns was over
threshold in coincidence (this coincidence also took into account the
signal moving from one row or column to the next in sequence).  Two
trigger bits, in rows or columns or both, was required to trigger the
readout of an event.  The FADC system has the feature that light from
different tubes can be combined by time.

With its fine angular resolution, measuring the pointing direction of
each PMT becomes important.  This measurement can be done by recording
the images stars on a screen placed over the face of the PMT cluster
\cite{Bergman-2001-ICRC-27-639} or by measuring the change in the
noise rate of each individual PMT as stars move in and out of its
field of view \cite{Sadowski-2002-APP-18-237}.

Because HiRes can look so much further through the atmosphere than
Fly's Eye could, it was imperative that there was a mechanism in place
to measure the transmission of UV light through the atmosphere.  This
was accomplished by means of a bistatic LIDAR system
\cite{Abbasi-2006-APP-25-74,Abbasi-2006-APP-25-93}.  A steerable laser
at each site shoot a planned series of light pulses through the
atmosphere, and the light scattered from these pulses was collected by
the detector at the other site.  These shots look very much like the
cosmic rays signals except that they move up through the atmosphere.
By measuring the amount of light scattered at various angles one can
determine the amount scattering by air molecules (Rayleigh
scattering), which is fairly constant, and the amount of scattering by
aerosols, which can vary from night to night.  The phase function of
the scattering by aerosols is also measured.  The earliest data
collected by HiRes-I was taken before any LIDAR system was in
operation.  For this data one must rely on average measurements.
However, the LIDAR system was in operation for several years, so a
good characterization of the atmospheric clarity was obtained.  The
atmosphere in Dugway is in fact quite clear, given that it is a
desert.

\subsection{The Pierre Auger Observatory}
\begin{figure}[h] 
  \begin{center}
    \includegraphics[width=\columnwidth]{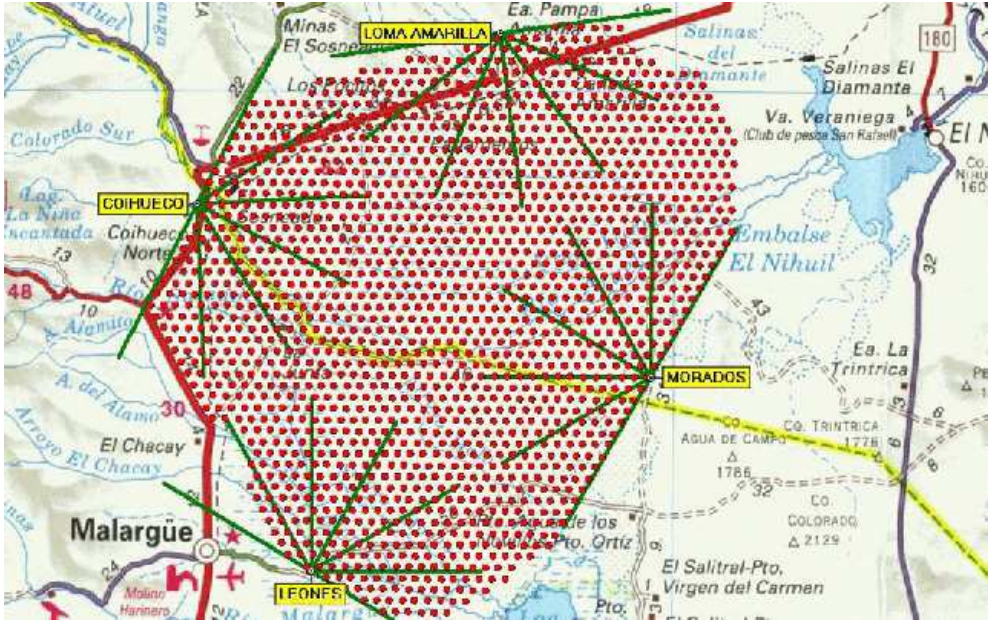}
  \end{center}
  \caption{The layout of the Auger surface array detectors (red dots)
    and the overlooking fluorescence detectors.  Figure taken from
    \citeasnoun{Mantsch-2005-ICRC-29-10-115}.}
  \label{auger-layout}
\end{figure}

The Pierre Auger Observatory \cite{AugerC-2004-NIMA-523-50} is a
planned, hybrid detector with sites in both the Northern and Southern
hemispheres.  The Southern detector is currently being deployed in
Malargue, Argentina, at (35$^\circ$ S, 69$^\circ$ W).  When completed
the surface array component of the detector will consist of 1600 water
tanks (10 m$^2$ in area, 1.2 m deep) on a triangular grid, with a
separation of 1.2 km.  The total area covered by this array will be
3000 km $^2$, about 30 times that of AGASA.  (See
Figure~\ref{auger-layout}.)  The surface array will be overlooked by a
set of four fluorescence stations placed around the edges of the array
and covering its entire area.  The fluorescence detectors consist of
mirrors viewing 30$^\circ$ in both elevation and azimuth.

\section{The UHECR Spectrum: Flux Measurements}
When \citeasnoun{Yoshida-Dai-1998-JPG-24-905} published their review,
there was the suggestion that many experiments saw a hint of the
expected GZK Cutoff.  That suggestion has become a near certainty with
the HiRes claim to have observed the cutoff
\cite{Bergman-2007-NPBps-165-19,Abbasi-2007-PRL}.  However, the
consensus was radically different for several years, when AGASA
claimed to have observed as many as 16 events about 100 EeV
\cite{Sakaki-2001-ICRC-27-333}.  This discrepancy points to the
importance of understanding the energy resolution of any experiment
when looking for features in the spectrum.

Rather than list the recent measurements of each experiment, we
thought it would be more illuminating to discuss in turn each of the
features of the flux spectrum, giving the evidence for each feature,
and the degree of confidence we have in their existence.  The features
of the spectrum are best described as changes in the spectral slope.
As such, one must understand not only the energy at which the change
occurs, but the degree to which we know the spectral slope between
features.  Because of differences between the normalization of
different experiments, it is often difficult to compare the absolute
energy of the various features.  It is much easier to compare the
spectral slopes.  The ratios of the energies of the features (the
difference in the logarithm of the energies) are also of prime
importance as it doesn't depend on the absolute energy scale.  This
gives special importance to experiments which can measure the energy
of more than one feature.

We discuss three features of the ultra high energy cosmic ray
spectrum: the Second Knee, the Ankle and the GZK Cutoff.  The Second
Knee is a softening of the spectrum (the spectral slope becoming
steeper) in the $10^{17}$ eV decade.  The Ankle is a hardening of the
spectrum (the spectral slope becoming less steep) in the $10^{18}$ eV
decade.  The GZK Cutoff is an expected, drastic reduction in the UHECR
flux above $\sim10^{19.8}$ eV due to photopion production of protons
on the cosmic microwave background radiation.

\subsection{The Second Knee}
\label{review-flux-second-knee}

Four experiments have shown evidence for the Second Knee: Akeno
\cite{Nagano-1992-JPG-18-423}, Haverah Park
\cite{Lawrence-1991-JPG-17-733,Ave-2003-APP-19-47}, Fly's Eye
\cite{Bird-1993-PRL-71-3401,Bird-1994-ApJ-424-491}, and the HiRes
Prototype/MIA experiment \cite{AbuZayyad-2001-ApJ-557-686}.  This
includes experiments which have measured a \emph{break} in the
spectrum in the $10^{17}$ eV decade, having measured different
spectral slopes above and below the break energy.  Of these, it
unclear whether one should still include the Haverah Park data below
the break because of the recent re-analysis \cite{Ave-2003-APP-19-47}.
The re-analysis only measures the spectral slope above the break and
doesn't treat data from the infill array used to measure the spectral
slope below the break at the lowest energies.  It is also difficult to
consider the HiRes Prototype/MIA measurement, as there is not a
separate measurement of the spectral slopes above and below the Second
Knee, and the measurement of the slope below the Second Knee has no
associated error.

In addition to the experiments listed above, Yakutsk
\cite{Egorova-2004-NPBps-136-3} has data above the Second Knee (but
provides no fit to the spectral slope), while HiRes
\cite{Abbasi-2007-PRL} measures the spectral slope above the Second
Knee in its monocular analysis (as does Haverah Park in its recent
analysis).  These measurements can be used to increase confidence in
the knowledge of the spectral slope above the Second Knee.  (The
latest HiRes measurement in \citeasnoun{Abbasi-2007-PRL} includes and
supersedes the data shown in several earlier papers
\cite{Abbasi-2004-PRL-92-151101,Abbasi-2005-APP-23-157,Abbasi-2005-PLB-619-271}.
A detailed account of the HiRes systematic uncertainties can be found
in \cite{Abbasi-2007-APP-astroph-0607094}.)

The measured spectral slopes and break point energies are shown in
Table~\ref{second-knee-values}.  There is good agreement between all
the experimental results for the spectral slopes, giving very high
confidence in the existence of a break, with the difference between
spectral having a significance of $7.5\sigma$.

\begin{table}
  \begin{tabular}{|llll|}
    \hline\hline
    Experiment & \multicolumn{1}{l}{Slope}
               & \multicolumn{1}{l}{Break Point}
               & \multicolumn{1}{l|}{Slope} \\
               & \multicolumn{1}{l}{Below}
               & \multicolumn{1}{l}{
                 $\log_{10}\left(\frac{E}{\rm eV}\right)$}
               & \multicolumn{1}{l|}{Above} \\
    \hline
    Akeno \cite{Nagano-1992-JPG-18-423} & 
      $3.02\pm0.03$ & 17.8 & $3.24\pm0.18$ \\
    Fly's Eye \cite{Bird-1993-PRL-71-3401} & 
      $3.01\pm0.06$ & 17.6 & $3.27\pm0.02$ \\ \hline
    Haverah Park \cite{Ave-2003-APP-19-47} &
                    &      & $3.33\pm0.04$ \\
    HiRes \cite{Abbasi-2007-PRL} &
                    &      & $3.32\pm0.03$ \\ \hline
    Average Slopes &
      $3.02\pm0.03$ &      & $3.29\pm0.02$ \\ \hline
    \hline
  \end{tabular}
  \caption{The measured slope parameters and break point energies for
    the Second Knee.}
  \label{second-knee-values}
\end{table}

We have performed our own fits to these experimental results.  The
fits use the binned maximum likelihood method comparing the actual
number of events in bin to the expected number given a flux and
exposure \cite{Yao-2006-JPG-33-302}.  The likelihood in this method is
normalized in such a way as to give a goodness-of-fit metric which
approaches the $\chi^2$ distribution in the large statistics limit.
The fitting function is a broken power law, where the break points are
allowed to vary as some of the fit parameters.  The fitting function
is continuous, with only one normalization parameter.

\begin{figure}
  \begin{minipage}[t]{0.49\columnwidth}
    \includegraphics[width=\columnwidth]{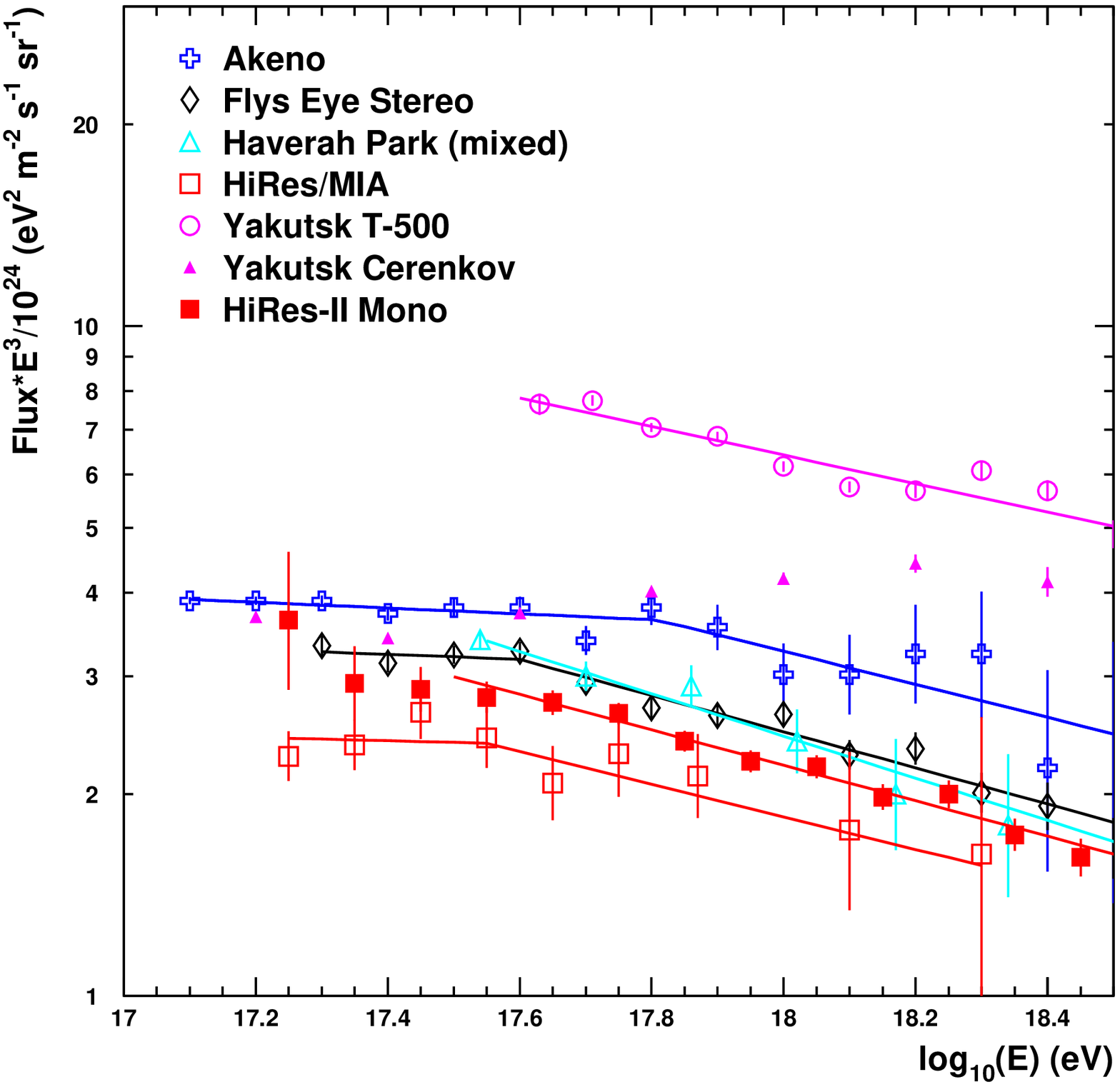}
  \end{minipage}
  \begin{minipage}[t]{0.49\columnwidth}
    \includegraphics[width=\columnwidth]{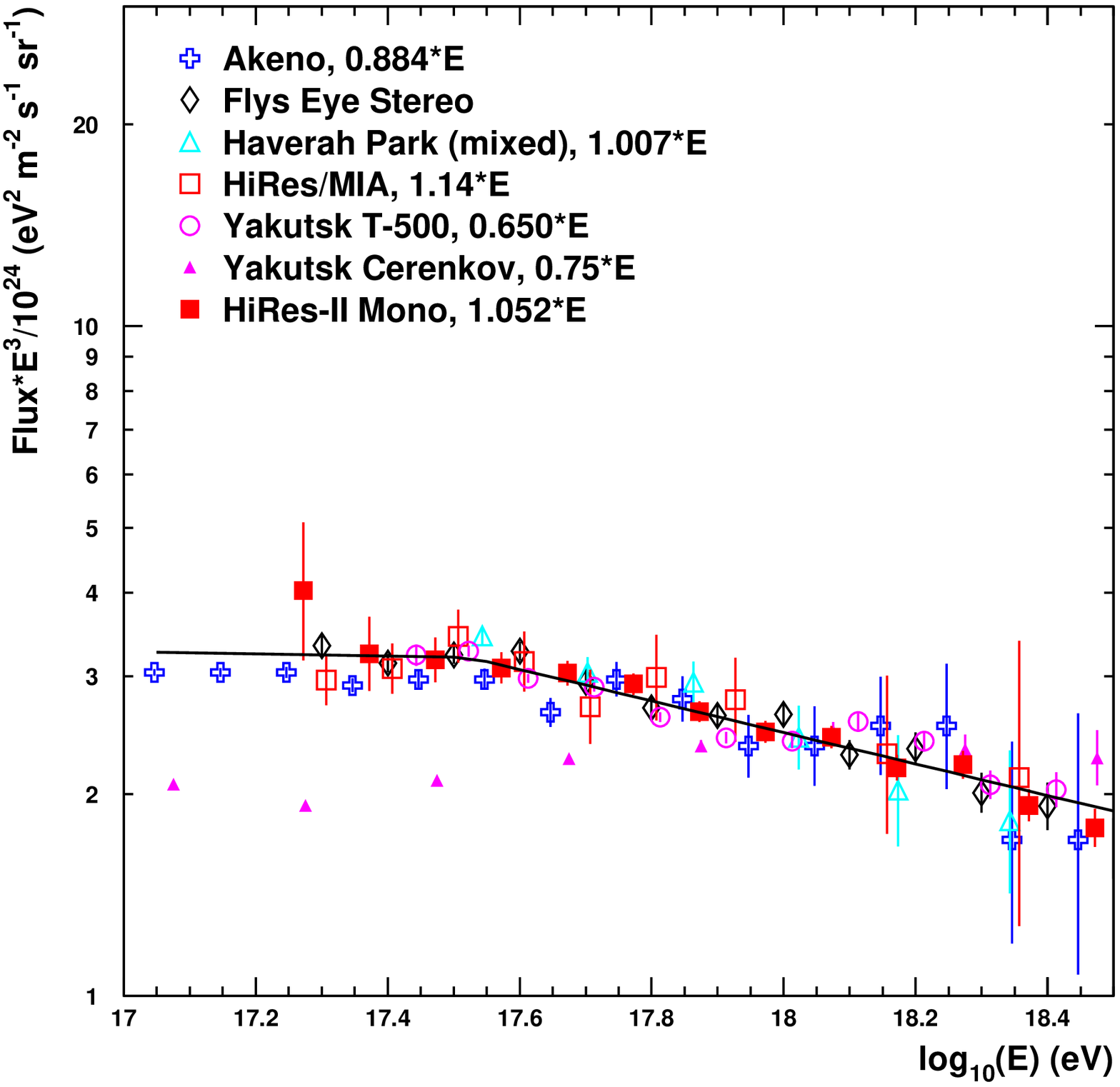}
  \end{minipage}
    \caption{Left: Flux measurements in the Second Knee energy range.  The
      shown fits are our calculation. Right: Flux measurements in the
      Second Knee energy range, scaled so that the flux agrees with
      the Fly's Eye result at $10^{18}$ eV.  The scaled data points
      were fit to a broken power law spectrum in a global fit, with
      the result shown.}
    \label{second-knee-measurements}
\end{figure}

The results from our fits to the experiments which make measurements
in the Second Knee energy range are shown in
Table~\ref{second-knee-fit-values}.  A few specific comments on the
fits are important.  The fit to the HiRes Prototype/MIA spectrum,
required special handling.  \citeasnoun{AbuZayyad-2001-ApJ-557-686}
mention that a fit to the first six points of the spectrum give a
spectral slope of 3.01.  We performed a fit to the same points and
found a spectral slope of $3.02\pm0.11$.  We then \emph{fixed} this
slope, and performed a broken power law fit to all the points.  In
fitting the Haverah Park spectrum, we only fit the lowest 7 points, to
avoid being biased by the Ankle.  For the fits to the Yakutsk T-500
\cite{Egorova-2004-NPBps-136-3} and HiRes-II Monocular
\cite{Abbasi-2005-PLB-619-271,Abbasi-2007-PRL} spectra, the fit
included another break point and spectral index for the Ankle and
beyond.  These parameters will be discussed in
Section~\ref{review-flux-ankle}.  The fit to the HiRes-II spectrum
does not include the points below $10^{17.5}$ eV, which agrees with
the way the HiRes Collaboration performs its own fits.  These spectra
along with the results of our fits are shown on the left side of
Figure~\ref{second-knee-measurements}.  The fit power law is only
drawn in the energy range that was used in the fitting.

The one recent measurement which does not agree \emph{in shape} with
the other measurements in the Second Knee region is that of the
\v{Cerenkov} array at Yakutsk
\cite{Ivanov-2003-NPBps-122-226,Knurenko-2007-NPBps}.  The Second Knee
in this analysis appears at a very high energy, $\sim10^{18.2}$ eV,
which is inconsistent with the standard Yakutsk ground array
measurement \cite{Egorova-2004-NPBps-136-3}.  We have not fit this
data, but it is shown in Figure~\ref{second-knee-measurements}.

To combine all the data into a global fit, we scaled each experiment
so that the fit fluxes all agree at $10^{18}$ eV.  This may be
equivalent to removing several degrees of freedom from the combined
fit, but does not necessarily give the best $\chi^2$.  We have not
adjusted the number of degrees of freedom to account for the shifting.
The global Second Knee Fit included all the data above, except the
Yakutsk measurement using the \v{C}erenkov Array.  The fit does
include most points left out of the individual fits, but does no
include the highest energy bin of the HiRes-II monocular sample which
may be above the GZK cutoff.  The results of the fit are also include
in Table~\ref{second-knee-fit-values}.  The scaled flux measurements
and the result of the global fit are shown on the right side of
Figure~\ref{second-knee-measurements}.

\begin{table}
  \begin{tabular}{|lllll|}
    \hline\hline
    Experiment & $\chi^2/$DOF & Slope & Break Point & Slope \\
    ({\it reference}) &       & Below & $\log_{10}\left(\frac{E}{\rm eV}\right)$
                                                    & Above \\ \hline\hline
    Akeno  & 8.3/13 & $3.04\pm0.02$ & $17.8\pm0.2$   & $3.25\pm0.12$ \\
      \cite{Nagano-1992-JPG-18-423} &&&& \\ \hline
    Fly's Eye & 13.7/18 &
      $3.04\pm0.05$ & $17.60\pm0.06$ & $3.27\pm0.02$ \\
      \cite{Bird-1993-PRL-71-3401} &&&&\\ \hline
    HiRes/MIA & 2.5/5 &
      $3.02$        & $17.6\pm0.2$   & $3.23\pm0.14$ \\
       \cite{AbuZayyad-2001-ApJ-557-686} &&&&\\ \hline
    Haverah Park & 1.4/5 &
                    &                & $3.32\pm0.05$ \\
       \cite{Ave-2003-APP-19-47} &&&&\\ \hline
    Yakutsk T-500  & 45.2/15 &
                    &                & $3.213\pm0.012$ \\
      \cite{Egorova-2004-NPBps-136-3} &&&&\\ \hline
    HiRes  & 8.55/15 &
                    &                & $3.26\pm0.02$ \\ 
      \cite{Abbasi-2007-PRL} &&&&\\ \hline\hline
    Global Fit & 109.4/93 &
      $3.02\pm0.01$ & $17.52\pm0.02$ & $3.235\pm0.008$ \\ 
      (at Fly's Eye $E$ scale) &&&&\\ \hline\hline
  \end{tabular}
  \caption{Our broken power law fits to spectrum measurements in the
    Second Knee energy range.  The fit parameters include a
    normalization (not shown), slope parameters above and below the
    break and the break point energy for the Second Knee.}
  \label{second-knee-fit-values}
\end{table}

One can see that there is, in fact, a consensus among the various
experiments on the existence of the Second Knee, and on the specific
spectral slopes at energies above and below the second knee.  The only
major discrepancies are in the energy normalization between the
experiments.

\subsection{The Ankle}
\label{review-flux-ankle}

Before considering all the experiments which have observed the Ankle,
we note that only one experiment has observed both the Second Knee and
the Ankle within a single measurement: Fly's Eye.  Observing both
features, allows a measurement of the ratio of the energies of the two
features (or the difference in the logarithm of the energies).  The
Fly's Eye collaboration fit their stereo spectrum to a broken power
law, with breaks at $10^{17.6}$ eV for the Second Knee, and
$10^{18.5}$ eV for the Ankle
\cite{Bird-1993-PRL-71-3401,Bird-1994-ApJ-424-491}.  These break
points were picked by eye and not allowed to float in the fit.  In our
fit to the Fly's Eye stereo spectrum (see
Tables~\ref{second-knee-fit-values} and ~\ref{ankle-fit-values}), we
find the break points at ($\log E$, $E$ in eV) $17.60\pm0.06$ and
$18.52\pm0.09$.  The difference, $\Delta\log E = 0.92\pm0.11$
corresponds to a ratio $E_2/E_1 = 8.3^{+2.5}_{-1.8}$.

In addition to the Fly's Eye Experiment, two other experiments have
strong observations of the Ankle: Yakutsk
\cite{Egorova-2004-NPBps-136-3} and HiRes
\cite{Abbasi-2005-PLB-619-271,Abbasi-2007-PRL}.  Of these, only HiRes
provides measurements of the spectral slopes.  Haverah Park observed
the Ankle in its old analysis \cite{Lawrence-1991-JPG-17-733}, but
only the section between the Second Knee and Ankle was treated in
their recent re-analysis \cite{Ave-2003-APP-19-47}.  We use only the
results from the later analysis, which does provide a spectral slope
for energies below the Ankle.  AGASA has also observed a break in
their spectrum \cite{Takeda-2003-APP-19-447}, but at a very high
energy, higher than one would expect from other experiments given the
overall flux level.  The AGASA collaboration has distanced itself from
interpreting the break as a measurement of the Ankle, because the break
also coincides with the energy at which the AGASA detector becomes
fully efficient \cite{Teshima-2002-personal}.  We only use AGASA flux
measurements above $10^{19}$ eV in our fits.  The The Pierre Auger
Observatory also measures the flux above the Ankle
\cite{Sommers-2005-ICRC-abs1}, but does not provide a measurement of
the spectral index.

The measured spectral slopes and break points from these experiments
are shown in Table~\ref{ankle-values}. 

\begin{table}
  \begin{tabular}{|llll|}
    \hline\hline
    Experiment & Slope & Break Point & Slope \\
               & Below & $\log_{10}\left(\frac{E}{\rm eV}\right)$
                                     & Above \\
    \hline\hline
    Fly's Eye \cite{Bird-1993-PRL-71-3401} & 
      $3.27\pm0.02$ & 18.5         & $2.71\pm0.10$ \\
    HiRes \cite{Abbasi-2007-PRL} &
      $3.25\pm0.01$ &$18.65\pm0.04$& $2.81\pm0.03$ \\ \hline
    Haverah Park \cite{Ave-2003-APP-19-47} &
      $3.33\pm0.04$ & & \\ \hline\hline
    Average Slopes &
      $3.26\pm0.01$ &              & $2.80\pm0.03$ \\ \hline
    \hline
  \end{tabular}
  \caption{The measured slope parameters and break point energies for
    the Ankle.}
  \label{ankle-values}
\end{table}

\begin{figure}
  \begin{minipage}[t]{0.49\columnwidth}
    \includegraphics[width=\columnwidth]{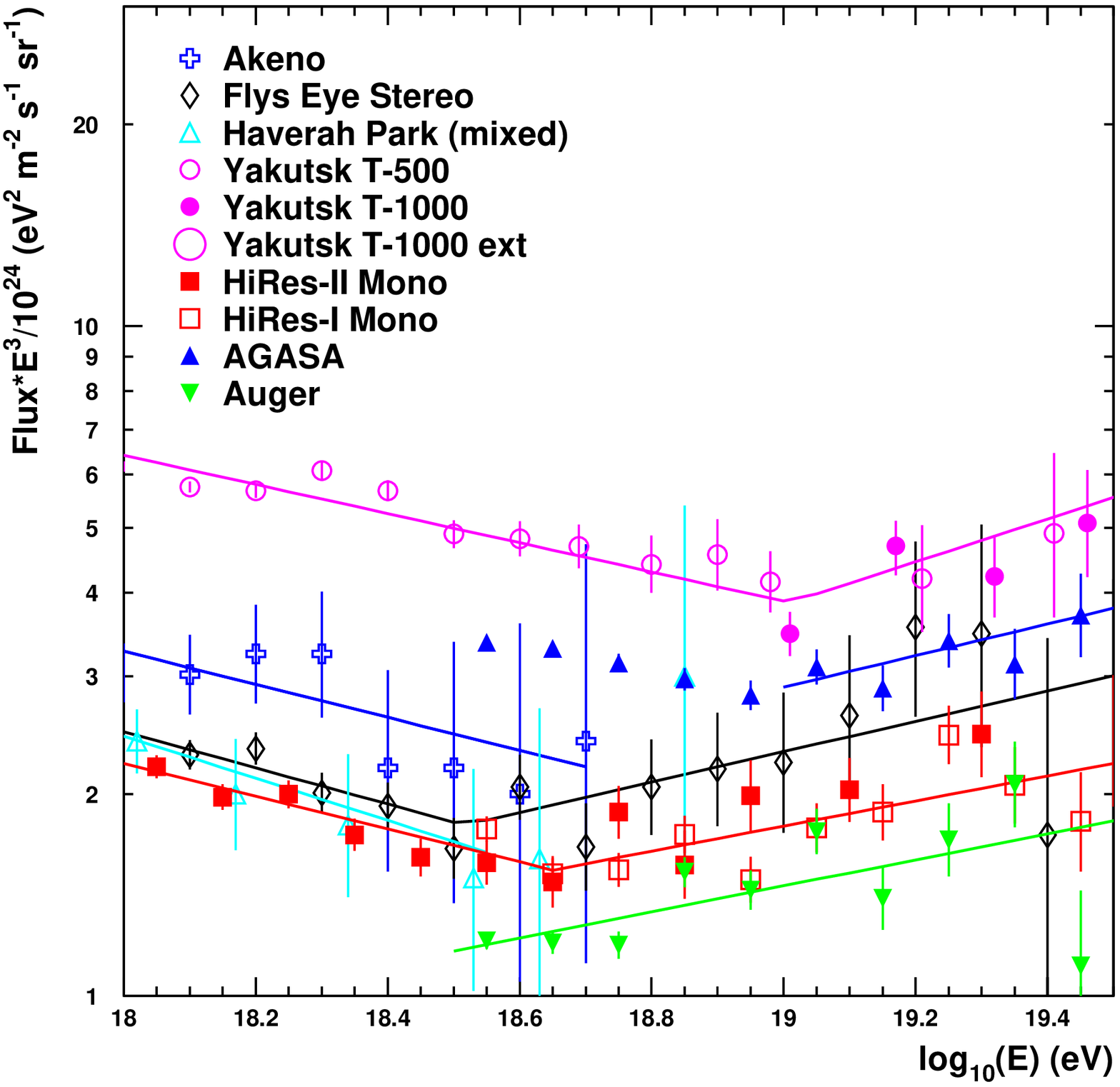}
  \end{minipage}
  \begin{minipage}[t]{0.49\columnwidth}
    \includegraphics[width=\columnwidth]{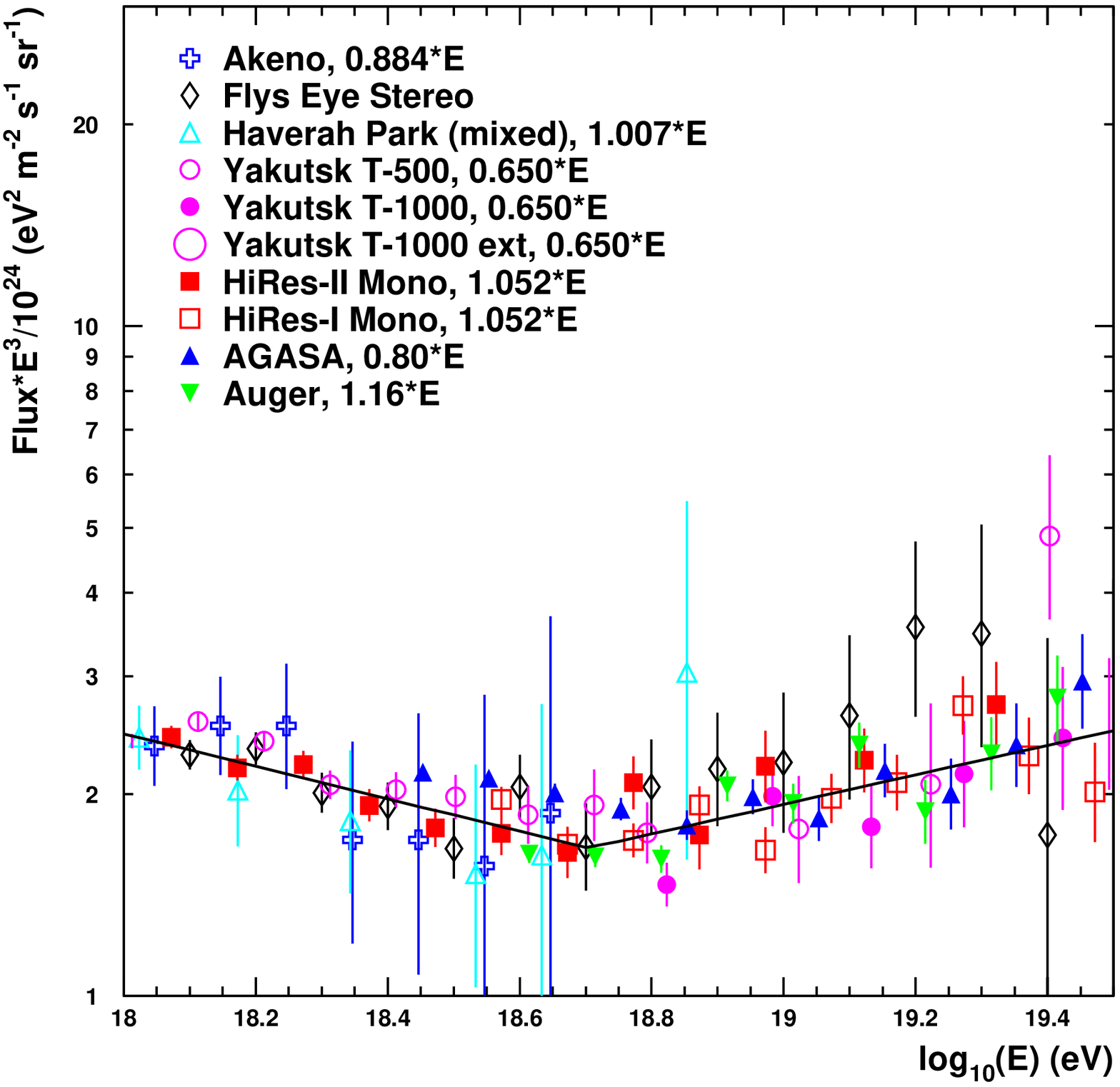}
  \end{minipage}
    \caption{Left: Flux measurements in the Ankle energy range.  The
      shown fits are our calculation. Right: Flux measurements in the
      Ankle energy range, scaled so that the flux agrees with the
      Fly's Eye result at $10^{18}$ eV (or $10^{19}$ eV for those
      experiments with no measurement at $10^{18}$ eV).  The scaled
      data points were fit to a broken power law spectrum in a global
      fit, with the result shown.}
    \label{ankle-measurements}
\end{figure}

As we did for the Second Knee, we have performed our own fits to all
the reported spectra.  Some of these fits are identical to the ones
listed previously, and the fitting methodology is the same in all
cases.  Two important differences are that the spectra from both
HiRes-I and HiRes-II in monocular mode have been fit in a combined
fit, as has the data from all three triggers of the Yakutsk
experiment.  No scaling was done to match the different measurements
from these experiments as the data is presumed to be already matched
in energy scale.  In both these fits, as well as in fits to AGASA and
Auger spectra, measurements above $10^{19.8}$ eV have been excluded,
so as not to complicate the fits with data above the GZK cutoff.  In
addition, HiRes-II data was only fit above $10^{17.5}$ eV, and AGASA
data was only fit above $10^{19}$ eV.  The results of all these fits
are displayed on the left side of Figure~\ref{ankle-measurements}.  In
the figure, on can clearly see that there is a good consensus on the
spectral slopes above and below the Ankle energy.  The parameters
found by the fits are given in Table~\ref{ankle-fit-values}.

\begin{table}
  \begin{tabular}{|lllll|}
    \hline\hline
    Experiment & $\chi^2/$DOF & Slope & Break Point & Slope \\
    ({\it reference}) &       & Below & $\log_{10}\left(\frac{E}{\rm eV}\right)$
                                                    & Above \\ \hline\hline
    Akeno        & 8.3/13  & $3.25\pm0.12$ && \\
      \cite{Nagano-1992-JPG-18-423}   &&&& \\ \hline
    Fly's Eye    & 13.7/18 & $3.27\pm0.02$ && \\
      \cite{Bird-1993-PRL-71-3401}    &&&& \\ \hline
    Haverah Park & 1.4/5   & $3.32\pm0.05$ && \\
       \cite{Ave-2003-APP-19-47}      &&&& \\ \hline
    Yakutsk      & 50.3/22 & $3.22\pm0.01$ & $19.01\pm0.07$ & $2.68\pm0.06$ \\
      \cite{Egorova-2004-NPBps-136-3} &&&& \\ \hline
    HiRes        & 29.6/28 & $3.22\pm0.03$ & $18.65\pm0.04$ & $2.81\pm0.03$ \\ 
      \cite{Abbasi-2007-PRL}  &&&& \\ \hline
    AGASA        & 6.7/6   &               &                & $2.76\pm0.08$ \\
      \cite{Takeda-2003-APP-19-447}   &&&& \\ \hline
    Auger        & 28.9/11 &               &                & $2.81\pm0.03$ \\
      \cite{Sommers-2005-ICRC-abs1}   &&&& \\ \hline\hline
    Global Fit  & 184.2/125 & $3.242\pm0.008$ & $18.70\pm0.02$ & $2.78\pm0.02$ \\ 
      (at Fly's Eye $E$ scale)        &&&& \\ \hline\hline
  \end{tabular}
  \caption{Our broken power law fits to spectrum measurements in the
    Ankle energy range.  The fit parameters include a
    normalization (not shown), slope parameters above and below the
    break and the break point energy for the Ankle.}
  \label{ankle-fit-values}
\end{table}

We again combined all the data into a global fit by scaling energies.
The scalings were identical to those used in
Section~\ref{review-flux-second-knee}, where there was overlap.  Data
from HiRes-I was scaled at the same level as data from HiRes-II, since
these two data sets are already matched in energy.  The same is true
of the three Yakutsk data sets.  The data from AGASA and Auger are
more problematic since they have no connection to a measured flux at
$10^{18}$ eV.  For these two experiments, we required that the flux at
$10^{19}$ eV match the energy scale of the HiRes Experiment.  This
matching was only required to be within $\sim5$\%.  While a very tight
requirement on match flux could be made at $10^{18}$ eV, the different
position of the fit Ankle break points, gives a range of fit flux
values at $10^{19}$ eV.  Thus, only a loose matching was made for the
AGASA and Auger data.  After scaling, the energy of the Ankle in our
fits ranges from $10^{18.5}$ eV for Fly's Eye, to $10^{18.8}$ eV for
Yakutsk.  The results of the combined fit to the data (which include a
floating break point for the Second Knee) is shown on the right side
of Figure~\ref{ankle-fit-values}, with the fitted parameters given in
Table~\ref{ankle-fit-values}.

\subsection{The GZK Cutoff}
\begin{figure}
  \includegraphics[width=0.49\columnwidth]{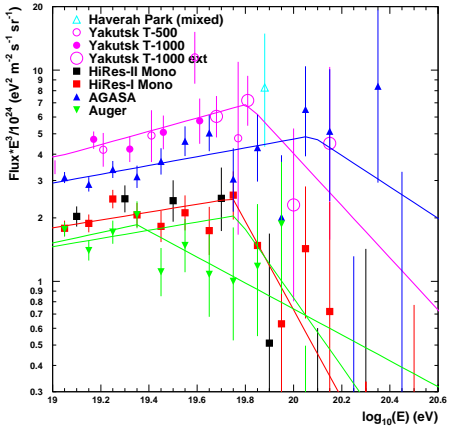}
    \caption{Flux measurements in the GZK Cutoff energy range.  The
      shown fits are our calculation.  }
    \label{gzk-measurements}
\end{figure}

\begin{figure}
  \includegraphics[width=\columnwidth]{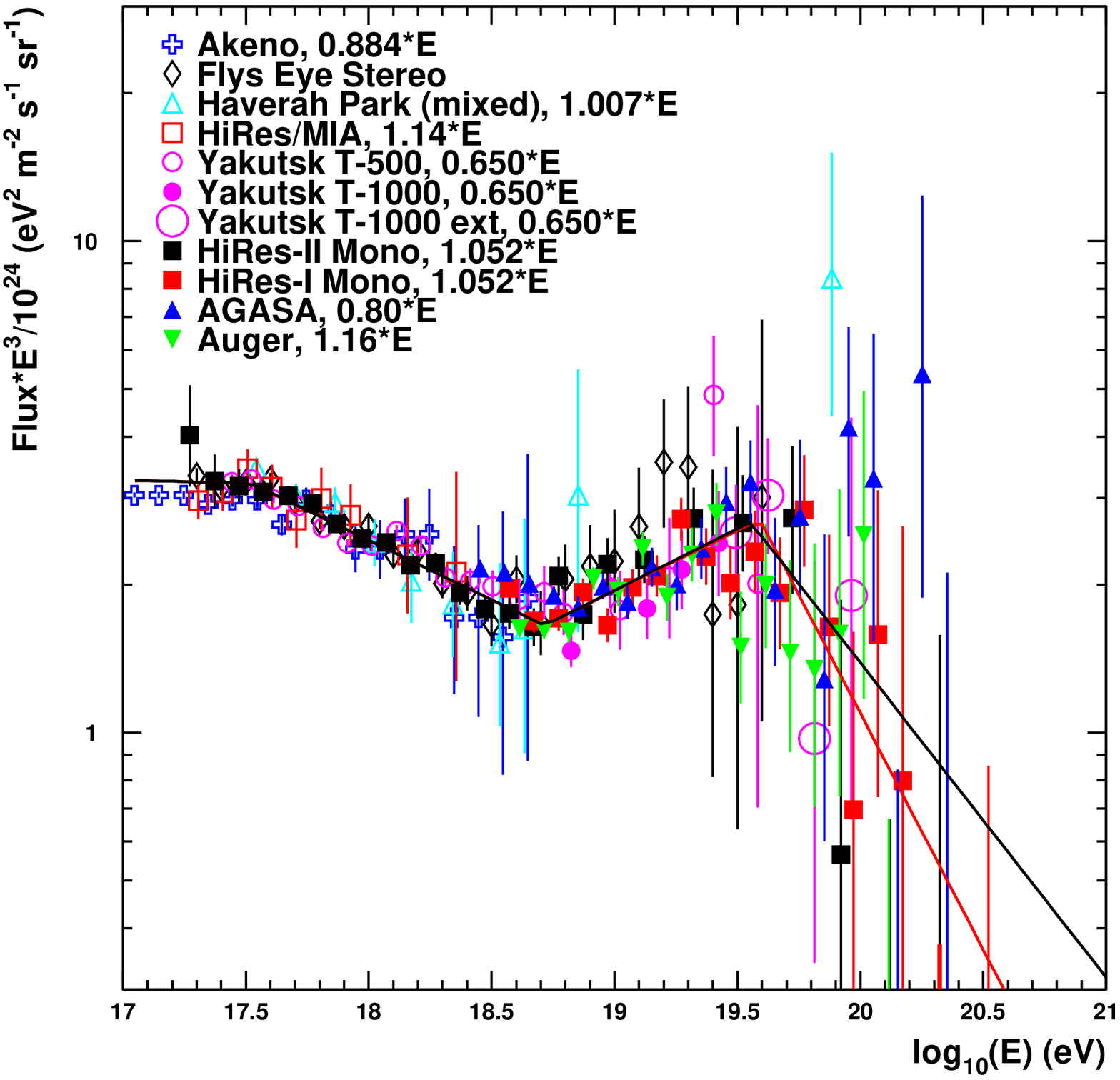}
    \caption{All the flux measurements discussed in this paper fit to
      a broken power law.  The black line shows the result of the fit
      including the AGASA data, while the red line shows the fits
      without including the AGASA data.}
    \label{comprehensive-measurements-fit}
\end{figure}

Four experiments currently claim to have significant aperture above
the expected energy of the GZK Cutoff of $6\times10^{19}$ eV
\cite{Greisen-1966-PRL-16-748,Zatsepin-1966-JETPL-4-78}: AGASA
\cite{Takeda-2003-APP-19-447}, Yakutsk
\cite{Egorova-2004-NPBps-136-3}, HiRes
\cite{Abbasi-2005-PLB-619-271,Abbasi-2007-PRL} and Auger
\cite{Sommers-2005-ICRC-abs1}.  Of these AGASA claims to see a
continuation of the spectrum above the GZK Cutoff with no change in
spectral index.  Yakutsk and HiRes on the other hand claim that their
spectra are consistent with the presence of the cutoff, and HiRes has
claimed an observation of the feature.  Auger in its preliminary
spectrum makes no claim in either direction.  The Haverah Park
experiment had previously claimed to see four events above $10^{20}$
eV \cite{Lawrence-1991-JPG-17-733}, but upon re-analysis, all four of
these events were found to be less than $10^{20}$ eV
\cite{Ave-2003-APP-19-47}.  (The average energy of the four events is
$10^{19.88}$ eV in the re-analysis.  This point is displayed in
Figures~\ref{gzk-measurements}
and~\ref{comprehensive-measurements-fit}, but is not included in any
fits.)

As we have for the other features, we fit each of these results with
our own fit.  In these fits, the empty bins with finite exposure play
a very important role.  The binned maximum likelihood method allows
one to include these bins in the fit.  We first fit each experiment
with no allowed break point for the GZK Cutoff (but including one for
the Ankle in the case of Yakutsk and HiRes).  We then fit allowing a
floating break point for the GZK Cutoff.  In all cases we found a
break, but the significances vary.  The results of these fits are
presented in Table~\ref{gzk-fit-values} and
Figure~\ref{gzk-measurements} (only for the fits with the floating GZK
Cutoff breakpoint).  No account was taken of overlapping exposure in
different measurements for either the Yakutsk or HiRes spectra.

\begin{table}
  \begin{tabular}{|lllll|}
    \hline\hline
    Experiment & $\chi^2/$DOF & Slope & Break Point & Slope \\
    ({\it reference}) &       & Below & $\log_{10}\left(\frac{E}{\rm eV}\right)$
                                                    & Above \\ \hline\hline
    Yakutsk & 
      55.6/24 & $2.73\pm0.06$ && \\
    \cite{Egorova-2004-NPBps-136-3}& 
      51.6/22 & $2.68\pm0.06$ & $19.81\pm0.10$ & $4.2\pm0.9$ \\ \hline
    HiRes  & 
      64.3/37 & $2.88\pm0.03$ && \\
    \cite{Abbasi-2007-PRL} &
      34.6/35 & $2.81\pm0.03$ & $19.75\pm0.04$ & $5.1\pm0.7$ \\ \hline
    AGASA &
      16.1/11 & $2.81\pm0.07$ && \\
    \cite{Takeda-2003-APP-19-447} &
      15.2/9  & $2.79\pm0.07$ & $20.1\pm0.4$   & $3.7\pm2.0$ \\ \hline
    Auger &
      34.8/12 & $2.82\pm0.03$ && \\
    (\emph{local minimum}) & 
      31.4/10 & $2.81\pm0.03$ & $19.8\pm0.2$   & $4.6\pm2.5$ \\
    \cite{Sommers-2005-ICRC-abs1} &
      21.1/10 & $2.76\pm0.03$ & $19.35\pm0.07$ & $3.6\pm0.3$ \\ \hline\hline
  \end{tabular}
  \caption{Our broken power law fits to spectrum measurements in the
    GZK Cutoff energy range.  The fit parameters include a
    normalization (not shown), slope parameters above and below the
    break and the break point energy for the GZK Cutoff.  The first
    fit for each experiment shows the slope and $\chi^2$ for a fit
    with no allowed GZK Cutoff break.  A local  $\chi^2$ minimum was
    found fitting the Auger data with a break point near the HiRes
    value.}
  \label{gzk-fit-values}
\end{table}

It is clear that HiRes sees a break with a very large significance
(the $\chi^2$ falling from 64.3 to 34.6 while adding only two
degrees-of-freedom), whereas none of the other breaks are nearly as
significant.  The fact that any breakpoint is found in the AGASA data
is interesting, and results largely from the empty bins at
$10^{20.25}$ eV and $10^{20.45}$ eV where one still expects a
significant number of events given the AGASA exposure.  The local
minimum found in fits to the Auger spectrum points, perhaps, to the
fact that fitting this region with a broken power law is reaching its
limits.  Both Auger and HiRes-I spectra seem to indicate a rounded
cutoff.  This could be from resolution however, as HiRes-II seems to
see a very sharp cutoff.

We end the section on flux measurements with a comprehensive fit to
all the data presented above (with the noted exceptions).  The fit has
three floating breakpoints, four spectral slopes and one
normalization.  The parameters are listed in
Table~\ref{comprehensive-fit-values}.  As we did above for the
individual fits in the GZK Cutoff energy range, we first fit without
a breakpoint for the GZK cutoff, then with it.  The improvement in
$\chi^2$ tells us about the significance of the added breakpoint.  We
also did these fits with and without the AGASA data, data which seems
to be at odds with the other experiments in this energy range.  Both
these fits are shown in Figure~\ref{comprehensive-measurements-fit}.
In either case the $\chi^2$ drops by about 41 while adding two degrees
of freedom.  This corresponds to a bit more than 6$\sigma$
significance.  However, while the preponderance of evidence points to
the existence of the GZK cutoff, the calculated significances are
somewhat suspect because of the {\it ad hoc} nature of the scalings
involved.

A final measure of the significance of the break at the GZK Cutoff
energy can be found by comparing observed and expected numbers of
events above $10^{19.8}$ in fits with and without the GZK Cutoff
breakpoint.  With the scaled AGASA data included, one finds 42 events
with energies above $10^{19.8}$ eV, where one would expect 85 from a
fit with no high energy breakpoint.  The expected number drops to 45.4
when one adds the break point.  The statistical probability of
expecting 85 events and observing 42 or less is $1.8\times10^{-7}$, or
just over 5$\sigma$.  If one removes the AGASA data, one observes 28
events, while expecting 67 with no high energy breakpoint or 30.1 with
it.  The probability of observing 28 events when expecting 67 is
$6\times10^{-8}$, or 5.3$\sigma$.

\begin{table}
  \begin{tabular}{|c|cc|cc|}
    \hline\hline
Parameter       & \multicolumn{2}{c|}{with AGASA}& \multicolumn{2}{c|}{without AGASA}\\
                & no GZK BP     & GZK BP        & no GZK BP     & GZk BP        \\ 
\hline
$\chi^2$/DOF    &  264.5/156    &  223.0/154    &  235.1/141    &  193.7/139    \\ \hline
$J_{18}/10^{24}$& $2.44\pm0.01$ & $2.43\pm0.01$ & $2.44\pm0.01$ & $2.43\pm0.01$ \\
$\gamma_1$      & $2.99\pm0.01$ & $2.99\pm0.01$ & $2.99\pm0.01$ & $2.99\pm0.01$ \\
$\log_{10}E_1$  &$17.52\pm0.02$ &$17.52\pm0.02$ &$17.52\pm0.02$ &$17.52\pm0.02$ \\
$\gamma_2$      &$3.228\pm0.007$&$3.230\pm0.007$&$3.228\pm0.007$&$3.230\pm0.007$\\
$\log_{10}E_2$  &$18.66\pm0.03$ &$18.67\pm0.02$ &$18.66\pm0.03$ &$18.69\pm0.02$ \\
$\gamma_3$      & $2.83\pm0.02$ & $2.74\pm0.03$ & $2.85\pm0.02$ & $2.77\pm0.03$ \\
$\log_{10}E_3$  &               &$19.56\pm0.05$ &               &$19.60\pm0.05$ \\
$\gamma_4$      &               & $3.64\pm0.19$ &               & $4.00\pm0.26$ \\ \hline
$N^{\rm obs}_{lE>19.8}$  & 42   & 42   & 28   & 28   \\
$N^{\rm pred}_{lE>19.8}$ & 85.0 & 45.4 & 67.0 & 30.1 \\

\hline\hline
  \end{tabular}
  \caption{Our broken power law fits to all the scaled spectrum
    measurements.  $J_{18}$ is the fit flux at an energy of $10^{18}$
    eV in units of (eV m$^2$ s sr)$^{-1}$.  The breakpoint energies
    are all in units of eV.}
  \label{comprehensive-fit-values}
\end{table}

In conclusion, there is strong evidence for each of the three
acknowledged features in the UHECR spectrum: the Second Knee, the
Ankle and the GZK Cutoff.  The most compelling evidence for the first
two features comes from the large degree of consensus on what the
spectral indexes are at energies above and below each feature.  And
where the spectral index changes there \emph{must} be a feature.  The
exact energy of either feature is not so well in agreement, especially
for the Ankle.  This comes primarily from the difficulty in
determining the absolute calorimetric scale of any experiment.  In
addition, even after trying to adjust the scales of the different
experiments to get identical fluxes at some energy ($10^{18}$ eV in
our case), the energy of Ankle still varies by a factor of two.  This
may point to the Ankle not being a particularly sharp break.

At this point, only the HiRes experiment has presented compelling
evidence of the GZK cutoff.  Moreover, they observe \emph{both} the
Ankle and the GZK Cutoff and so can measure the ratio of the two
energies ($10^{19.75-18.65}=12.6$).  While Yakutsk and Auger don't
have enough data to claim an observation of the Cutoff yet, their data
does add to the significance of the observation by HiRes.

\section{The UHECR Spectrum: Composition Measurements}
Unlike the situation in measuring the flux of cosmic rays, there is
little consensus among experiments on the primary composition of
cosmic rays.  Some of the disagreements, no doubt, come from the
difficulty of the measurements and their indirect nature.  In order to
have an aperture of a large enough area, one is forced to study
Extensive Air Showers (EAS), rather than the primary cosmic rays
themselves.  For measurements of the energy (and therefore the flux)
there is a strong correlation between the energy deposited in the
atmosphere by the EAS and the kinetic energy of the primary cosmic
ray.  It is only in the shape of the shower that one can try to get at
the composition of the primary cosmic rays.

EAS's grow as the primary cosmic ray collides with an atom in the
atmosphere producing many secondary particles which divide the energy
of the primary between them.  The secondaries will also collide with
atoms in the atmosphere and form a cascade of more and more particles
(with less and less energy) as the shower develops.  When the
particles of a shower become too low in energy, they no longer
contribute to the growth of the shower, but are rather absorbed by the
atmosphere.  In this way the shower grows geometrically for a certain
distance through the atmosphere but the growth first begins to slow,
and eventually the shower will begin to shrink.  The depth in the
atmosphere where the shower is at it's largest is usually called
$X_{\rm max}$.  $X_{\rm max}$ will depend logarithmically on the energy
because that energy can be spread among more particles.  The change in
the average $X_{\rm max}$ with $\log E$ is called the elongation rate.

Nuclei heavier than hydrogen (i.e. more than one nucleon) develop, to
first order, as if each constituent nucleon created its own shower.
This means that one would expect iron with 56 nucleons to generate a
shower with a similar $X_{\rm max}$ to that created by a proton 56
times less energetic.  Unfortunately, EAS development isn't so simple
and there are large fluctuations in $X_{\rm max}$ from
shower-to-shower for a given type of primary.  These fluctuations
should be smaller for heavier primaries because one is essentially
averaging over many showers.  At any given energy however, the
distributions of $X_{\rm max}$ for proton and iron showers overlap by
enough that one cannot expect to determine the primary particle type
on an event-by-event basis.

Still the average $X_{\rm max}$ at a given energy and the width of the
distribution should tell one something about the average composition
of UHECR's at that energy.  Unfortunately, there is no consensus on
the absolute value of the average $X_{\rm max}$ to expect for a given
primary type: different models give different results at the level of
20\% of the difference between protons and iron.  Two recent
interaction models used extensively are the QGSJet model
\cite{Kalmykov-1997-NPBps-52b-17} and Sibyll
\cite{Fletcher-1994-PRD-50-5719}.  In either case one must also model
the development of the shower, using programs such as Corsika
\cite{Heck-1998-FZKA-6019} or Aires
\cite{Sciutto-1999-astro-ph-9911331}.  The models agree better in the
determination of the elongation rate, which is expected to be about 60
g/cm$^2$/decade for showers generated by either iron primaries or
proton primaries.  One would also expect the width of the $X_{\rm
  max}$ distribution to be robustly determined by the models, but we
have not seen any predictions of this (and only one published
experimental measurement \cite{Cassiday-1990-ApJ-356-669}).

\begin{figure}
  \includegraphics[width=\columnwidth]{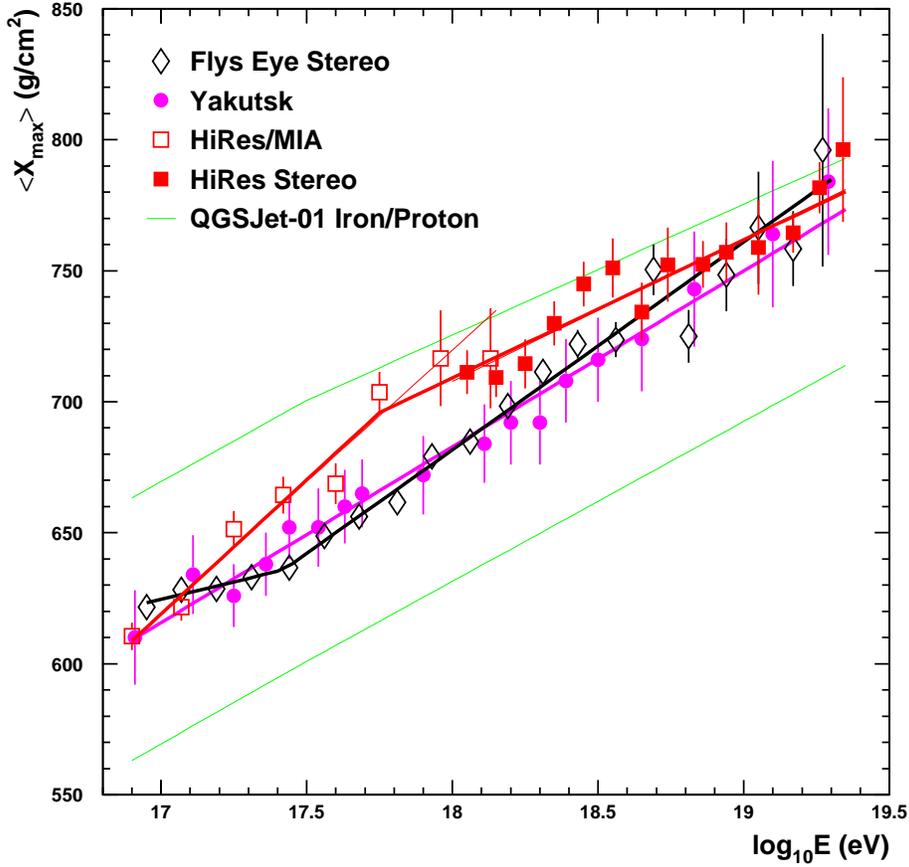}
    \caption{Measurements of the average $X_{\rm max}$ as a function
      of energy.}
    \label{xmax-measurements}
\end{figure}

Only fluorescence detectors can determine $X_{\rm max}$ directly for
each shower, though Yakutsk, with its \v{C}erenkov detectors can make
a measurement as well.  Four experiments have made measurements of the
average $X_{\rm max}$ as a function of energy: Fly's Eye
\cite{Cassiday-1990-ApJ-356-669,Bird-1993-PRL-71-3401}, Yakutsk
\cite{Egorova-2001-JPSJ-70-supB-9}, HiRes/MIA
\cite{AbuZayyad-2000-PRL-19-4276,AbuZayyad-2001-ApJ-557-686} and HiRes
(in stereo) \cite{Abbasi-2005-ApJ-622-910}.  These measurements are
shown in Figure~\ref{xmax-measurements}.  While the figure includes
lines for the QGSJet expectation for iron and proton primaries (the
proton line is higher), one should pay less attention to the absolute
position of the lines than to their slope and separation.

We have fit the data of each experiment (and for HiRes/MIA and HiRes
stereo combined) for the elongation rate, allowing for possibly one
break point in the fit.  The results are shown in
Table~\ref{xmax-fit-values}.  These fits are simple $\chi^2$ fits
where the break point (if any) is one of the fit parameters and the
fit line is required to be continuous in that case.  We allow the
floating break point for fitting the Fly's Eye data following the fit
shown in \citeasnoun{Bird-1993-PRL-71-3401} and for the combined
HiRes/MIA and HiRes stereo data following what is shown in
\citeasnoun{Abbasi-2005-ApJ-622-910}, though they did not in fact
perform a combined fit.

\begin{table}
  \begin{tabular}{|lllll|}
    \hline\hline
    Experiment & $\chi^2/$DOF & Elongation & Break Point & Elongation \\
    ({\it reference}) &       & Rate & $\log_{10}\left(\frac{E}{\rm eV}\right)$
                                                    & Rate \\ \hline\hline
    Yakutsk      &  2.6/17 &&& $67\pm6$ \\
      \cite{Egorova-2001-JPSJ-70-supB-9} &&&& \\ \hline
    Fly's Eye    & 16.6/16 & $27\pm14$& $17.42\pm0.08$ & $79\pm3$ \\
      \cite{Bird-1993-PRL-71-3401}       &&&& \\ \hline
    HiRes/MIA    & 6.0/6   &&& $100\pm8$ \\
       \cite{AbuZayyad-2001-ApJ-557-686} &&&& \\ \hline
    HiRes stereo & 7.4/12  &&& $54\pm6$ \\ 
      \cite{Abbasi-2005-ApJ-622-910}     &&&& \\ \hline
    HiRes/MIA, HiRes & 12.9/18 & $103\pm7$ & $17.75\pm0.10$ & $53\pm5$ \\
    Combined Fit &&&& \\ \hline\hline
  \end{tabular}
  \caption{Our fits of the elongation rates (and break points) to
    experimental data of average $X_{\rm max}$ vs $\log E$.  The fit
    elongation rates have units of g/cm$^2$/decade.}
  \label{xmax-fit-values}
\end{table}

As we stated at the beginning of this section, there is little
consensus in either the average $X_{\rm max}$ values or in the
elongation rates.  The Fly's Eye and Yakutsk data seem to indicate a
slowly changing composition becoming lighter because the elongation
rate is greater than the expected value of $\sim$60 g/cm$^2$/decade
for constant composition.  The change occurs from $10^{17.5}$ eV to
over least $10^{19}$ eV.  They also agree with each other well in the
actual average $X_{\rm max}$ values measured.  These results,
especially that from Fly's Eye which had wide visibility, were used to
argue for a changing composition in conjunction with the changing
spectral slope of the Ankle.

The HiRes Prototype/MIA experiment and the HiRes stereo measurement,
however, tell a different story.  HiRes/MIA measures a more quickly
changing composition, again getting lighter, but one that is complete
by an energy of $10^{18}$ eV.  The HiRes stereo measurement confirms
this by measuring a constant, light composition above $10^{18}$ eV.
This result supports the explanation of the Ankle as being the result
of energy losses of extragalactic protons due to electron-pair
creation
\cite{Berezinsky-Grigoreva-1988-AA-199-1,Berezinsky-2006-PRD-74-043005}. 

It is important to keep in mind that the Fly's Eye measurement quotes
a systematic uncertainty in $X_{\rm max}$ of 20 g/cm$^2$
\cite{Cassiday-1990-ApJ-356-669}, so that the difference in the
measured average $X_{\rm max}$ values between the experiments in not
inconsistent.  In addition, the HiRes Prototype and HiRes proper had
much finer angular resolution than Fly's Eye, which translates into
finer measurements of shower profiles.  Thus, while the statistical
uncertainty in the HiRes/MIA measurements is larger than that of Fly's
Eye, the systematic error may be smaller.

Trying to measure the composition with a surface detector is more
indirect than with fluorescence, and more dependent or shower
modeling.  In the Haverah Park experiment, the risetime of the shower
front as function of distance from the shower core was used to
estimate the composition \cite{Ave-2003-APP-19-61}.  One expects that
the closer the shower maximum is to the ground (i.e. that larger the
value of $X_{\rm max}$) the longer will be the risetime of the shower
front.  This is just the result of the range of pathlengths involved.
Haverah Park find a constant, fairly heavy composition between
$10^{17.2}$ eV and $10^{18}$ eV.

In the AGASA experiment, the ratio of number of muons to the number of
electrons at 600 m from the shower core was used to estimate the
composition
\cite{Shinozaki-Teshima-2004-NPBps-136-18,Hayashida-1995-JPG-21-1101}.
They find that the composition is getting lighter between $10^{17.5}$
eV and $10^{19}$ eV, but at a slower rate than observed by Fly's Eye.

The HiRes Prototype/MIA experiment performed a similar measurement to
that of AGASA: in addition to measuring the average $X_{\rm max}$,
they also measured the density of muons at 600 m from the shower core.
They found a composition which was getting lighter between $10^{17}$
eV and $10^{18}$ eV agreeing with the trend in the $X_{\rm max}$
measurement, but compared to predictions of QGSJet they found a
composition heavier than iron even at $10^{18}$ eV (the situation is
even worse using Sibyll).  This points to a problem with the current
interaction models: they don't produce enough muons.  This problem may
have been corrected in the new interaction code, EPOS
\cite{Pierog-Werner-2006-astro-ph-0611311}.

\section{The UHECR Spectrum and Composition: Synthesis}
As one looks at the various measurements of the UHECR flux, one gets
the impression that we are beginning to understand the spectrum of
cosmic rays at these energies.  Even if the experimental measurements
don't agree in absolute normalization, there in consensus on the
existence of specific breaks in the spectrum and what the spectral
indexes are on either side of the breaks.  This impression is,
however, reversed the moment one begins to consider composition
measurements.  Not only is there no agreement on where showers of a
given energy develop, an average, in the atmosphere, but there is no
agreement on the trend.  Is the composition of UHECR's getting lighter
or staying the same?  Experiments disagree.

We haven't discussed theoretical or phenomenological models for the
acceleration and propagation of UHECR's in this review.  They are, of
course, crucial for understanding the flux of cosmic rays, but even
the best model has trouble dealing with contradictory data.  This is
especially evident in models of the Ankle region, where two different
paradigms for the origin of the Ankle continue to coexist because of
the different elongation rate measurements.

To allow for more robust comparisons between data and models one must
obtain data that provides more constraints on the models.  In
measurements of the flux of cosmic rays this means that one would like
to have a measurement of two or even all three of the spectrum
features within one experiment.  This is a difficult undertaking, as
the features are roughly each an order-of-magnitude apart in energy,
and measuring all three features requires a useful dynamic range of
over two and a half orders-of-magnitude.  The Telescope Array (TA)
experiment \cite{Kawai-2005-ICRC-29-8-141}, which is currently being
deployed in Delta, UT, USA, is one experiment which may be able to
cover this whole energy range if its Low Energy extension is funded.
In any case, it is important to realize that we must understand more
than just the GZK Cutoff.

In measurements of composition, having more constraints also means
having a large dynamic range in energy, of course, but it also could
mean having more ways, orthogonal way, of determining the primary
particle type.  The large hybrid experiments, Auger and TA, will
hopefully provide this, by measuring both longitudinal ($X_{\rm max}$)
and lateral distributions for each shower.

\section{Anisotropy Measurements of UHECR's}

While a great deal may be inferred about the sources of UHECR from
spectral and composition studies, the evidence thereby obtained is
indirect. A complete picture would ideally combine an acceleration
model or models with a consistent primary particle arrival direction
distribution.

As we discuss below, the detection of anisotropies in UHECR event
arrival directions is a challenging enterprise, which to date has
yielded results which are ambiguous at best. But even a completely
isotropic cosmic ray sky has something to tell us about source
models. For example, at energies below a few $\times 10^{19}$~eV
isotropy can be understood as the effect of diffusion by galactic
magnetic fields, provided the primary particles are not neutral. At
higher energies for which magnetic bending effects are negligible,
isotropy indicates that sources are abundant and distributed
throughout the Universe.

We begin our discussion of anisotropy studies by reviewing the
arrival-direction resolutions of the experiments whose data we
consider. We will look at searches for both large-scale features,
including full-sky dipole moments, and small-scale or point-like
excesses of UHECR. We conclude with some general comments about
lessons learned in anisotropy studies over the past decade.

\subsection{Experimental Arrival-Direction Resolutions}
\label{sec-anis-resolution}
It is appropriate that we summarize the arrival direction resolutions
of the various experiments we will consider in this review, namely
AGASA, HiRes, Auger, SUGAR and Yakutsk. By ``resolution'', we
generally mean the size of the opening angle which would contain 68\%
of events arriving from a point source of cosmic rays. However in some
detectors, in particular the monocular fluorescence detectors covered
in this review, arrival direction uncertainties cannot easily be
expressed with a single number.

\begin{figure}
  \begin{minipage}[t]{0.54\columnwidth}
    \includegraphics[width=\columnwidth]{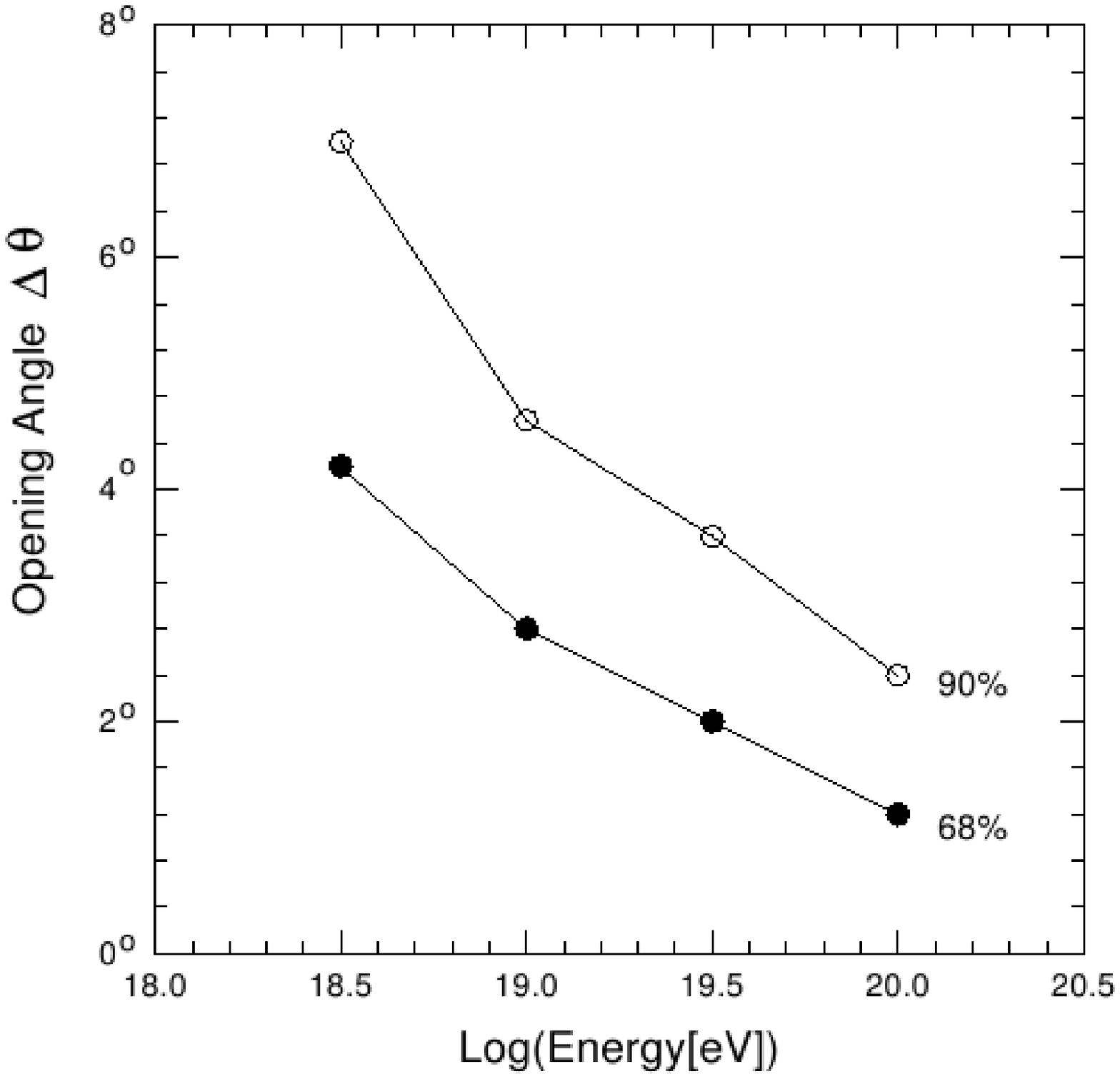}
  \end{minipage}
  \begin{minipage}[t]{0.45\columnwidth}
    \includegraphics[width=\columnwidth]{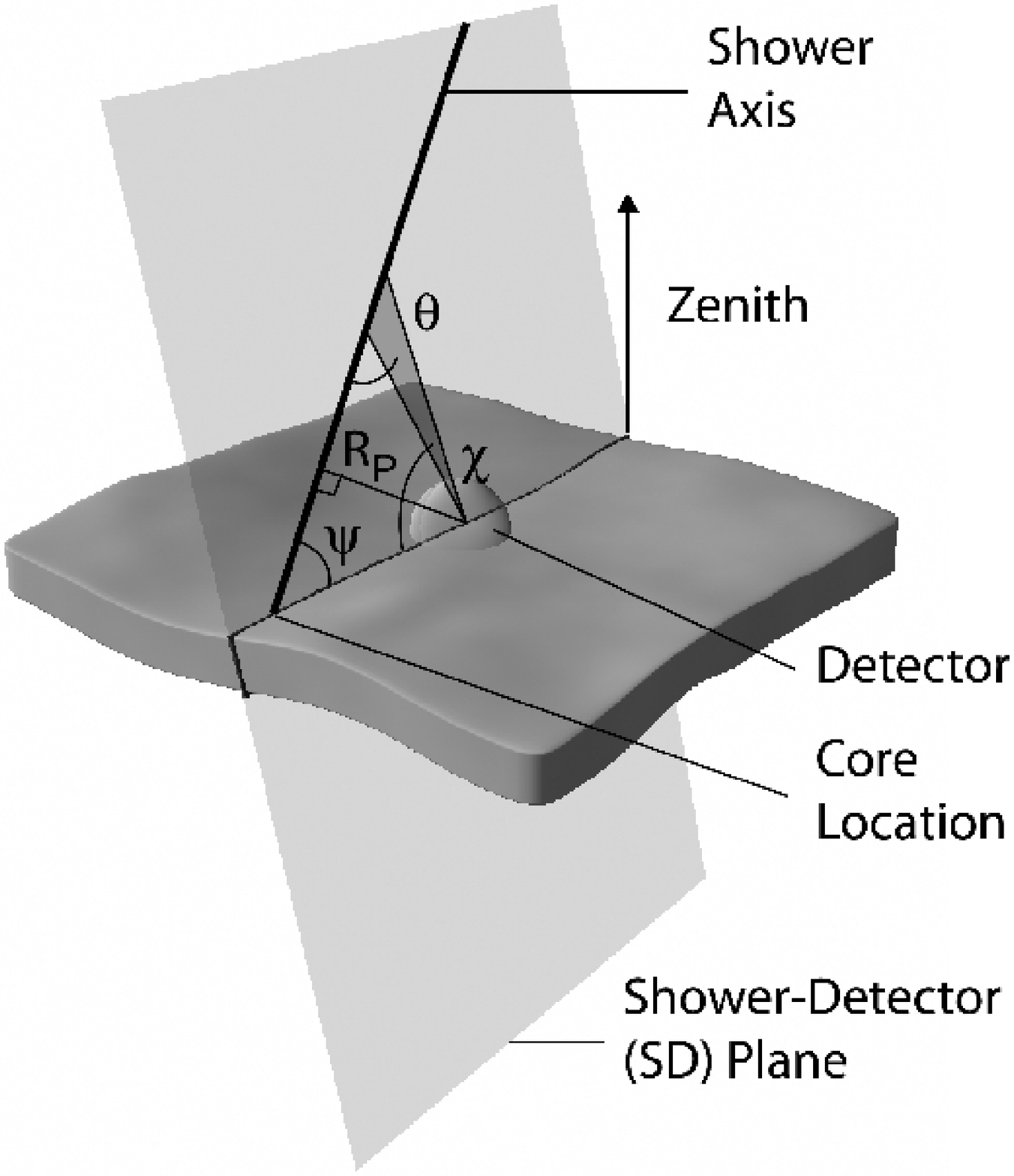}
  \end{minipage}
  \caption{Left: Arrival direction uncertainty for the AGASA
    experiment. Closed and open circles are the opening angles
    encompassing 68\% and 90\% of the data, respectively. Source:
    \citeasnoun{Takeda-1999-ApJ-522-225}. Right: The geometry of
    reconstruction for monocular air fluorescence detector. $R_P$ is
    the shower's distance of closest approach, $\Psi$ is the angle of
    the shower within the shower-detector plane (uncertainty
    parametrized in Equation~\ref{eq-psi}), and $\chi$ is the angular
    tracklength of the shower.}
  \label{resolution_figs}
\end{figure}

The arrival direction uncertainties of the AGASA surface detector (SD)
array are ``circular'' (uniform in two angular dimensions), and
decrease with energy, as shown in Figure~\ref{resolution_figs}
\cite{Takeda-1999-ApJ-522-225}. In this review we consider AGASA data
including events with energies as low as $10^{17}$~eV. In the
literature, angular resolutions of $3^{\circ}$ are reported at
$10^{18}$~eV \cite{Hayashida-1999-APP-10-303}) but comparison with
Figure~\ref{resolution_figs} suggests that a different definition of
resolution is being employed.

For the HiRes air fluorescence observatory, there are large
differences in angular resolution in monocular and stereo detection
modes. Monocular reconstruction by air fluorescence is subject to
asymmetric arrival direction uncertainties. Consider
Figure~\ref{resolution_figs} (right). The angle of the shower-detector
plane (SDP) is generally well reconstructed, for the HiRes--I
monocular detector uncertainties in the SDP are parametrized as
\begin{equation}
\label{eq-sdp} \sigma_{SDP} =
88.2^{\circ}e^{-\Delta\chi/1.9595}+0.37^{\circ}
\end{equation}
where $\Delta\chi$ is the angular tracklength of the shower in
degrees. The angle of the track within the SDP, $\Psi$, is less well
reconstructed and is parameterized by
\begin{equation}
\label{eq-psi} \sigma_{\Psi} =
18.4^{\circ}e^{-\log_{10}(E)/0.69085}+4.1^{\circ}
\end{equation}
where the energy $E$ is expressed in EeV ($10^{18}$~eV). The
parameterizations of equations~\ref{eq-sdp} and~\ref{eq-psi} are
carried out in ~\citeasnoun{Stokes-2005-Utah-Thesis}.

The HiRes observatory operating in stereo derives its
arrival-direction reconstruction from the intersection of two
well-measured SDPs. Unlike the case for surface detectors, where
higher energy events generally have better hit statistics and hence
improved angular resolution, fluorescence observatories tend to see
angular resolutions worsen with energy because at higher energies,
showers are on average farther away.
\citeasnoun{Abbasi-2004-ApJ-610-L73} report 68\% opening angles of
$0.57^{\circ}$, $0.61^{\circ}$, and $0.69^{\circ}$ for showers
observed at $10^{19}$~eV, $4 \times 10^{19}$~eV, and $10^{20}$~eV
respectively.

\begin{figure}[h]
  \begin{center}
    \includegraphics[width=0.50\columnwidth]{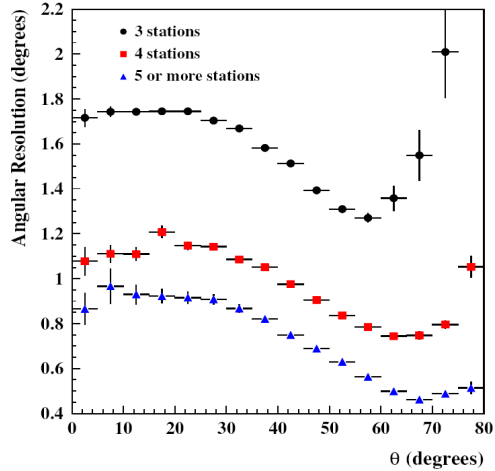}
  \end{center}
  \caption{\label{auger_ang_res} Angular resolution (68\% opening
    angles) for the Auger Surface Detector (SD)as a function of zenith
    angle $\theta$ for various SD station multiplicities. Circles:
    3~stations or approximately $E < 4 \hspace{0.1cm} {\rm EeV}$.
    Squares: 4~stations or approximately $3< E < 10 \hspace{0.1cm}
    {\rm EeV}$. Triangles: More than 4~stations or approximately $E >
    8 \hspace{0.1cm} {\rm EeV}$. Source:
    \citeasnoun{Bonifazi-ICRC-29-7-17}.}
\end{figure}

\citeasnoun{Bonifazi-ICRC-29-7-17} describe the angular resolution of
the Pierre Auger observatory, for both surface detector (SD) and
hybrid (fluorescence and SD) detection modes. The best reconstruction
occurs in hybrid mode, in which quoted uncertainties (68\% opening
angles) of $0.6^{\circ}$ are determined by observing
Rayleigh-scattered light from laser shots. Angular resolutions for
SD-only events are determined by comparing SD reconstruction with the
hybrid results, and are found to depend on both the zenith angle
(Figure~\ref{auger_ang_res}) and SD station hit multiplicity.
Generally SD angular resolution is found to be better than
$2.2^{\circ}$ for 3-hit events (approximately $E < 4 \hspace{0.1cm}
{\rm EeV}$), better than $1.7^{\circ}$ for 4-hit events (approximately
$3< E < 10 \hspace{0.1cm} {\rm EeV}$), and better than $1.4^{\circ}$
for events with multiplicity of 5 or more (approximately $E > 8
\hspace{0.1cm} {\rm EeV}$).

The SUGAR air shower detector, consisting of buried liquid
scintillator detectors at 50~meter spacing, quotes a directional
uncertainty of $3^{\circ}\sec{\theta}$ for airshowers at zenith angle
$\theta$ \cite{Bellido-2001-APP-15-167}. Again, this number
corresponds to the space angle that would include 68\% of events from
a point source. The parameterization should be regarded as an average
over energies between $10^{17.9}$ and $10^{18.5}$~eV.

The error in arrival direction for the Yakutsk array is $3^{\circ}$
for the primary cosmic rays with energies above $4 \times 10^{19}$~eV
discussed in this review \cite{Uchihori-2000-APP-13-151}.

\subsection{Large Scale and Dipole Anisotropy Searches}
\label{sec-anis-dipole}
At lower energies, for which magnetic diffusion may be expected to
play a major role in determining event arrival directions, the most
likely scenario for observing anisotropy is in extended excesses over
the sky.  (An exception applies to models in which the primary
particles are neutral; We consider this possibility below in
Section~\ref{sec-pointlike_low_e}.)  In some cases, such excesses may
be manifested as dipole moments in the full-sky arrival direction
distribution.

Due to near-uniform exposure of ground arrays in Right Ascension (RA),
the study of harmonic structure --- particularly the dipole moment ---
in RA has historically been an important tool in the search for
large-scale anisotropy
\cite{Linsley-1975-PRL-34-1530,Sokolsky-1992-PRp-217-225}. Since in an
isotropic sky events will be distributed uniformly in RA ($\psi$), the
dipole fluctuation in intensity $I$ is well parametrized by
\begin{equation}
\label{eq-radipole}
I(\psi) = I_0(1+r \cos{(\psi - \phi)})
\end{equation}
where $r$ is known as the harmonic amplitude and $\phi$ is the phase
of the dipole anisotropy.

The variables $r$ and $\phi$ are commonly extracted by treating them
as the magnitude and phase of a Rayleigh vector $\vec{r} = (x,y)$,
with elements determined from the $n$ observed events in the dataset
according to
\begin{equation}
x = \frac{2}{n} \sum_{i=1}^n \cos(\psi_i), \hspace{1.0cm} y =
\frac{2}{n} \sum_{i=1}^n \sin(\psi_i) 
\end{equation}
where $\psi_i$ is the RA of the $i$th event.

The chance probability of observing in an isotropic dataset a harmonic
amplitude of strength $\geq r$ is given by the Rayleigh formula
\begin{equation}
P(\geq r) = e^{-nr^2/4}
\end{equation}
As pointed out by \citeasnoun{Linsley-1975-PRL-34-1530} however, only
for $r \gg 2/\sqrt{n}$ ({\em i.e.} large anisotropy and/or large $n$)
can $r$ be interpreted as a credible estimator of the true anisotropy.

\begin{figure}
  \begin{minipage}[t]{0.45\columnwidth}
    \includegraphics[width=\columnwidth]{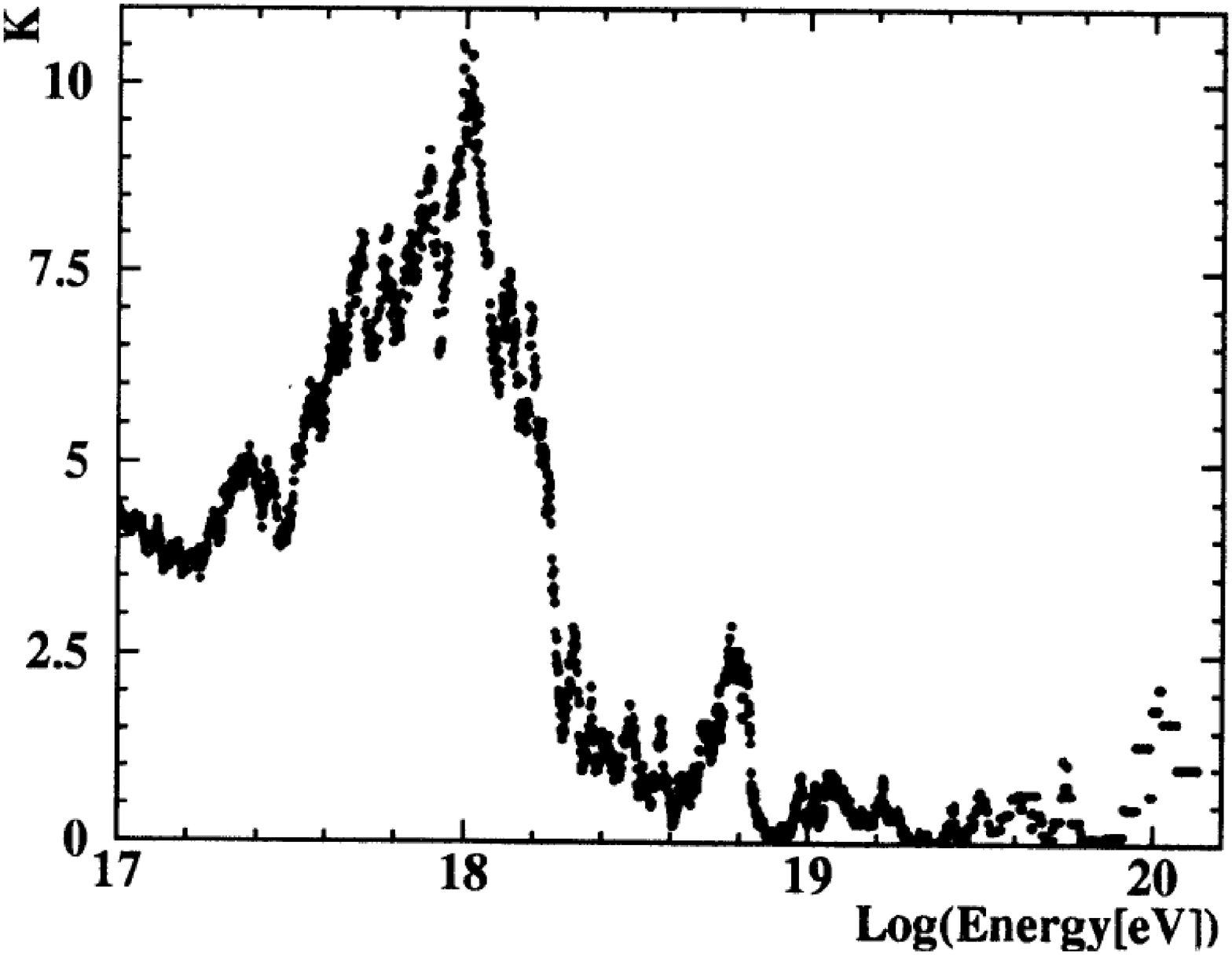}
  \end{minipage}
  \begin{minipage}[t]{0.55\columnwidth}
    \includegraphics[width=\columnwidth]{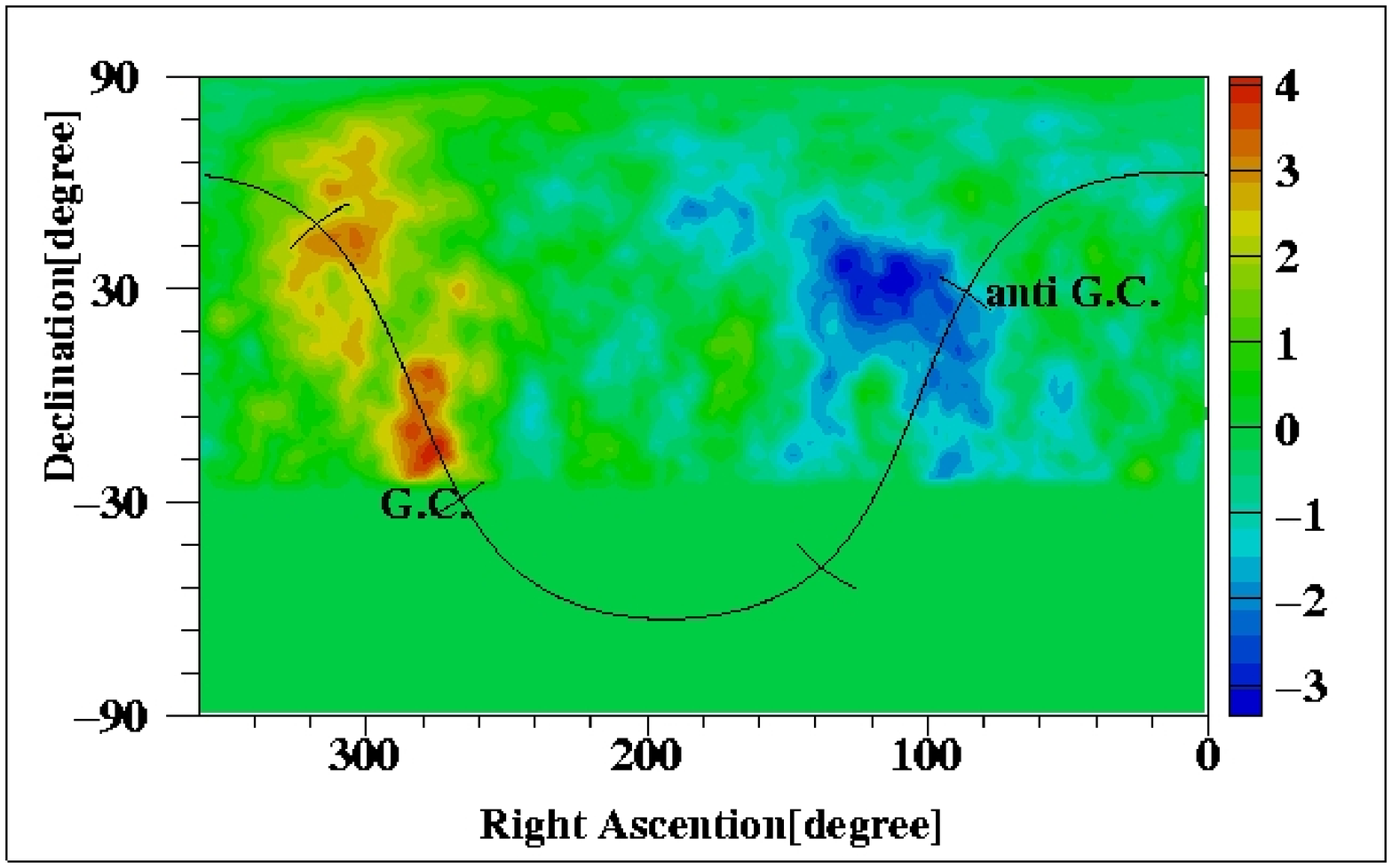}
  \end{minipage}
  \caption{Left: $K = nr^2/4$ for energy thresholds between $10^{17}$
    and $10^{20}$~eV, for AGASA data taken prior to July 1995. Right:
    Significance map of excess and deficit events for energies $17.9 <
    \log{E({\rm eV})} < 18.3$. Events are summed within a $20^{\circ}$
    search circle, and compared with expectations from an isotropic
    background. Source: \citeasnoun{Hayashida-1999-APP-10-303} and the
    AGASA web site \citeasnoun{AGASA-web-2003}}
  \label{agasa_dipole}
\end{figure}

\citeasnoun{Hayashida-1999-APP-10-303} report the results of a
large-scale anisotropy search using the AGASA data around 1~EeV. First
harmonic (dipole) amplitudes in RA are calculated for data by energy
threshold, for thresholds between $10^{17}$ and $10^{20}$~eV
(Figure~\ref{agasa_dipole}), for events having a zenith angle of $\leq
60^{\circ}$. There are approximately 114,000 events in the sample used
in this study.  Noting the high-significance effect near a threshold
of $10^{18}$~eV, the authors perform logarithmically-binned first
harmonic analyses and find a harmonic amplitude $r = 4.1\%$ for the
$1.0 < E(EeV) < 2.0$ bin, which contains 18,274 events. The chance
probability of this effect in this bin is 0.00035.  Taking into
account the number of logarithmic ``energy bins'' of 0.5 decade width
in the AGASA data (six), the authors arrive at a chance probability of
$\sim 0.21\%$ for this amplitude to be a statistical fluctuation. The
authors do not claim to take into account any scan over bin sizes that
may have been performed to maximize the observed signal. Considering
the significance of fluctuations over the full sky map
(Figure~\ref{agasa_dipole}), \citeasnoun{Hayashida-1999-APP-10-303}
interpret the observed dipole behavior as evidence for a excess in the
direction of the galactic plane: In a $20^{\circ}$ radius circle, 308
events are observed where 242.5 are expected (27\% excess).

\citeasnoun{Abbasi-2004-APP-21-111} present the results of searches
for dipole anisotropy in the HiRes--I monocular data using a somewhat
different technique, because the RA method is limited by its
insensitivity to dipoles with predominantly north-south orientation.
In this method, specific candidate dipole centers --- the galactic
center, Cen A \cite{Farrar-Piran-2000-astro-ph-0010370}, and M87
\cite{Biermann-2000-astro-ph-0008063} --- are identified {\em a
  priori} on the basis of theoretical models. Opening angle
distributions are plotted, and compared to those observed in simulated
data sets with dipole anisotropies of varying strengths.

\begin{figure}[h]
  \begin{minipage}[t]{0.50\columnwidth}
    \includegraphics[width=\columnwidth]{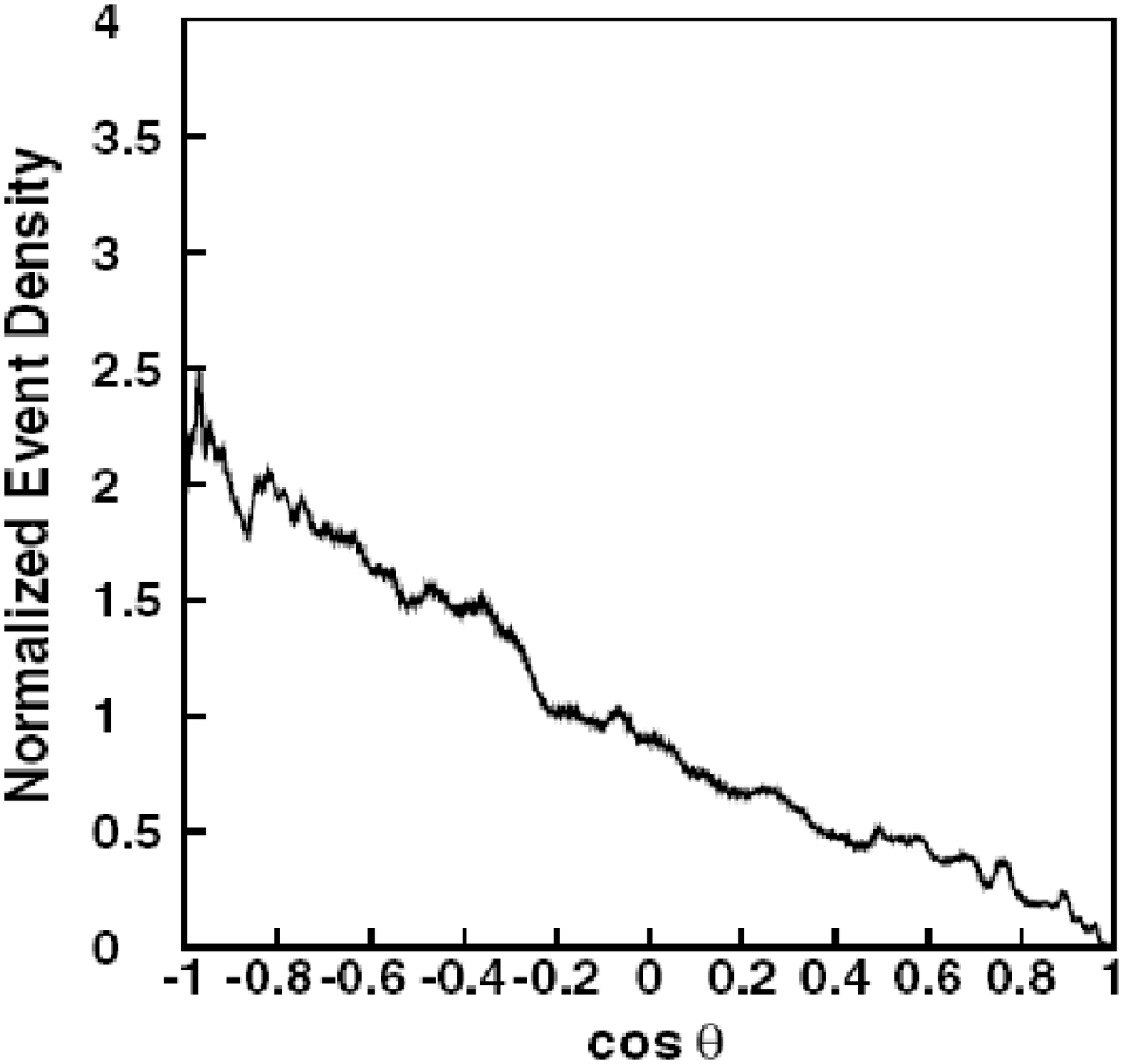}
  \end{minipage}
  \begin{minipage}[t]{0.50\columnwidth}
    \includegraphics[width=\columnwidth]{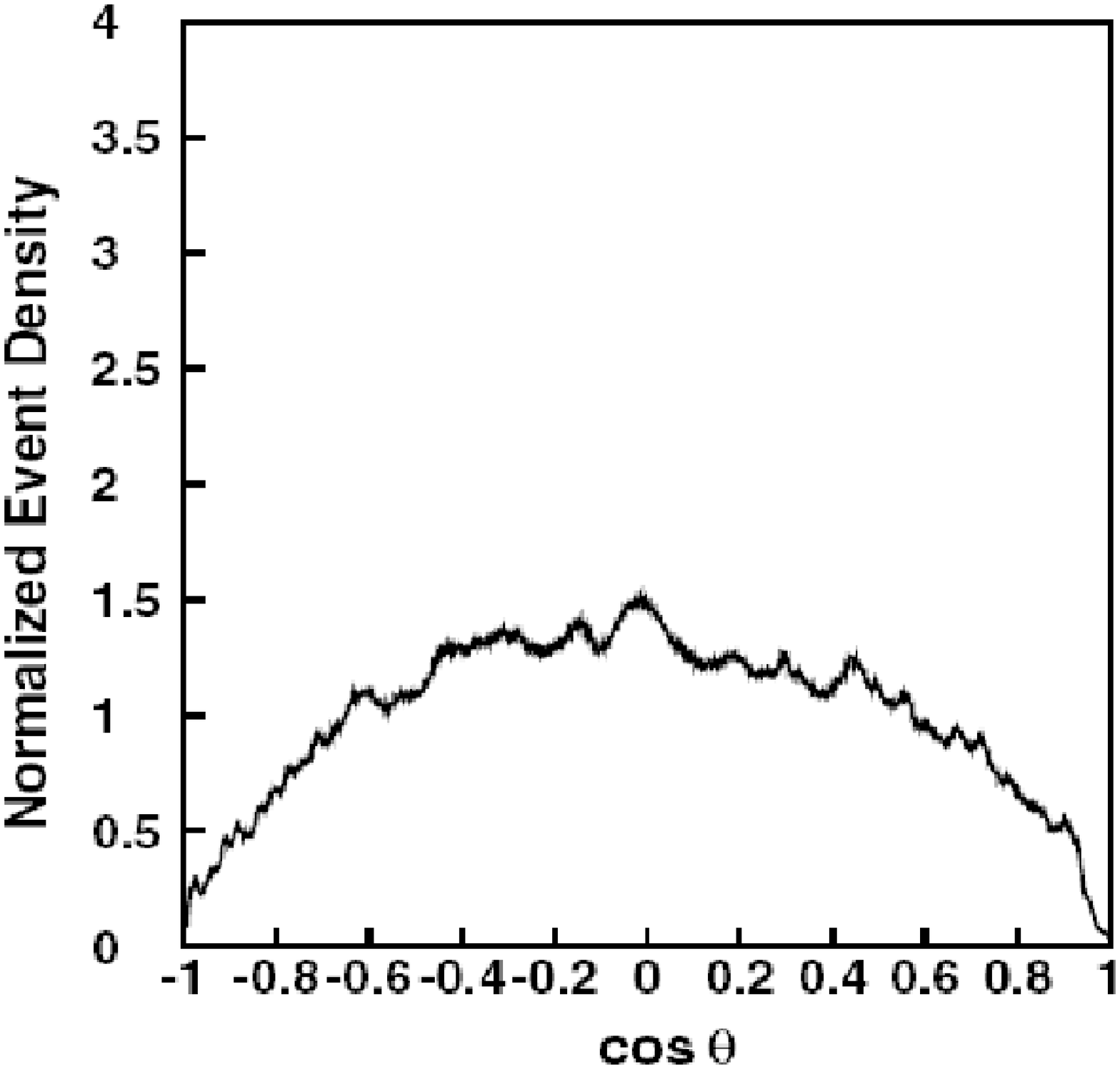}
  \end{minipage}
  \caption{Left: Galactic center dipole function, with angular
    resolution included (as described in the text) for HiRes--I
    monocular data set. Right: Galactic center dipole function for
    simulated data set with $\alpha = 1$. Source:
    \citeasnoun{Abbasi-2004-APP-21-111}.}
  \label{hires_dipole_costheta}
\end{figure}

In this analysis, the dipole anisotropy is parametrized as
\begin{equation}
\label{eq-stokesdipole}
I(\theta) = I_0(1+\alpha \cos{\theta})
\end{equation}
where $\theta$ is the opening angle between a given direction and the
global maximum of the distribution, and $\alpha$ is a measure of the
strength of the dipole. Note that $\alpha$ is {\em not} equivalent to
the harmonic amplitude $r$: $\cos{\theta}$ rather than $\theta$ is
flat in the case of isotropy under this parameterization, and $\alpha$
is the slope of $I$ versus $\cos{\theta}$. In general, $r$ is related
to $\alpha$ by
\begin{equation}
\label{eq-ralpha}
r = \frac{\pi}{4} \alpha \cos{\delta} 
\end{equation}
for a dipole with amplitude $\alpha$ centered at declination $\delta$.
Also note that there is no phase in this parameterization, the peak of
the harmonic being defined as the coordinates of the {\em a priori}
specified source.

The HiRes--I monocular dataset used by
\citeasnoun{Abbasi-2004-APP-21-111} consists of 1,526 events observed
between May 1997 and February 2003. The analysis technique used
consists of histogramming the cosine of the space angle between all
events and the maximum of the hypothetical dipole model
(Figure~\ref{hires_dipole_costheta}). As described in the previous
section, airshowers reconstructed by the monocular air fluorescence
technique feature highly asymmetric (elliptical) uncertainties. To
take this into account, each ``event'' is actually represented by a
large number of points spread according to the uncertainty functions
(Equations \ref{eq-sdp} and \ref{eq-psi}), in the production of this
histogram.

By comparing the ``dipole functions'' thus obtained (specifically, the
value of $\langle \cos{\theta} \rangle$ of the functions) to the
dipole functions for simulated data sets with varying $\alpha$, best
fit values for $\alpha$ are obtained. This analysis is performed for
each of the three {\em a priori} source candidates, with the following
results: Galactic dipole $-0.085 < \alpha < 0.090$ (90\% c.l.), Cen A
$-0.090 < \alpha < 0.085$ (90\% c.l.), and M87 $-0.080 < \alpha <
0.070$ (90\% c.l.).

A negative $\alpha$ in this formulation corresponds to a different
source model entirely from that of a positive $\alpha$. Specifically,
the maximum of the first harmonic for negative $\alpha$ occurs at
$180^{\circ}$ from the positive $\alpha$ maximum. While this
formulation is perhaps computationally easier, we feel that it muddies
the interpretation of confidence limits somewhat and that therefore a
formulation with positive-definite $\alpha$ is preferred.
Nevertheless, comparisons can be made between this and prior results.

The HiRes-I limit on $\alpha$ (galactic dipole) translates to a limit
of $-0.058 < r < 0.062$(90\% c.l.) (using Equation~\ref{eq-ralpha})
for the harmonic amplitude. This result therefore does not exclude the
prior AGASA result. In any case, the range of energies probed by the
HiRes--I monocular dataset does not overlap the energy range in which
the AGASA excess was claimed.

\begin{figure}[h]
  \begin{center}
    \includegraphics[width=0.67\columnwidth]{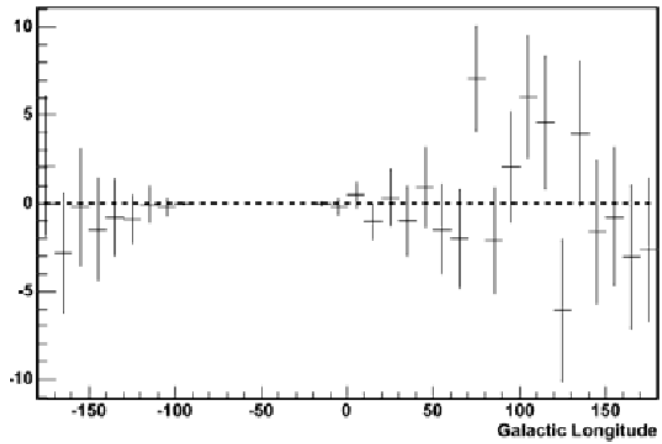}
  \end{center}
  \caption{\label{hires-stereo-galctr} The number of excess HiRes
    stereo events as a function of Galactic longitude for a $\pm
    10^{\circ}$ region around the galactic equator
    \cite{Sinnis-2005-ICRC-abs1}.}
\end{figure}

The HiRes stereoscopic dataset extends to lower energies than HiRes-I
monocular, and is thus better suited to compare to the AGASA result. A
complete harmonic analysis of the HiRes stereo data has not been
reported in the literature, however an analysis of event excesses as a
function of galactic longitude is reported by
\citeasnoun{Sinnis-2005-ICRC-abs1}.

The data analyzed was collected between September 1999 and February
2004, and consists of 4,651 events passing quality and weather cuts
and possessing a zenith angle less than $70^{\circ}$. 1,438 events are
in the AGASA energy band between 1-2~EeV.
Figure~\ref{hires-stereo-galctr} shows the distribution in Galactic
longitude of the number of excess events above background, for a $\pm
10^{\circ}$ region along the Galactic equator. In the
$0^{\circ}$-$100^{\circ}$ longitude region, HiRes measures 59 events
against an expected background of 55.8 events, and places a 90\% c.l.
upper limit of 14.4 events (26\%) excess. They conclude that no
evidence for Galactic center or Cygnus region excesses exist in the
HiRes data, however this result is not inconsistent with the claimed
AGASA excess.

\begin{figure}[h]
  \begin{center}
    \includegraphics[width=0.80\columnwidth]{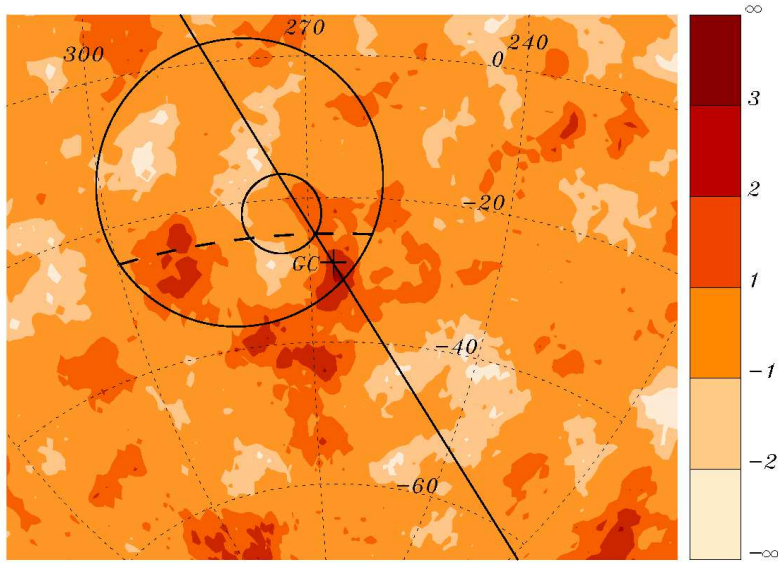}
  \end{center}
  \caption{\label{auger-galctr} Map of Li-Ma overdensity significances
    near galactic center (GC) region of skymap. Overdensities are
    calculated over windows of $5^{\circ}$ radius, for events with
    energy between $10^{17.9}$ and $10^{18.5}$~eV. The GC is indicated
    by a cross, lying along the galactic plane (solid line). The
    regions of the AGASA (large circle) and SUGAR (small circle)
    excesses are also shown. From
    \citeasnoun{Abraham-2007-APP-27-244}.}
\end{figure}

The first anisotropy results by the Pierre Auger observatory are
reported in \citeasnoun{Abraham-2007-APP-27-244}. This paper
aims at providing a direct test of both the AGASA and SUGAR (below,
Section~\ref{sec-pointlike_low_e}) excesses in the vicinity of the
galactic center. Figure~\ref{auger-galctr} is the Auger map of the
galactic center, depicting Li-Ma~\cite{Li-Ma-1983-ApJ-272-317}
significances. These significances are calculated over windows of
$5^{\circ}$ radius, for events with energy between $10^{17.9}$ and
$10^{18.5}$~eV. With approximately four times AGASA statistics (2,116
events) in a $20^{\circ}$ circle centered on the AGASA excess
location, no significant excess is seen. For the range $10^{17.9} < E
< 10^{18.5}$~eV, Auger reports $n_{obs}/n_{exp} = 1.01 \pm 0.02$, in
contrast to the 27\% excess reported by AGASA. Similar results are
also reported for other nearby energy bands, as a check against energy
scale discrepancies between the two experiments.

\subsection{Searches for Small Scale and Point-Like Anisotropy}
\label{sec-anis-point}
Point-like arrival direction excesses may reasonably be expected under
two scenarios: Charged primaries with sufficient energy that they
experience little magnetic deflection in the journey from their
source, or neutral particles sufficiently long-lived to survive the
journey. We consider searches appropriate to both models in the
following discussion.

Over the past decade, considerable attention has been focused on
reports of arrival direction ``clustering'' at the highest energies by
the AGASA experiment. We begin our look at point-like excesses with
these reports. As there has been much debate over the statistical
significance of the AGASA clustering signal, we also will consider two
reanalyses of AGASA data by independent groups. Then, we describe
searches for similar effects in other experiments.

We consider the results of searches for point-like excesses at lower
energies, motivated by an apparent excess in the galactic center found
in a SUGAR reanalysis and by neutron-source models. Finally we
consider searches for correlation of UHECR with the BL-Lacertae class
of Quasi-Stellar Objects (QSOs).

\subsubsection{The AGASA Clustering Signal} 

``Clustering'' of arrival directions for cosmic rays with energies
above $4 \times 10^{19}$~eV was first reported by AGASA in
\citeasnoun{Hayashida-1996-PRL-77-1000}. In a sample of 36 events,
three pairs (labeled C1, C2, and C3) were reported to have an angular
separation of less than $2.5^{\circ}$ in a dataset with arrival
direction uncertainty of about $1.6^{\circ}$
(Figure~\ref{resolution_figs}).  Using simulated isotropic data sets,
a chance probability ($P_{ch}$) of such grouping was estimated to be
2.9\%.

Additional clustering above $4 \times 10^{19}$~eV using an expanded
data set of 47 events was reported in
\citeasnoun{Takeda-1999-ApJ-522-225}.  An additional pair (C4) was
found, and cluster C2 was promoted to the status of a ``triplet'' with
the addition of a third event within $2.5^{\circ}$.  By lowering the
energy cutoff to $3.89 \times 10^{19}$~eV (and gaining a 48$th$
event), the authors were able to identify a fifth cluster C5 as well.
In this paper, the authors estimated the chance probability of
observing the same or a greater number of doublets (excluding C5) in
an isotropic data set to be 0.32\% (counting the triplet as three
doublets) and the same or a greater number of triplets in an isotropic
data set to be 0.87\%.

A sixth cluster (C6) was reported in
\citeasnoun{Hayashida-2000-astroph-0008102}, an update of the AGASA
energy spectrum and cluster map. This update included arrival
direction information from an additional 10 events above $4 \times
10^{19}$~eV. Using the same technique as
\citeasnoun{Takeda-1999-ApJ-522-225}, the triplet chance probability
was revised to 1\%. No revised probability was given for the doublets.

\begin{figure}[h]
  \begin{center}
    \includegraphics[width=1.00\columnwidth]{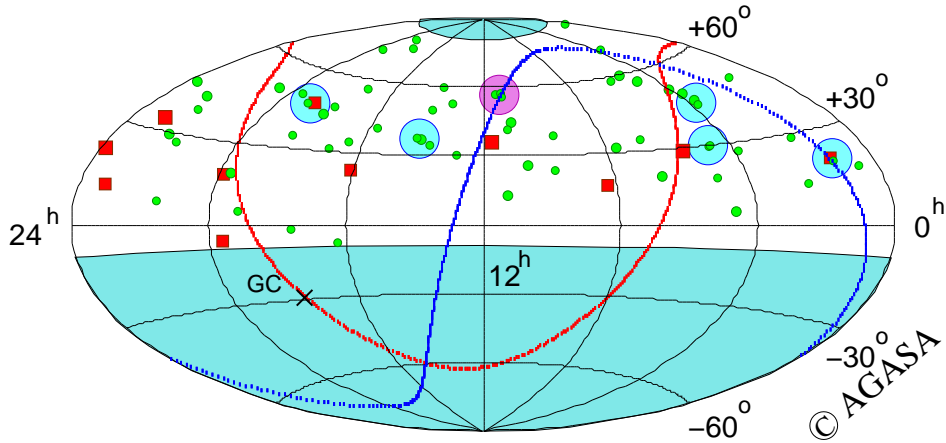}
  \end{center}
  \caption{\label{agasa_clusters_2003} AGASA events with energy above
    $4 \times 10^{19}$~eV superimposed on a skymap, in equatorial
    coordinates. Green circles represent events with energies of
    $(4-10) \times 10^{19}$~eV, red squares represent events with
    energies of $\geq 10^{20}$~eV. Clusters C1-C7 (C5 excluded) are
    identified, doublets with large blue circles and the triplet with
    a violet circle. The galactic (supergalactic) plane is represented
    by a red (dark blue) curve. Figure taken from the AGASA web site
    \cite{AGASA-web-2003}.}
\end{figure}

At the 27$th$ ICRC, \citeasnoun{Takeda-2001-ICRC-27-341} reported an
analysis based on a total of 59 events, with five doublets and a
triplet. It is not clear if this sample consists of one or two new
events added to the \citeasnoun{Hayashida-2000-astroph-0008102}
sample: Energies below $4 \times 10^{19}$~eV are not mentioned, and in
the most recent skymaps (Figure~\ref{agasa_clusters_2003}) a seventh
cluster (C7) replaces C5. The arrival time and energy of one of the C7
events has not been made publicly available, however we have extracted
the coordinates from Figure~\ref{agasa_clusters_2003}. The available
information on all events comprising the seven AGASA clusters is
summarized in Table~\ref{agasa_clusters_table}.

\begin{table}[h]
\begin{center}
\begin{tabular}{lcccccc} 
         & RA           & DEC            & Date      & UT    & $\times 10^{19}$~eV \\
\hline
\\
C1             &  $1^h$ $15^m$ & $21.1^{\circ}$ & 93/12/03 & 21:32:47 & 21.3 \\
               &  $1^h$ $14^m$ & $20.0^{\circ}$ & 95/10/29 & 00:32:16 & 5.07 \\
\\
C2             & $11^h$ $29^m$ & $57.1^{\circ}$ & 92/08/01 & 13:00:47 & 5.50 \\
               & $11^h$ $14^m$ & $57.6^{\circ}$ & 95/01/26 & 03:27:16 & 7.76 \\
               & $11^h$ $13^m$ & $56.0^{\circ}$ & 98/04/04 & 20:07:03 & 5.35 \\
\\
C3             & $18^h$ $59^m$ & $47.8^{\circ}$ & 91/04/20 & 08:24:49 & 4.35 \\
               & $18^h$ $45^m$ & $48.3^{\circ}$ & 94/07/06 & 20:34:54 & 13.4 \\
\\
C4             &  $4^h$ $38^m$ & $30.1^{\circ}$ & 86/01/05 & 19:31:03 & 5.47 \\
               &  $4^h$ $41^m$ & $29.9^{\circ}$ & 95/11/15 & 04:27:45 & 4.89 \\
\\
C5$^{\dagger}$ & $16^h$ $06^m$ & $23.0^{\circ}$ & 96/01/11 & 09:01:21 & 14.4 \\
               & $15^h$ $58^m$ & $23.7^{\circ}$ & 97/04/10 & 02:48:48 & 3.89 \\
\\
C6             & $14^h$ $17^m$ & $37.7^{\circ}$ & 96/12/24 & 07:36:36 & 4.97 \\
               & $14^h$ $08^m$ & $37.1^{\circ}$ & 00/05/26 & 18:38:16 & 4.89 \\
\\
C7$^{\ddagger}$&  $3^h$ $45^m$ & $44.9^{\circ}$ & 98/10/27 & 00:45:37 & 6.11 \\
            &  $3^h$ $41^m$ & $46.6^{\circ}$ &  unknown &  unknown & unknown \\
\\
\hline
\end{tabular}
\caption{\label{agasa_clusters_table} Clusters of cosmic rays observed
  in a set of 59 events collected by AGASA. Equatorial coordinates
  have been taken from \citeasnoun{Hayashida-2000-astroph-0008102} and
  Figure~\ref{agasa_clusters_2003}. $^{\dagger}$Cluster C5, consisting
  of one event with energy below the typical energy threshold of $4
  \times 10^{19}$~eV was generally omitted in later publications.
  $^{\ddagger}$The arrival time and energy of the second event
  comprising C7 has not been made public.}
\end{center}
\end{table}

At the 27$th$ ICRC, AGASA reported $P_{ch} = 0.05\%$ for doublets and
$P_{ch} = 1.66\%$ for the triplet. The results of scans of the
threshold energy and angular separation criteria were presented. The
original selection criteria that energies exceed $>4 \times
10^{19}$~eV and angular separations be less than $2.5^{\circ}$ were
found to be close to the values giving the most significant clustering
signal. Also, the point spread function --- representing the observed
distribution of events from a point-source, due to detector resolution
effects --- was used to demonstrate that these clusters are consistent
with pointlike sources at AGASA resolving power. Integral energy
spectra were shown for all AGASA events and events comprising the
clusters. The clusters are shown to have a relatively hard spectrum
with index $-0.8 \pm 0.5 ({\rm stat}) \pm 0.5 ({\rm syst})$.

The most recent update of the cluster analysis by AGASA was presented
by \citeasnoun{Teshima-2003-ICRC-28-437} at the 28$th$ ICRC. This
analysis featured the same 8 pairs ($8 = 1 \times 3 + 5 \times 1$) as
previous analyses (C5 is omitted), and the calculated background was
reported as 1.7 pairs. It was also reported that simulation studies
indicate a chance probability of observing 8 pairs when 1.7 are
expected is $< 10^{-4}$. This is below the Poisson fluctuation
probability of 0.0003. Two-dimensional autocorrelation plots --- scans
for excesses over separation angles --- were also presented, with some
correlation of the two coordinates at the $3 \sigma$ level indicating
possible magnetic field deflections consistent with $B_z \sim 0.3
\mu$G in the galactic halo.

\subsubsection{Independent Commentary on AGASA Clusters}
\label{sec-agasa-independent}

Before turning to comparisons of the AGASA clustering signal with
independent experimental measurements, we acknowledge the ongoing
debate about the significance of the AGASA result by considering two
articles which attempt independent statistical evaluations.  Both
papers express concern that the choice of energy threshold (and in one
case, the angular separation) cut used by AGASA were chosen {\em a
  posteriori} in order to maximize the significance of the observed
signal.

\begin{figure}
  \begin{minipage}[t]{0.49\columnwidth}
    \includegraphics[width=\columnwidth]{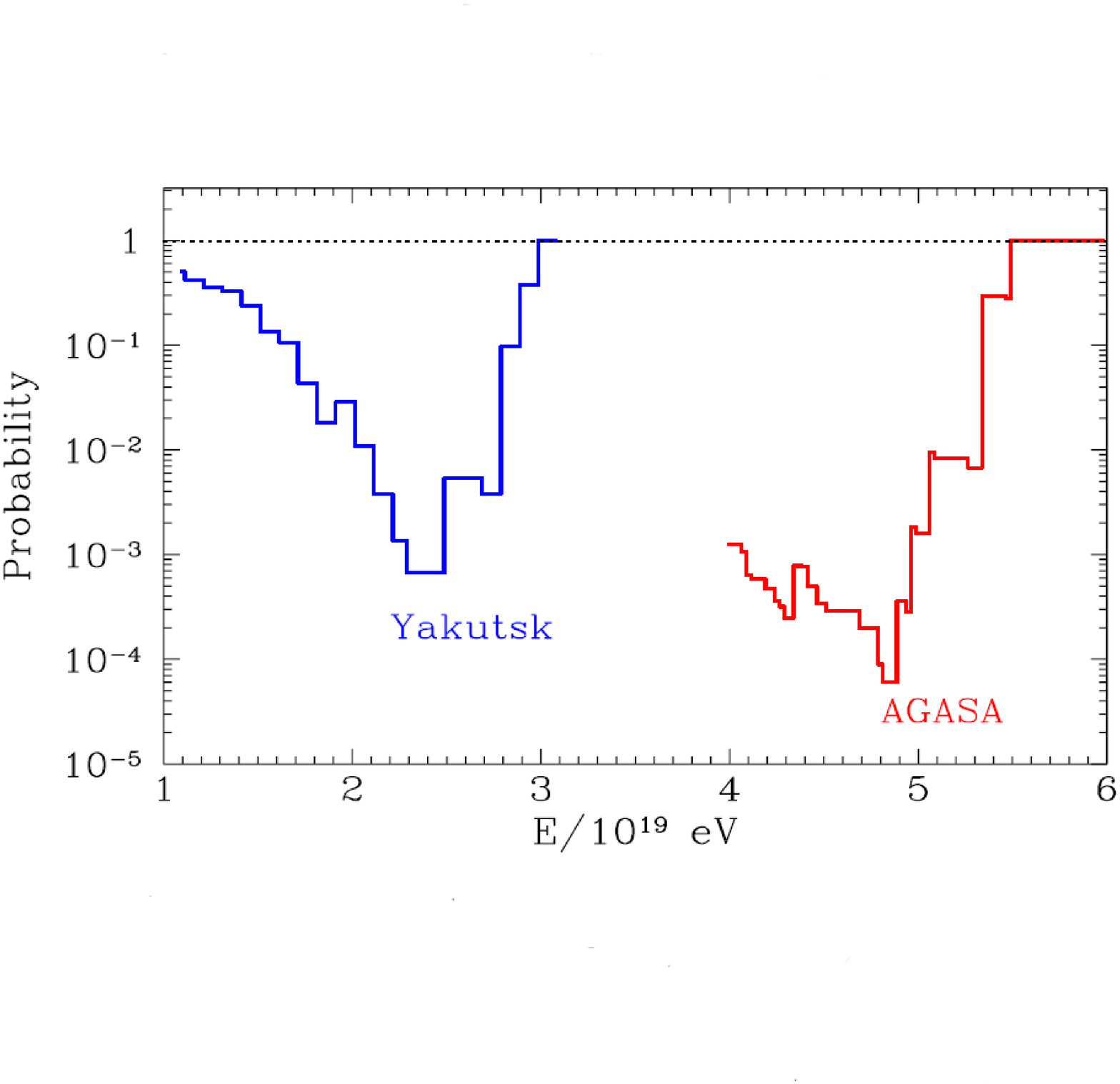}
  \end{minipage}
  \begin{minipage}[t]{0.49\columnwidth}
    \includegraphics[width=\columnwidth]{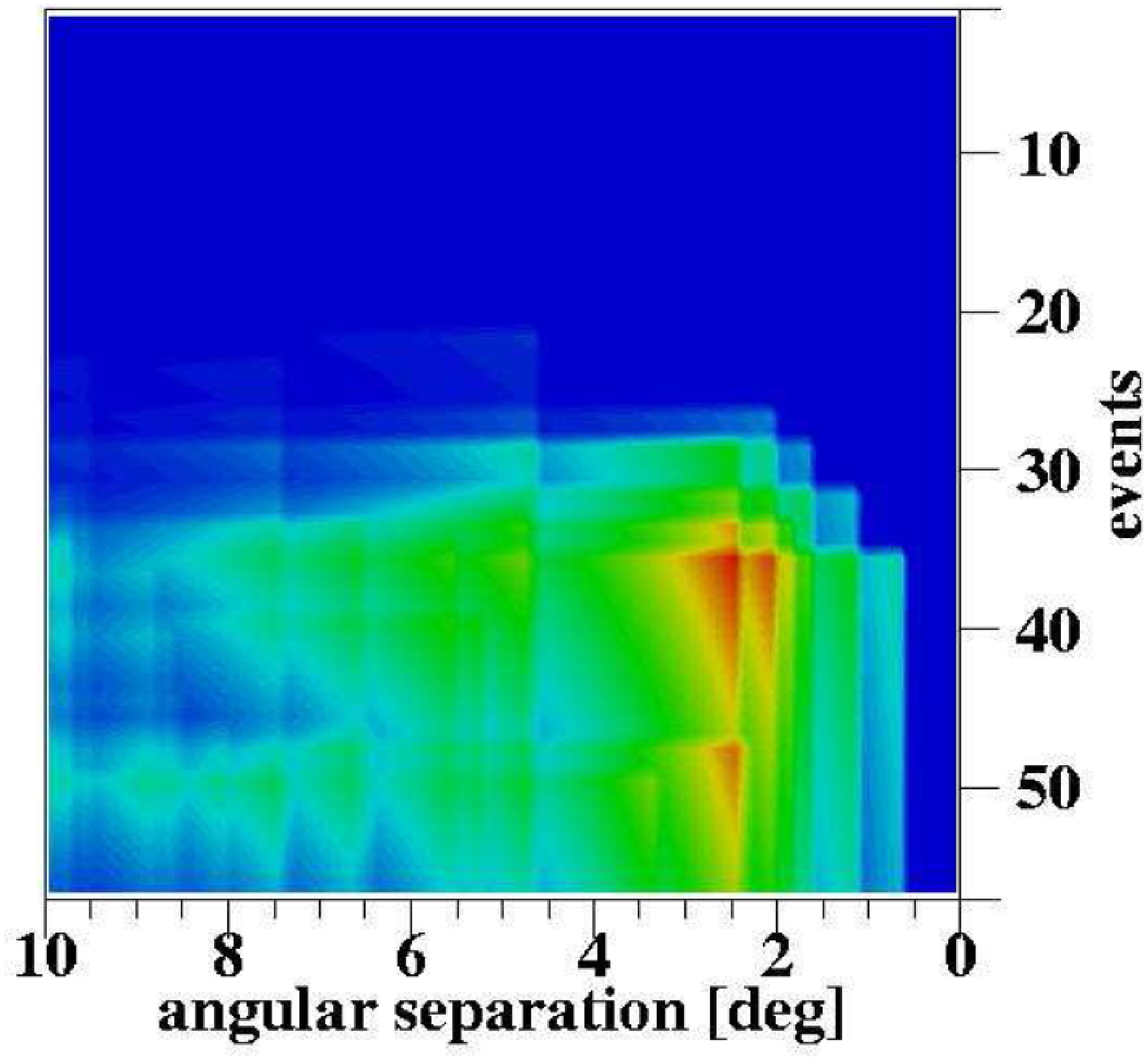}
  \end{minipage}
  \caption{Left: Scan of chance probability of observing the same or a
    greater number of pairs as seen in the Yakutsk and AGASA
    \cite{Hayashida-2000-astroph-0008102} data as a function of energy
    threshold, for energies exceeding $4 \times 10^{19}$~eV.
    \cite{Tinyakov-Tkachev-2001-JETPL-74-1}.  Right: Scan of chance
    probability of observing the same or a greater number of pairs as
    seen in the AGASA data as a function of both energy threshold
    (event number is assigned by energy ordering) and angular
    separation cut \cite{Finley-Westerhoff-2004-APP-21-359}.  Event
    number 36 (the cutoff for the minimum chance probability)
    corresponds to an energy cut of $4.89 \times 10^{19}$~eV. The
    color scale is logarithmic, with red corresponding to chance
    probabilities of order few $\times 10^{-4}$. In both scans, the
    ``triplet'' C2 is treated as three doublets. }
  \label{cluster_scans}
\end{figure}

\citeasnoun{Tinyakov-Tkachev-2001-JETPL-74-1} perform a reanalysis of
the data above $4 \times 10^{19}$~eV released in
\citeasnoun{Hayashida-2000-astroph-0008102}, consisting of four
doublets and one triplet. (Clusters C5 and C7 from
Table~\ref{agasa_clusters_table} are not included in their analysis.)
Taking the angular cut of $2.5^{\circ}$ as valid based on AGASA's
reported angular resolution
\cite{Hayashida-1996-PRL-77-1000,Takeda-1999-ApJ-522-225} they perform
a scan over energy thresholds to determine the cutoff with the lowest
chance probability.  Since this scan is performed systematically, it
can be used to estimate the statistical penalty to be paid for the
{\em de facto}, but heretofore unacknowledged, scan used by AGASA in
placing their energy cutoff at $4 \times 10^{19}$~eV. The results of
this scan are shown in Figure~\ref{cluster_scans}. The minimum chance
probability is $< 10^{-4}$, however the authors estimate a scanning
penalty factor of 3 and thus conclude that the chance probability of
the AGASA result is approximately $3 \times 10^{-4}$.

\citeasnoun{Finley-Westerhoff-2004-APP-21-359} take the additional
step of questioning whether the AGASA angular separation cut was {\em
  a priori} justified. The authors point out that the AGASA angular
uncertainty is energy-dependent (see
\citeasnoun{Takeda-1999-ApJ-522-225} and
Figure~\ref{resolution_figs}), varying from $2.0^{\circ}$ at
$10^{19.5}$ to $1.2^{\circ}$ at $10^{20}$~eV for the angular size of
the 68\% ($1 \sigma$) error circles. Therefore the choice of
$2.5^{\circ}$ as the cutoff was probably chosen so as to maximize the
signal in the original AGASA clustering paper. The authors chose to
scan over both energy and angular cuts (Figure~\ref{cluster_scans}) in
order to estimate the true significance of the result. While the
results of the scan are consistent with the ``most significant
energy'' threshold reported by
\citeasnoun{Tinyakov-Tkachev-2001-JETPL-74-1}, the penalty for
scanning over angles results in a chance probability of 0.35\%,
approximately 10~times larger.

As an additional check on the true sensitivity of the result,
\citeasnoun{Finley-Westerhoff-2004-APP-21-359} point out that it is
appropriate to perform an independent analysis of AGASA data taken
after that used in tuning the energy and angular cuts, {\em i.e.} that
data for which the $2.5^{\circ}$ and $4 \times 10^{19}$~eV cuts can be
regarded as {\em a priori}. Including just events observed after
October 1995 (the end date for events included in the original cluster
publication) just one pair is found, with a chance probability of
28\%. Since doublet C4 contains one event on either side of the
October 1995 cutoff, they then analyze the chance probability in a way
that allows for cross-correlations between the two data sets. Allowing
for this, the chance probability is 8\%.

\subsubsection{Search for Clustering Above $4 \times 10^{19}$~eV in Other
Experiments} 

It is clear that in order to resolve questions about the validity of
the AGASA cluster claims, confirmation or refutation by an independent
experiment is required. A data set with statistics comparable to the
AGASA sample was obtained by the Yakutsk array over 30 years of
running. Through 2003, Yakutsk detected a total of 29 extensive air
showers with energy above $4 \times 10^{19}$~eV. Two doublets are
found in this data, with distance between the showers less than
$5^{\circ}$. (The angular resolution of the Yakutsk array data is
$3^{\circ}$ (Section~\ref{sec-anis-resolution}) for airshowers at
these energies.  The chance probability of these clusters is estimated
to be 10\% \cite{Mikhailov-2004-RCRC-28-1306}, which we interpret as
being consistent with isotropy.

\cite{Tinyakov-Tkachev-2001-JETPL-74-1} include Yakutsk data with
energies above $1 \times 10^{19}$~eV in their reanalysis, described
above in Section~\ref{sec-agasa-independent} and
Figure~\ref{cluster_scans}.  They find 8 doubles (pairs of events
within $4^{\circ}$) out of 26 events above a threshold energy of $2.4
\times 10^{19}$~eV. Accounting for the scanning statistical penalty,
they claim a chance probability of 0.002 for this clustering signal.

\begin{figure}[h]
  \begin{center}
    \includegraphics[width=1.00\columnwidth]{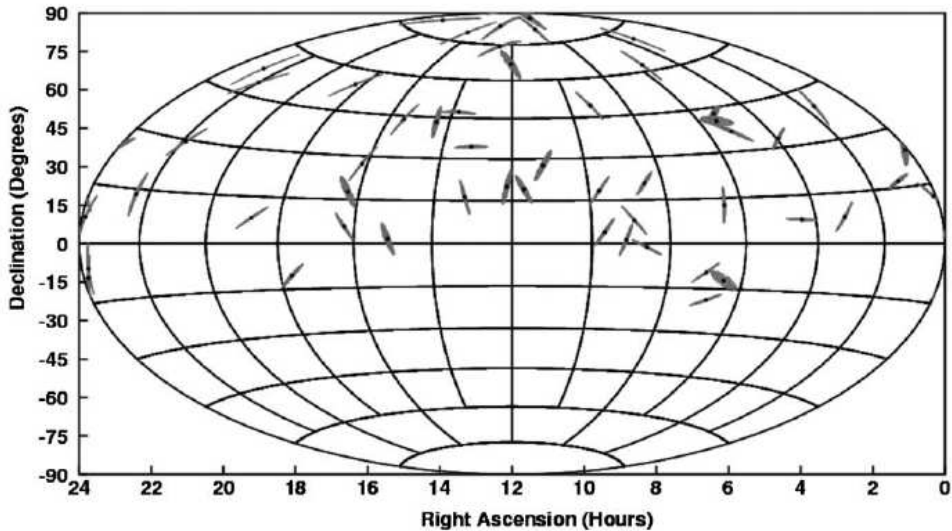}
  \end{center}
  \caption{\label{hires_mono_skymap_2004} The arrival directions of
    the HiRes--I monocular events with reconstructed energies above
    $10^{19.5}$~eV, and their $1\sigma$ angular resolution. Figure
    from \cite{Abbasi-2004-APP-22-139}.}
\end{figure}

As of this writing, the data set collected by HiRes contains the best
available statistics above $10^{19}$~eV.
\citeasnoun{Abbasi-2004-APP-22-139} report on a search for arrival
direction clustering in the HiRes--I monocular dataset above
$10^{19.5}$~eV (Figure~\ref{hires_mono_skymap_2004}). The data
collected through 2003 contains 52 events with energy greater than
$10^{19.5}$~eV, comparable in size to the data set used to establish
the AGASA cluster claims.

\begin{figure}
  \begin{minipage}[t]{0.49\columnwidth}
    \includegraphics[width=\columnwidth]{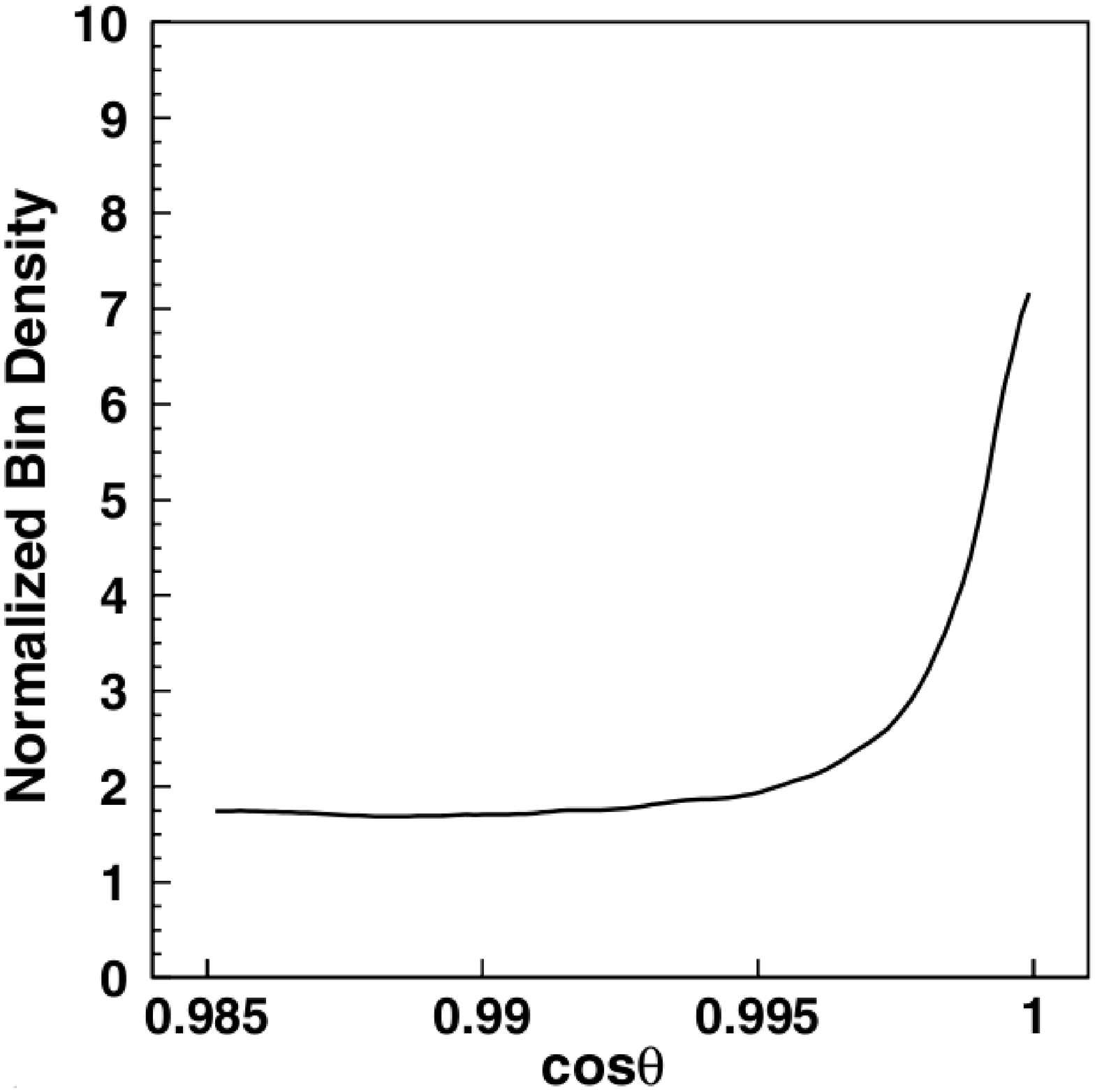}
  \end{minipage}
  \begin{minipage}[t]{0.49\columnwidth}
    \includegraphics[width=\columnwidth]{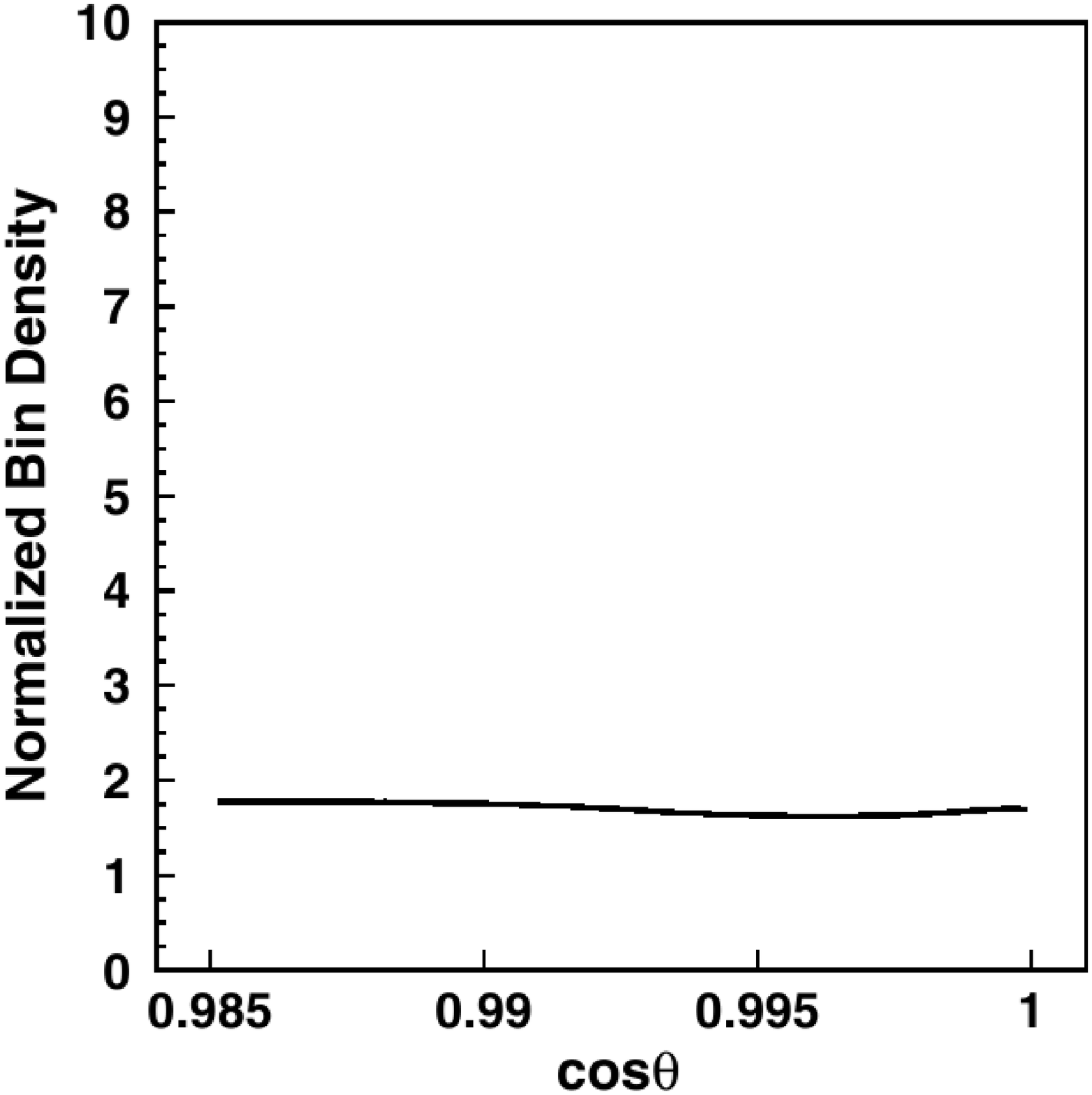}
  \end{minipage}
  \caption{Left: Autocorrelation function for AGASA data above
    $4\times10^{19}$~eV \cite{Hayashida-2000-astroph-0008102}.  Right:
    Autocorrelation function for HiRes--I monocular data above
    $10^{19.5}$~eV. The upturn of the autocorrelation function at
    small opening angles is indicative of arrival direction
    clustering. Both analyses performed in
    \citeasnoun{Abbasi-2004-APP-22-139}.}
  \label{hires_autocorr}
\end{figure}

This analysis is done by calculating an autocorrelation function,
effectively a scan over event separation angles. In the monocular
analysis, no scan is performed over energy threshold. Rather, the
threshold is fixed at $10^{19.5}$~eV. As in the full-sky dipole
analysis described above, each event is represented by a set of points
distributed according to the angular resolution functions
(Equations~\ref{eq-sdp} and \ref{eq-psi}) in order to account for the
effects of asymmetric error ellipses. For each point for a given
event, the cosine of the opening angle with every point from all other
events in the skymap is histogrammed. The results are shown in
Figure~\ref{hires_autocorr}.

The significance of this result is established by comparison of the
mean value of the $\cos{\theta}$ distribution for $\theta <
10^{\circ}$ with the same quantity for simulated datasets with varying
numbers of clusters. The authors conclude that the HiRes--I data is
consistent with no arrival direction clustering, and place a 90\% c.l.
upper limit of 3.5 doublets above background for the HiRes--I
monocular data set.

\begin{figure}[h]
  \begin{center}
    \includegraphics[width=1.00\columnwidth]{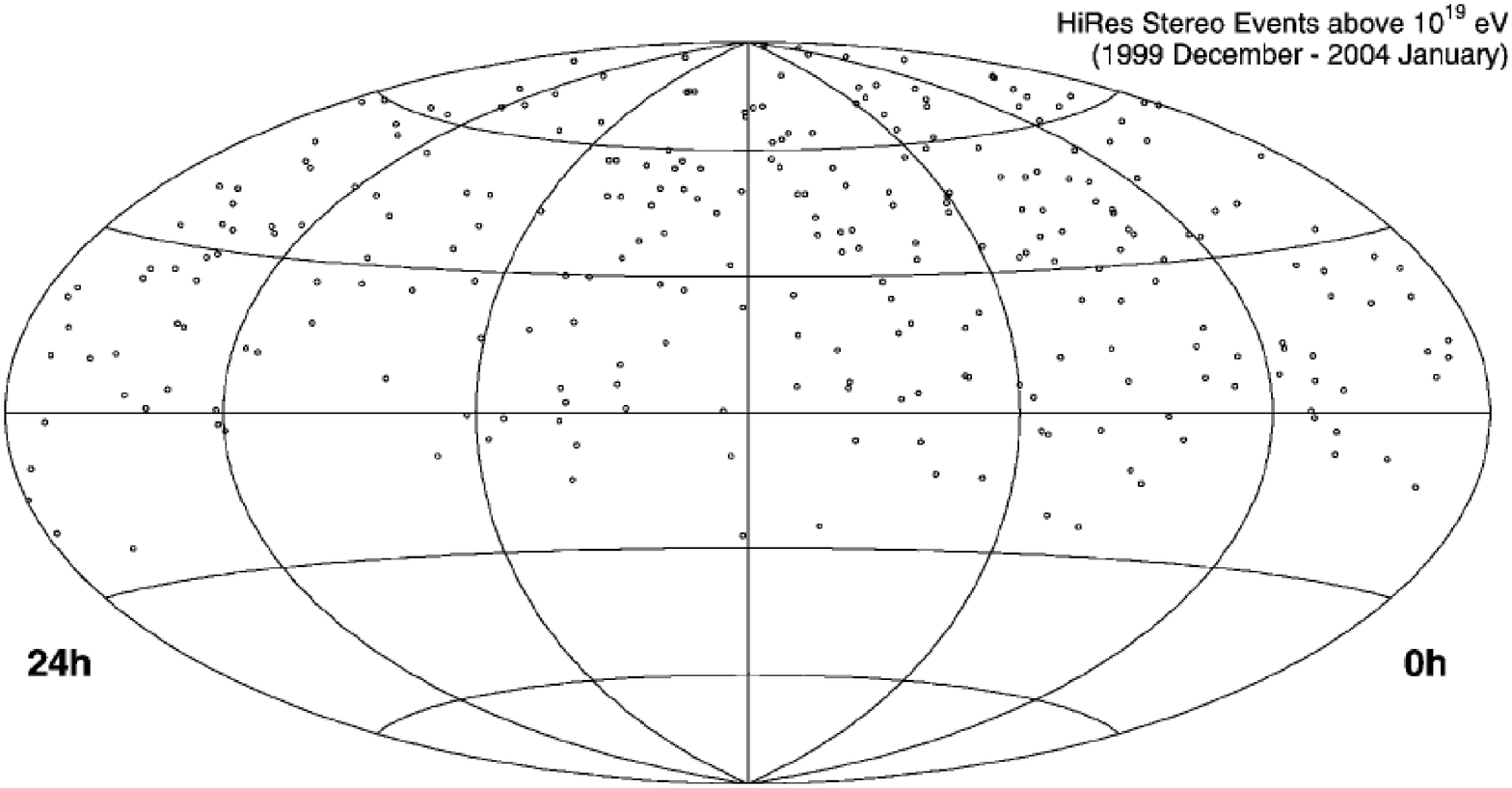}
  \end{center}
  \caption{\label{hires_stereo_skymap_2004} Skymap (equatorial
    coordinates) of 271 HiRes stereo events above $10^{19}$~eV. Shown
    is the typical error radius of $0.6^{\circ}$. From
    \citeasnoun{Abbasi-2004-ApJ-610-L73}.}
\end{figure}

The HiRes stereo data set, while statistically smaller than the
monocular data set, features significantly smaller, symmetric
uncertainties in event arrival direction
(Figure~\ref{hires_stereo_skymap_2004}). The search for arrival
direction clustering in the HiRes stereo data was carried out by
\citeasnoun{Abbasi-2004-ApJ-610-L73}, using a procedure essentially
identical to the energy--opening angle scan conducted by
\citeasnoun{Finley-Westerhoff-2004-APP-21-359}
(Section~\ref{sec-agasa-independent}).

\begin{figure}[h]
  \begin{center}
    \includegraphics[width=0.67\columnwidth]{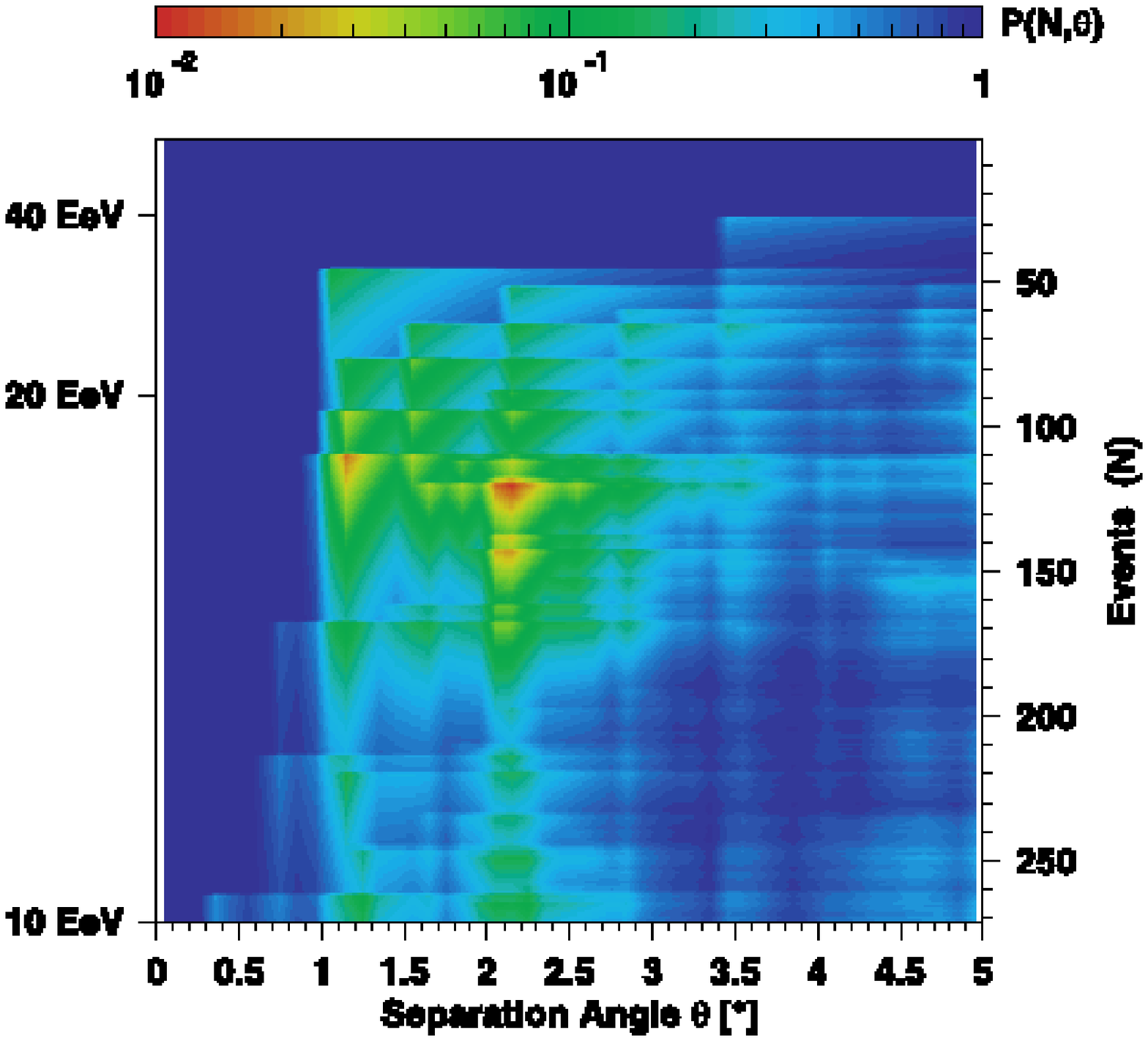}
  \end{center}
  \caption{\label{hires-stereo-scan} Autocorrelation scan of the HiRes
    data set above $10^{19}$~eV. $P(N,\theta)$ is the probability of
    obtaining the same or greater number of pairs as is actually
    observed in the data using a maximum separation angle $\theta$ and
    searching among the $N$ highest energy events. These probabilities
    do not include the statistical penalty due to scanning. From
    \citeasnoun{Abbasi-2004-ApJ-610-L73}.}
\end{figure}

The strongest clustering signal, $P_{min} = 1.9\%$, was observed at an
energy threshold $E_c = 1.69 \times 10^{19}$~eV
(Figure~\ref{hires-stereo-scan}). The signal observed corresponded to
$n_p = 10$~pairs separated by less than $\theta_c = 2.2^{\circ}$
within a set of $N_c = 120$ events. Taking into account the
statistical penalty for performing the scan, a statistical
significance $P_{ch} = 52\%$ is attached to the clustering signal in
the HiRes stereo data. The authors conclude their data is consistent
with the null hypothesis for arrival direction clustering above that
expected by chance.

We conclude our discussion of tests of the AGASA clustering signal
with a noteworthy AGASA/HiRes stereo combined data set analysis.
\cite{Abbasi-2005-ApJ-623-164} report the results of an unbinned
maximum likelihood analysis of the combined data sets above $4 \times
10^{19}$~eV. The method is as follows:

\begin{figure}[h]
  \begin{center}
    \includegraphics[width=1.00\columnwidth]{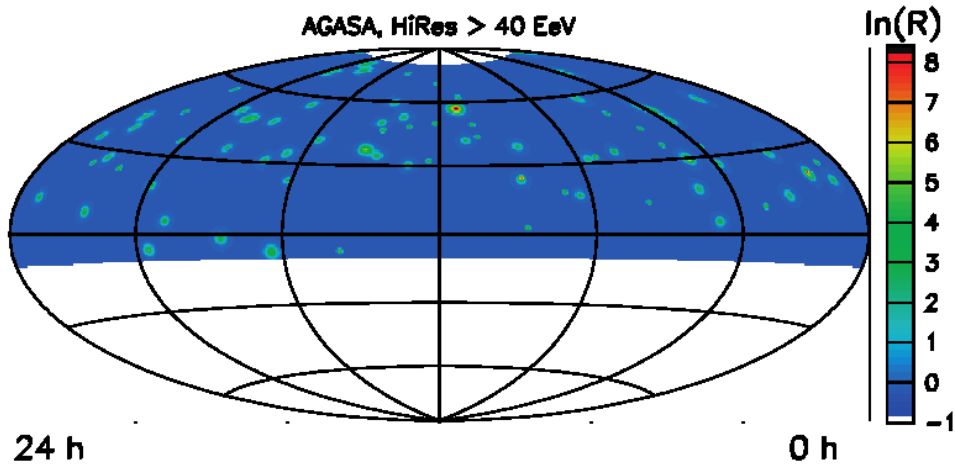}
  \end{center}
  \caption{\label{hires-agasa-maxlikelihood} Likelihood ratio $\ln
    {\cal R}$, maximized with respect to $n_s$ (number of source
    events), as a function of RA and DEC of the source position for
    the combined set of AGASA and HiRes events in excess of $4\times
    10^{19}$~eV. Local maxima occur wherever events or clusters of
    evens are located on the sky. The global maximum, {\em i.e.} the
    most likely position of a point-source is at RA = $169.3^{\circ}$
    and DEC = $57.0^{\circ}$  From~\cite{Abbasi-2005-ApJ-623-164}.}
\end{figure}

Define $Q_i(x_i,x_s)$ as the probability for an event observed at
coordinate $x_i$ to have a true arrival direction at $x_s$, and
$R_i(x)$ as the probability distribution for the event to be observed
anywhere in the sky. $R_i$ depends on the detector acceptance and
exposure. The probability associated with a given event, under the
point source hypothesis, is the weighted sum $P_i$ of the source and
background probabilities.
\begin{equation}
P_i(x,x_s) = \frac{n_s}{N}Q_i(x,x_s) + \frac{N-n_s}{N}R_i(x)
\end{equation}
The product of the $P_i$ for all events gives the likelihood $L$ for a
particular choice of the number of source events $n_s$. The best
estimate for $n_s$ is the value which maximizes $L$:
\begin{equation}
L(n_s,x_s) = \prod^{N}_{i=1} P_i(x,x_s)
\end{equation}
The authors actually maximize $\ln {\cal R}$, the log of the ratio of
the likelihood of $n_s$ relative to the likelihood of the null
hypothesis $n_s = 0$
\begin{equation}
\ln {\cal R} = \ln{\frac{L(n_s,x_s)}{L(0,x_s)}} 
\end{equation}
$\ln{\cal R}$ is the measure of deviation from the null hypothesis of
no source events.

Figure~\ref{hires-agasa-maxlikelihood} shows the distribution of the
likelihood ratio $\ln{\cal R}$ for the combined AGASA/HiRes stereo
data sets. The significance of fluctuations in this skymap is is
determined by scanning over Monte Carlo data sets and counting the
fraction with $\ln{\cal R}_{MC} > \ln{\cal R}_{DATA}$. The highest
value of $\ln{\cal R} = 8.54$, corresponding to $N_S = 2.9$, is at the
location of the AGASA triplet. The fraction of Monte Carlo sets with
greater $\ln{R}$ is 28\%. The authors conclude that no significant
point source is found in the combined set of HiRes stereo and AGASA
events above 40~EeV.

\subsubsection{Search for Pointlike Excesses at Lower Energy}
\label{sec-pointlike_low_e}

Although at energies substantially lower than $10^{19}$~eV magnetic
deflections are expected to diffuse arrival directions for charged
particles, there are factors motivating the search for pointlike
excesses. In a reanalysis of data from the SUGAR array designed to
test the apparent AGASA excess in this region
(Section~\ref{sec-anis-dipole}), \citeasnoun{Bellido-2001-APP-15-167}
observed evidence for a point source $7.5^{\circ}$ from the galactic
center. Motivated in part by this observation, several theoretical
models have predicted fluxes of neutrons from compact sources, in
particular the galactic center
\cite{Medina-Tanco-2001-ICRC-27-531,Bossa-2003-JPG-29-1409,Biermann-2004-ApJ-604-L29,Crocker-2005-ApJ-622-892,Aharonian-2005-ApJ-619-306,Grasso-2005-APP-24-273}.

The data used in the SUGAR reanalysis consists of 3,732 events in the
energy range $10^{17.9} \rightarrow 10^{18.5}$~eV, roughly
corresponding to the range in which AGASA observes their most
significant excess.

\begin{figure}[h]
  \begin{center}
    \includegraphics[width=1.00\columnwidth]{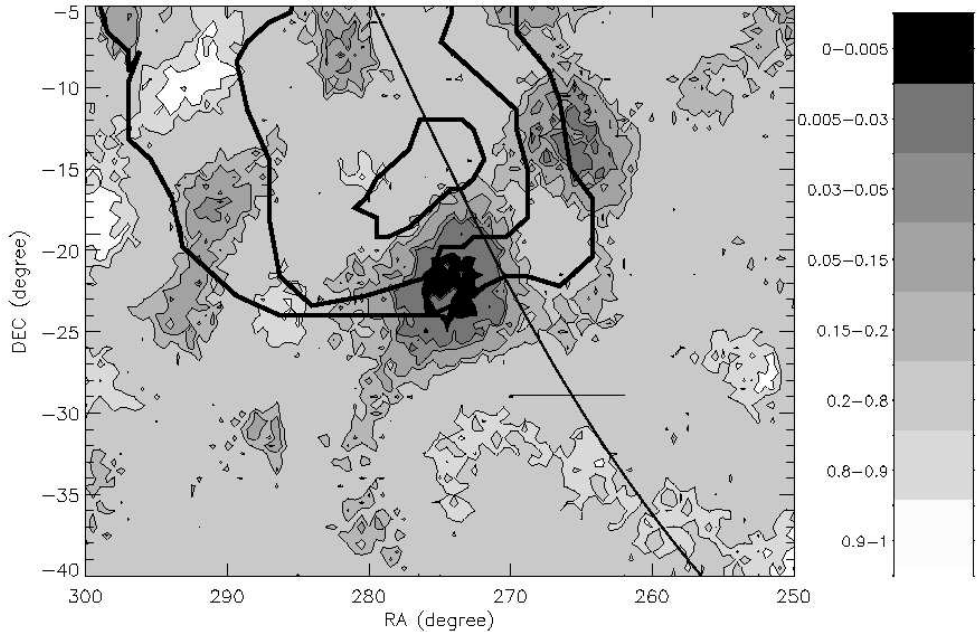}
  \end{center}
  \caption{\label{sugar-galctr} Significance of excess detected in the
    galactic center region, in reanalysis of SUGAR data
    \cite{Bellido-2001-APP-15-167}. Contours represent the chance
    probability of SUGAR detecting the observed density or greater.
    The galactic plane and galactic center are indicated by the thin
    curved line and cross, respectively, while the thick lines
    represent $2\sigma$, $3\sigma$ and $4\sigma$ contours from AGASA
    excess \cite{Hayashida-1999-APP-10-303}. The central excess is
    pointlike on the scale of the SUGAR angular resolution.}
\end{figure}

The analysis is carried out using a skymap technique, in which events
are represented on a map by a Gaussian function equal to their
directional uncertainty. The significance of fluctuations is
determined by comparing the skymap for the data with simulated skymaps
generated by the ``shuffling'' technique. In the shuffling technique,
most appropriate for ground array experiments with near-uniform
exposure in right ascension (RA), simulated skymaps are generated by
pairing a real arrival time from one event with a zenith and azimuth
angle from another event in the same set, and repeating until a new
data set is filled. The resultant ``significance map'', in the
vicinity of the galactic center is shown in Figure~\ref{sugar-galctr}.

A localized excess (pointlike within the SUGAR resolution) is observed
in the data, centered at RA $274^{\circ}$ DEC $-22^{\circ}$, close to
the position of the AGASA excess and the galactic center. No excess is
observed from the true center of the galaxy, although SUGAR has a
clear view of this region. The excess has an estimated chance
probability $P_{ch} = 0.5\%$, and corresponds to a point source flux
of $(2.7 \pm 0.9)$~km$^{-2}$yr$^{-1}$ between $10^{17.9}$ and
$10^{18.5}$~eV.

\begin{figure}[h]
  \begin{center}
    \includegraphics[width=0.80\columnwidth]{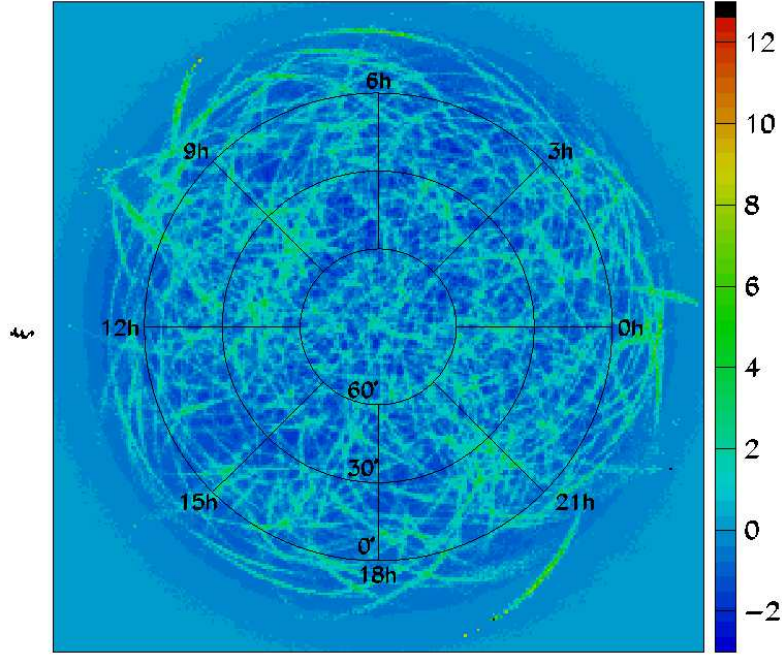}
  \end{center}
  \caption{\label{hires-monopt-xi} Significance map, plotted in polar
    projection, equatorial coordinates, for the HiRes--I monocular
    dataset. The quantity plotted is $\xi = (N_{DATA} - \langle N_{MC}
    \rangle )/\sigma_{MC}$, where each event is represented by an
    asymmetric probability density function (Equations~\ref{eq-sdp}
    and \ref{eq-psi}). From
    \citeasnoun{Abbasi-2007-APP-astro-ph-0507663}.}
\end{figure}

Two recent submissions describe further point source searches, one
each in the northern and southern hemispheres.
\citeasnoun{Abbasi-2007-APP-astro-ph-0507663} describe a point-source
search in northern hemisphere using the HiRes--I monocular data above
$10^{18.5}$~eV. In the HiRes technique, a binned sensitivity map
(Figure~\ref{hires-monopt-xi}) is created from the data (1,525 events)
and Monte Carlo simulated datasets. Each ``event'' in this map is
represented by a probability density function determined from
Equations~\ref{eq-sdp} and \ref{eq-psi}.

To search for point sources, \citeasnoun{Abbasi-2007-APP-astro-ph-0507663}
look for localized excesses by calculating the fraction of
``significant'' bins within a $2.5^{\circ}$ search circle. Simulated
sources are used to tune selection criteria, to maximize sensitivity
to true point sources while rejecting false positives due to
fluctuations in the isotropic background. This tuning is performed
``blind'', {\em i.e.}  without looking at the actual data. The
selection criteria are then applied to the data and no significant
pointlike excesses are found.

HiRes places an upper limit of 0.8~km$^{-2}$yr$^{-1}$ (90 \% c.l.) on
the flux from pointlike sources of hadronic cosmic rays in the
northern hemisphere. This is below the flux reported by the SUGAR
experiment for the pointlike source near the galactic center.

The Auger galactic center studies reported in
\citeasnoun{Abraham-2007-APP-27-244} and described above in
Section~\ref{sec-anis-dipole} also contain a direct test of the
reported SUGAR excess. As shown in Figure~\ref{auger-galctr}, Auger
sees no significant excess in the SUGAR signal region for events with
energy between $10^{17.9}$ and $10^{18.5}$~eV. Using a
``semi-analytic'' technique to evaluate significances, Auger places a
flux limit of $\Phi^{95} = \kappa$ 0.13~km$^{-2}$yr$^{-1}$ (95 \%
c.l.) on point-like neutron sources in the vicinity of the galactic
center, where $\kappa$ is a correction factor for overall flux
normalization: Adopting the HiRes (AGASA) spectral normalization at
3~EeV corresponds to $\kappa = 1.2$ ($\kappa = 2$). In any case, this
number is well below the point-like excess flux reported by SUGAR.

\subsubsection{Correlations of UHECR with BL-Lacertae Objects}

BL-Lacertae (BL-Lac) objects are a class of blazars, active galactic
nuclei in which the jet axis points towards the Earth. It is known
that blazars are sources of gamma rays with energies ranging from $>
100$~MeV \cite{Hartman-1999-ApJS-123-79} to over a TeV
\cite{Aharonian-2006-AA-455-461}. Such high-energy photons are
possible byproducts of UHECR acceleration, and hence BL-Lac objects
are an obvious choice for correlation studies with UHECRs.

Over 1,000 BL-Lac objects have been catalogued as of this writing. See
for example \citeasnoun{Veron-2006-AA-455-773}. This is far greater
than the world sample of UHECRs which have been observed with
sufficient energy that their arrival directions might be correlated
with distant point-like objects: Below a few $\times 10^{19}$~eV,
trajectories of charged particles are expected to be significantly
perturbed by interactions with galactic and extragalactic magnetic
fields, obscuring the primary cosmic ray's point of origin.

Such a wealth of potential sources comes with a potential trap,
however. By choosing subsets of BL-Lac source objects and varying cut
criteria on UHECR events, one may readily find correlations but at the
same time accumulate sampling penalties which make the true
significance difficult or impossible to know. The authors of the
earliest papers reporting correlations of BL-Lacs with UHECRs from
AGASA and Yakutsk datasets
\cite{Tinyakov-Tkachev-2001-JETPL-74-445,Tinyakov-Tkachev-2002-APP-18-165,Gorbunov-2002-APJ-577-L93}
perform such explicit tuning.  \citeasnoun{Evans-2003-PRD-67-103005}
attempts to recreate these analyses while applying the full
statistical penalty and concludes that the observed signal was
``entirely due to selection effects'', a claim that
\citeasnoun{Tinyakov-Tkachev-2004-PRD-69-128301} refute.  In any case,
reanalysis of statistically independent data sets from Haverah Park
and Volcano Ranch \cite{Torres-2003-ApJ-595-L13} fail to support the
earliest claims, as does the HiRes stereo data above 24~EeV
\cite{Abbasi-2006-ApJ-636-680}.

The most recent claims have been made with 271 events above 10~EeV in
the HiRes stereo sample. \citeasnoun{Gorbunov-2004-JETPL-80-145}
compare these events to 157 BL Lacs from the 10th Veron Catalog
\cite{Veron-2000-ESO-19,Veron-2001-AA-374-92} having optical magnitude
$m<18$. In a $0.8^{\circ}$ binned analysis, they find 10 BL-Lac/UHECR
pairs with a chance probability of 0.001.  An unbinned maximum
likelihood analysis of the same data by the HiRes collaboration
\cite{Abbasi-2006-ApJ-636-680} finds a correlation with a 0.0002
chance probability.

Due to the energy threshold used in this analysis, primary cosmic rays
originating from distant QSOs would almost certainly need to be
neutral to maintain their arrival-direction correlation after passing
through galactic and extragalactic magnetic fields. But neutrons
possess a lifetime of $3 \times 10^{12}$ seconds at 10~EeV and so
cannot have originated more than 100~kpc from Earth. The mean-free
path of photons at these energies is of order a few Mpc. Thus these
results, if true, are particularly intriguing. We await the results of
statistically independent comparisons from the next generation of
UHECR observatories.

\subsection{Summary of Anisotropy Results}
\label{sec-anis-summ}
Over the past decade, the search for anisotropy in the arrival
directions of the highest energy cosmic rays has been pursued
aggressively. During that time, suggestions of statistically
significant arrival direction excesses on all angular scales have
invariably failed the crucial test of confirmation by independent data
sets. We have not yet arrived at the era of cosmic ray astronomy.

At lower energies (below $10^{19}$~eV), in spite of substantial
statistics there is no confirmed large-scale structure to the cosmic
ray sky. While AGASA has seen evidence for an enhanced flux from the
galactic center, the HiRes and Auger observatories have seen no such
evidence in datasets with comparable event counts. Tantalizing
evidence in the reanalysis of SUGAR data suggested a low-energy
point-like excess near the galactic center, prompting the promulgation
of models featuring compact sources of neutrons.  But the Auger
observatory has set flux limits well below the SUGAR observation level
in the galactic center region, and HiRes has set new constraints on
neutron sources in the Northern Hemisphere.

At higher energies, above a few $\times 10^{19}$~eV, even charged
particles should experience negligible deflections by galactic and
extragalactic magnetic fields. Consequently, the arrival directions of
these primary cosmic rays should maintain their correlation with their
sources, leading to detectable small-scale anisotropies. Exciting
reports of arrival direction clustering by the AGASA experiment have
not been confirmed by HiRes. And evidence for UHECR correlations with
the plausible source candidate BL-Lacertae objects have not persisted
in independent data sets.

One important lesson for the community to take from the past decade of
anisotropy studies is just how susceptible we are to fooling ourselves
with statistics. The more ways one partitions the data, and the more
one tunes signal selection criteria, the more chances one has to
observe a {\em fluctuation} and be tempted to call it an {\em
effect}. We are encouraged to note that a discussion is emerging about
the pitfalls of hidden trials and the proper fair usage of
experimental data, which we hope will improve the efficiency and
credibility of anisotropy studies in the future.

\section*{Acknowledgements}
This work is supported by the US National Science Foundation under
grants PHY-0305516 and PHY-0636162. The authors thank M. Kirn,
L. Scott, B. Stokes, G. Thomson, and G. Yodh for useful discussion and
advice.

\section*{References}
\bibliography{UHECR-spectrum}
\end{document}